\documentclass[a4paper,11pt]{article}
\pdfoutput=1

\usepackage[usenames,dvipsnames,table]{xcolor}
\usepackage[utf8]{inputenc}
\usepackage{jheppub}
\usepackage{comment}
\usepackage{ytableau}
\usepackage{cancel} 
\usepackage{xcolor}
\usepackage{amsmath,bm}
\usepackage{mathtools}
\usepackage{float}
\usepackage{graphicx}
\usepackage{amsmath, amssymb}
\usepackage{youngtab}
\usepackage{changepage}
\usepackage{etoolbox}

\pretocmd{\eqref}{Eq.\ }{}{}

\usepackage{tikz-cd} 
\usepackage{circuitikz}
\usepgfmodule{shapes}
\usetikzlibrary{decorations.pathmorphing}
\usetikzlibrary{decorations.pathreplacing,decorations.markings,snakes}
\usetikzlibrary{backgrounds, arrows,calc,shapes,decorations.pathreplacing, automata,positioning}
\usetikzlibrary {arrows.meta}
\usetikzlibrary{shapes.misc}

\tikzset{cross/.style={cross out, draw=black, minimum size=5*(#1-\pgflinewidth), inner sep=0pt, outer sep=0pt}, 
cross/.default={2pt}}
\tikzset{snake it/.style={decorate, decoration=snake}}
\tikzset{mid arrow/.style={postaction={decorate,decoration={
        markings,
        mark = at position .55 with {\arrow[#1]{Straight Barb[width=5pt]}}
    }}}}   
\tikzset{mid arrowsm/.style={postaction={decorate,decoration={
        markings,
        mark = at position .55 with {\arrow[#1]{Straight Barb[width=3pt]}}
    }}}}     
\tikzset{middx arrowsm/.style={postaction={decorate,decoration={
        markings,
        mark = at position .7 with {\arrow[#1]{Straight Barb[width=3pt]}}
    }}}}      
\tikzset{midsx arrowsm/.style={postaction={decorate,decoration={
        markings,
        mark = at position .4 with {\arrow[#1]{Straight Barb[width=3pt]}}
    }}}}

\def\CC{{\cal C}}

\def\CH{{\cal H}}

\def\cN{{\cal N}}

\def\cS{{\cal S}}

\def\cT{{\cal T}}

\def\CZ{{\cal Z}}

\def\d{\delta}

\def\r{\rho}

\def\s{\sigma}

\def\D{\Delta}

\newenvironment{conjecture}[1][Conjecture :]{\begin{trivlist}
\item[\hskip \labelsep {\bfseries #1}]}{\end{trivlist}}

\newcommand{\nn}{\nonumber}
\newcommand{\PFS}[1]{\mathcal{Z}_{#1}}
\newcommand{\PFT}[2]{\mathcal{Z}_{T_{#1}^{#2}}}
\newcommand{\intd}{\int \; \text{d}}
\newcommand{\dd}[1]{\bm{\delta}_{#1}}
\newcommand{\DCB}[1]{\Delta^{\rm CB}_{#1}}
\newcommand{\mf}[1]{\mathfrak{#1}}
\newcommand{\TcS}[4]{\cT\left[\mf{su}(#1),\CC_{#2,#3};#4\right]}
\newcommand{\maxp}[1]{\left[1^{#1}\right]}

\title{Breaking bad theories of class $\cS$}

\author[a,b]{Riccardo Comi}
\author[a,b]{Sebastiano Garavaglia}
\author[a,b]{Simone Giacomelli}
\author[a,b]{Sara Pasquetti}
\author[a,b]{Palash Singh}
\affiliation[a]{Dipartimento di Fisica, Università di Milano--Bicocca, Piazza della Scienza 3, I-20126 Milano, Italy}
\affiliation[b]{INFN, Sezione di Milano--Bicocca, Piazza della Scienza 3, I-20126 Milano, Italy}
\emailAdd{r.comi2@campus.unimib.it}
\emailAdd{s.garavaglia18@campus.unimib.it}
\emailAdd{simone.giacomelli@unimib.it}
\emailAdd{sara.pasquetti@gmail.com}
\emailAdd{palash.singh@unimib.it}

\abstract{We study weakly-coupled descriptions/channel decompositions of the 4d $\cN=2$ theories of class $\cS$ of type $\mf{su}(N)$, from the perspective of the 3d $\cN=4$ mirror duals of their circle compactifications. This is a delicate problem when the channel decomposition produces pathological, or bad, 4d configurations that correspond to spheres with non-maximal punctures. The star-shaped quivers, describing the 3d mirrors associated with such bad 4d configurations, are bad 3d $\cN=4$ theories. Leveraging recent results regarding  3d bad theories, we identify a new and interesting family of bad theories, which we coin {\it broken} theories, that naturally arise in this context. Using these broken theories, we develop a systematic and analytic method that determines the generically non-Lagrangian matter sectors and the weakly-coupled gauge groups in such channel decompositions. We understand these weakly-coupled descriptions as emerging dynamically via Higgs mechanisms triggered by operators acquiring vacuum expectation values.}

\begin{document} 

\maketitle

\flushbottom

\section{Introduction}
\label{sec:intro}

Understanding the physics of four-dimensional $\mathcal{N}=2$ supersymmetric theories has been an important problem in the study of quantum field theories over the past couple of decades. An important and expansive family of such theories, obtained by compactifications of 6d $\cN=(2,0)$ SCFTs on punctured Riemann surfaces, are the theories of class $\cS$ \cite{Gaiotto:2009we,Gaiotto:2009hg}. The punctures on the Riemann surface correspond to the locations of certain codimension-two defects in 6d, amongst which a distinguished class is referred to as maximal punctures. These compactifications often lead to intricate infrared (IR) dynamics in 4d, especially when the compactification instead involves non-maximal punctures, which correspond to a more general class of defects.

While these theories are generically strongly coupled, many of them possess weakly-coupled descriptions associated with pair-of-pants decompositions of the Riemann surface. The identification of the gauge symmetries and the (generically non-Lagrangian) matter sector in such weakly-coupled descriptions in the presence of non-maximal punctures is a particularly subtle aspect of the 4d IR dynamics. In particular, many compactifications involving non-maximal punctures give rise to a pathological, or {\it bad}, 4d configuration.\footnote{It is important to note here that these bad configurations arise as the (generically non-Lagrangian) matter sectors in a weakly-coupled description, and are typically not studied in isolation.}

One fruitful strategy to study these (as well as the non-pathological) descriptions is to analyse the structure of the Seiberg--Witten (SW) curves associated with the theory. In particular, the behaviour of the SW differential near the punctures, {\it i.e.}, the structure of its poles, provides key information about the types of punctures and the symmetries that they preserve. The analysis of the spectrum of the Coulomb branch of the 4d $\cN=2$ moduli space obtained from this data led to the development of the tinkertoys program, {\it c.f.} \cite{Chacaltana:2010ks} and references therein. 

Within this framework, one views the theory as being assembled from a discrete set of allowed three-punctured spheres (trinions) viewed as the matter building blocks, with various ``tubes," or gauge groups, connecting them. This provides an atomic or bottom-up geometric approach to building typically strongly coupled 4d $\cN=2$ SCFTs associated to generic punctured Riemann surfaces. This approach attempts to systematise the construction of theories of class $\mathcal{S}$ from basic ingredients, and thus provides insight into the physics of the bad 4d configurations.

An alternative route to understanding the physics of theories involving non-maximal punctures is by studying the renormalisation group (RG) flows connecting these to compactifications involving only maximal punctures \cite{Benini:2009gi,Chacaltana:2012zy}. These are Higgs branch flows triggered by nilpotent VEVs that partially Higgs the flavour symmetries associated to maximal punctures, effectively ``closing" the maximal puncture to generate the non-maximal ones. In principle, one can follow these RG flows to uncover the IR dynamics; however, doing so in practice requires a detailed understanding of the Higgs branch chiral ring relations of the theories with all maximal punctures, which is challenging. Notably, such partial Higgsings were studied in \cite{Gaiotto:2012uq} from the perspective of the superconformal index \cite{Romelsberger:2005eg,Kinney:2005ej}, leading to insights into some bad 4d configurations.

A third approach to analyse these pathological compactifications is from the perspective of the star-shaped quiver associated to the theories of class $\cS$ \cite{Benini:2010uu}. These are 3d $\cN=4$ SCFTs that are mirror dual to the direct circle compactifications of the theories of class $\cS$. It was observed in \cite{Gaiotto:2011xs} that these bad 4d configurations lead to 3d $\cN=4$ {\it bad} SCFTs, in the sense of the good, bad, and ugly trichotomy proposed in \cite{Gaiotto:2008ak}. The 3d bad criterion reflects the existence of monopole operators that violate the unitarity bound and thus decouple from the theory in the IR.\footnote{Note that the 3d monopole operators describe the Coulomb branch of the 3d mirror, and hence correspond to 4d Higgs branch operators.} This approach was also explored in \cite{Nanopoulos:2010bv}, where the badness of the 3d mirror is cured by the addition of extra matter. This leads to a good 3d theory that was then used to probe some properties of the original SCFT.

In this work, we follow this third approach, leveraging the recent progress that has helped further our understanding of 3d $\mathcal{N}=4$ bad theories \cite{Giacomelli:2023zkk, Giacomelli:2024laq}. While this has been a long-standing and challenging problem that has been addressed from various perspectives over the years \cite{Gaiotto:2008sa, Nanopoulos:2010bv, Gaiotto:2011xs, Gaiotto:2012uq, Kim:2012uz, Yaakov:2013fza, Bashkirov:2013dda, Bullimore:2015lsa, Hwang:2015wna, Hwang:2017kmk, Assel:2017jgo, Assel:2018exy, Bourget:2021jwo}, the analysis of their partition functions in \cite{Giacomelli:2023zkk,Giacomelli:2024laq} has revealed an intricate structure. In particular, it was shown that the partition functions of bad theories are a sum of distributions supported on specific loci on the Fayet--Illiopoulos (FI) parameter space, where the IR physics corresponds to interacting good SCFTs accompanied by decoupled free sectors. Thus, the partition function of bad theories can be interpreted as a sum of IR {\it frames}, each corresponding to a particular IR SCFT, free sector, and parameter space locus associated with an {\it a priori} different vacuum.

Physically, the appearance of the Dirac delta distributions can be interpreted as certain monopole operators acquiring VEVs, spontaneously breaking the UV global symmetries and selecting a specific IR frame. Even though these results are shown at the level of the $S^3_b$ partition function \cite{Hama:2011ea}, we naturally expect similar results for other 3-manifolds on which we can perform supersymmetric localisation, such as the superconformal index \cite{Bhattacharya:2008zy}. One further expects that the full path integral of the theory should take the form of a functional distribution in a sense that generalises the notable result for a pure $U(1)$ gauge theory \cite{Kapustin:1999ha}. 

In this paper, we identify a novel interesting class of bad theories, that we name {\it broken theories}, which exhibit a special structure: their partition function is a distributional sum over identical frames, {\it i.e.}, every frame describes the same good SCFT in the IR, the same constraints from the Dirac deltas, and the same chiral multiplets. Therefore, broken theories possess many copies of the same vacuum that are reached by a collection of monopole operators, which transform non-trivially under the UV global symmetry, acquiring a VEV. The multiplicity of the identical frames has a natural interpretation as the many equivalent ways for the monopoles to acquire the VEV, which are all related by global symmetry transformations.

Therefore, a broken theory possesses a ``single vacuum" up to the action of the global symmetry, which distinguishes it from a general bad theory that typically contains many inequivalent vacua. This immediately implies that when the global symmetry of a broken theory is gauged, the VEV spontaneously Higgses the gauge symmetry, resulting in a good theory in the IR. This is once again a non-trivial characteristic of broken theories that distinguishes them from a generic bad theory, which does not result in a good theory when gauged.

As an infinite class of examples, we conjecture that the 3d mirrors of pathological/bad 4d class $\cS$ configurations are broken theories. More precisely, the star-shaped quiver associated to the compactification of 6d $\cN=(2,0)$ theory on a sphere with a single maximal puncture, and many (``small enough") non-maximal punctures, as long as the 4d configuration itself is pathological, is a broken 3d $\cN=4$ theory.\footnote{The possibility for the Hall--Littlewood limit of the superconformal index of the 4d $\mathcal{N}=2$ theory to be a distribution was already pointed out in \cite{Gadde:2011uv}.} 

\begin{figure}[H]
    \centering
    \includegraphics[width=1.1\linewidth]{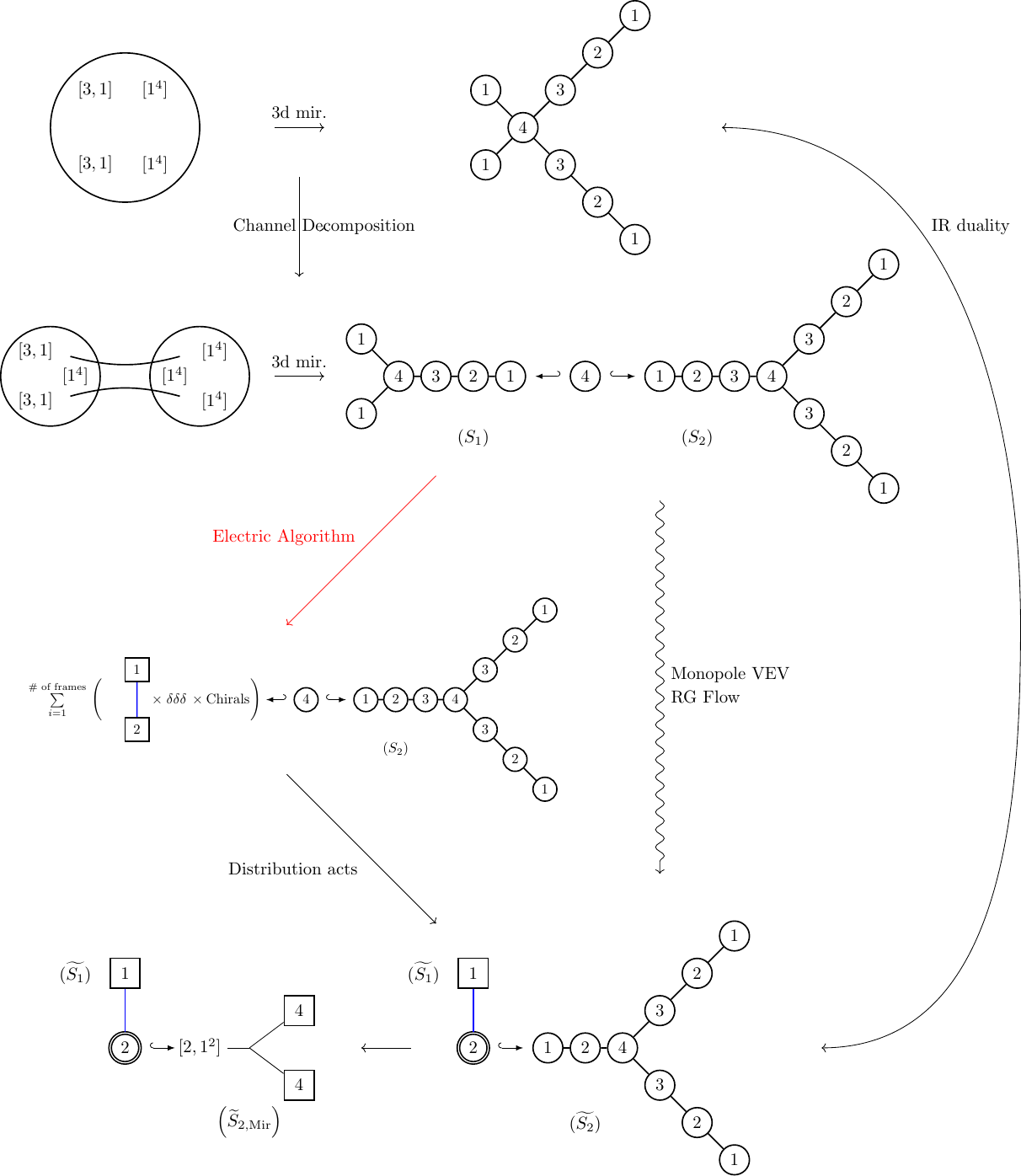}
    \caption{Sketch of the analysis performed on the channel decomposition of the theory of class $\cS$ of type $\mf{su}(4)$ associated to a sphere with two maximal and two minimal punctures. We depict the channel decomposition on line two, wherein the theory $S_1$ is bad. Running the electric algorithm on this produces the intermediate step, depicted on line three. This distribution then acts across the gauging to produce the final line. The result represents the effect of the monopole VEV, acquired in $S_1$, that partially Higgses the $U(4)$ gauge group along with the theory $S_2$. Note that we highlight twisted hypermultiplets, with respect to the hypers in the initial star-shaped theory, in blue instead of black, and depict the gauging of an emergent Coulomb branch symmetry by hook-arrows.}
    \label{fig:exintro}
\end{figure}

Hence, although these 4d configurations are ``bad" or distributional in isolation, they produce a new good theory once they are glued to another good theory via the gauging of the maximal puncture. Conversely, 
these theories can appear in specific channel decompositions of good theories as illustrated in Figure \ref{fig:exintro}.
Our goal is to explain how gluing such distributional theories leads (via Higgsing) to the expected gauge symmetries and the (generically non-Lagrangian) matter sector in such weakly coupled descriptions.

We derive this information from a three-dimensional perspective via a purely dynamical analysis. It is crucial to note here that our analysis provides a complete description of the weakly-coupled theory, arising from a pair-of-pants decomposition involving a bad 4d configuration of type $\mf{su}(N)$, along with the dynamical process that partially Higgses the full $U(N)$ gauge symmetry associated with the maximal puncture along the tube. We carry out this analysis by employing the powerful {\it electric dualisation algorithm}, which allows a detailed analysis of the partition function of a bad quiver gauge theory \cite{Giacomelli:2024laq}.

Before some general comments on our approach, we illustrate our analysis using a concrete example, which will be discussed in full, gory detail in the main text.

\paragraph{An illustrative example:} Consider the 6d $\cN=(2,0)$ theory of type $\mf{su}(4)$ compactified on a sphere with two maximal and two minimal punctures. We postpone the detailed characterisation of such theories and the associated channel decompositions to Sections \ref{sec:classS_review} and \ref{sec:channel_decompSSQ}. We depict this theory by the Riemann surface and the four punctures along with the associated 3d mirror star-shaped quiver in the first line of Figure \ref{fig:exintro}.

There are two equivalent pair-of-pants decompositions for the sphere with two maximal and two minimal punctures. The one that clubs a maximal and a minimal puncture together is well-known to yield a 4d $\cN=2$ $SU(4)$ SQCD with 8 fundamental hypermultiplets (and its 3d counterpart) \cite{Gaiotto:2009we}. We are instead interested in the other channel decomposition that clubs the two minimal punctures together, which is depicted in the second line of Figure \ref{fig:exintro}. Note that the 3d counterpart of this channel decomposition on the star-shaped quiver gives rise to a bad theory, $S_1$, and a good theory, $S_2$, as depicted in Figure \ref{fig:exintro}. These two star-shaped quivers are thus glued together by gauging the diagonal $U(4)$ emergent Coulomb branch symmetry on the two full $T[U(4)]$ tails.\footnote{For technical reasons, we prefer to work with $U(N)$ gauge groups along the tube, rather than $SU(N)$
and  start with the $S_1$ and $S_2$  theories where the redundant diagonal $U(1)$ gauge symmetry still needs to be decoupled. We then mod-out carefully the overall $U(1)$
as we explain in Subsection \ref{subsec:partfun_SSQ}.}

By running the electric dualisation algorithm, we determine that $S_1$ is a broken theory, as its partition function is a distribution containing 12 identical frames. Each individual frame contains a good theory $\widetilde S_1$, which is a single twisted hypermultiplet\footnote{As will be explained in the main text, when electric algorithm produces $\widetilde S_1$ as a free theory, or an interacting theory containing a free sector, then the free hypermultiplet are twisted w.r.t.\ to the hypermultiplets in the original star-shaped quiver. Thus, these correspond to the direct circle reduction of the the 4d hypermultiplets.} in the fundamental of an $SU(2)$ subgroup of the $U(4)$ global symmetry carried by the maximal puncture. Additionally, each frame contains a collection of chiral multiplets, and three Dirac deltas indicating that the monopole operators are acquiring a VEV which breaks $U(4)$ down to $SU(2)$, that the hypermultiplet in $\widetilde S_1$ is charged under.

The propagation of the monopole VEVs across the $U(4)$ gauging into $S_2$ can be tracked as the action of the distributional partition function of $S_1$ on the partition function of $S_2$.\footnote{Upon gauging the result of the electric algorithm onto the theory $S_2$, the multiplicity of the identical frames accounts for the change in the dimension of the Weyl group. Indeed, the ratio between the dimensions of the Weyl group of $U(4)$ and $SU(2)$ is equal to the number of frames, 12.} The result of this propagation is depicted in the last line of Figure \ref{fig:exintro}, wherein the theory $S_1$ is completely Higgsed down to a fundamental $SU(2)$ hypermultiplet. Meanwhile, the $U(4)$ gauge group is partially Higgsed down to $SU(2)$, alongwith the partial Higgsing of the $T[U(4)]$ tail of $S_2$ down to a $T_{[2,1^2]}[U(4)]$ tail. The gauged $SU(2)$ is the diagonal combination of the $SU(2)$ global symmetry of $\widetilde S_1$, and an $SU(2)$ subgroup of the Coulomb branch symmetry on the $T_{[2,1^2]}[U(4)]$ tail.

Going one step further, we can consider the 3d mirror dual of $\widetilde S_2$, which is a non-Lagrangian SCFT (also referred to as $R_{0,4}$ in the language of \cite{Chacaltana:2010ks}). This allows us to read off the direct circle reduction of the theory of class $\cS$ associated to a sphere with two maximal and two minimal punctures in the chosen channel decomposition. This corresponds to a single fundamental $SU(2)$ hypermultiplet (see Footnote 5) glued via an $SU(2)$ gauging to the non-Lagrangian circle reduction of the theory associated to a sphere with two maximal and a regular puncture. This can be considered as a generalisation of the Argyres--Seiberg duality \cite{Argyres:2007cn}. In more general examples, as we will see, the theory $\widetilde S_1$ will be an interacting star-shaped quiver. In such cases, the final step also involves taking the 3d mirror dual of $\widetilde S_1$ to read off the final direct reduction, or electric theory.

Finally, the big arrow in Figure \ref{fig:exintro} indicates an IR duality between the original star-shaped quiver and the final theory. This duality can also be proved backwards, starting from the final theory and obtaining the initial star-shaped quiver. This is achieved by studying the result of the gauging of the emergent Coulomb branch symmetry in the theory $\widetilde S_2$. This operation, in turn, can be analysed by a series of steps involving local mirror dualisations. We explain this in detail for a single example in Subsection \ref{subsec:su3_example}.

\paragraph{Key features/insights and outlook}

The analysis presented in this paper not only reproduces known results, mostly in \cite{Chacaltana:2010ks}, but also offers a systematic framework allowing us to precisely determine the matter sector in each weakly-coupled description. In particular, we can exactly determine the interacting part in $\widetilde S_1$, $\widetilde S_2$ and the representation of the twisted hypermultiplets. Notably, our approach provides the exact map of the fugacities for the global symmetries across each channel decomposition. 

One key advantage of our approach is that it streamlines the analysis of bad configurations corresponding to spheres with many punctures. Indeed, as we will see, we can bypass the decomposition into three-punctured spheres since we can directly analyse the surface by running the electric algorithm on the bad 3d mirror.

Our analysis furnishes a 3d perspective on channel decompositions that give rise to bad theories. This process, as already shown via the example, involves monopole operators in the 3d mirror acquiring a VEV. One might expect the dynamics of these monopole operators to be related to those of the Higgs branch operators in the 4d theory. This opens up a venue to studying the properties of the Higgs branches of these bad 4d $\cN=2$ theories. 

A core result proposed in this paper is the discovery and characterisation of the special class of 3d $\cN=4$ bad theories, which we have coined broken theories. As an infinite class of examples, we conjecture that all the star-shaped quivers with a full tail and any number of regular tails (as long as the central node is underbalanced) are broken. It would be interesting to further characterise these theories and develop a criterion that determines whether a bad theory is broken directly from the Lagrangian without running the electric algorithm; however, we leave this problem to future analysis. 

We expect the existence of many other broken theories apart from those described in this paper. For example, in an upcoming work \cite{badusp}, more broken theories will be analysed, extending and generalising the class described in this paper. We also expect them to play an important role in the problem of understanding the gauging of Coulomb branch symmetries of 3d $\mathcal{N}=4$ theories \cite{cbgauge}.

While we only consider theories of class $\cS$ of type $\mf{su}(N)$ in this work, we expect similar results to hold true for the theories of class $\cS$ of type $\mf{so}(2N)$ and exceptional types, as well as those including outer-automorphism twist lines on the Riemann surface.\footnote{Indeed, the tinkertoys program has been extended to all these cases \cite{Chacaltana:2011ze,Chacaltana:2012ch,Chacaltana:2013oka,Chacaltana:2014jba,Chacaltana:2015bna,Chacaltana:2016shw,Chacaltana:2017boe,Chacaltana:2018vhp}.} In particular, we expect a similar analysis using broken theories that will appear in certain channel decompositions of these more general theories, which will, however, involve a generalisation of the electric algorithm beyond the unitary case.

Our results for the final theories appearing in each channel can be straightforwardly uplifted to four dimensions, as already noted. In this sense, the final outcomes for each channel directly translate into 4d statements. What has not yet been established is how to uplift the intermediate steps involving the analysis of bad theories; in particular, the implementation of the electric algorithm that leads to specific distributions. A natural and compelling direction emerging from this work is to investigate whether the analysis of the $S^3_b$ partition function for bad 3d mirror theories, obtained via the algorithm, can be uplifted to the superconformal index (or suitable limits thereof) of the 4d $\mathcal{N}=2$ theories \cite{Gadde:2009kb,Gadde:2011uv,Gadde:2011ik,Gaiotto:2012xa}. Another interesting possibility would be to compare our analysis to the vertex operator algebra that can be associated with these bad theories in isolation \cite{Beem:2013sza,Beem:2014rza,Arakawa:2018egx}.\footnote{From the construction of the chiral algebras of class $\cS$, it is known that there is no formal obstruction to defining the vertex operator algebras for bad Riemann surfaces, such as a cylinder or even a cap.}

The possibility of uplifting the cases where the 3d theory is bad (but not broken) to 4d is perhaps even more intriguing; for instance, configurations lacking a full tail. These do not arise from a standard channel decomposition. In such cases, multiple duality frames exist, and it would be interesting to explore how this structure manifests in 4d. \newline

\noindent We leave these and more investigations for future work.

\paragraph{Organization of the paper}
In Section \ref{sec:classS_review}, we briefly review the construction and channel decomposition of the theories of class $\cS$, the methods of \cite{Chacaltana:2010ks}, and the 3d mirror of the theories of class $\cS$. Section \ref{sec:channel_decompSSQ} then reviews the technical details needed to study the $S^3_b$ partition function of the 3d $\mathcal{N}=4$ star-shaped quivers. In Section \ref{sec:brokenbad}, we review the electric dualisation algorithm for 3d $\mathcal{N}=4$ bad theories, and introduce the new family, {\it broken theories}. Section \ref{sec:goodSSQ_twofull} contains several low rank examples of the proposed approach to studying the channel decompositions of four-punctured spheres, containing two maximal and two regular punctures, that produce a bad theory. Finally, Section \ref{sec:further_results} highlights further interesting features of this approach, such as its applicability to bad theories with more than three punctures and channel decompositions that give rise to two bad theories. We supplement this with five appendices: \ref{app:S3b_partfn}, \ref{app:TSUN_TRHOSIGMA}, and \ref{app:higgsing_TSUN} are technical appendices that discuss our conventions for the $S^3_b$ partition functions, 3d linear unitary quivers, and relevant nilpotent Higgsings of these. Appendix \ref{app:roadmap_4dto3d} connects the 4d and 3d notion of badness, while Appendix \ref{app:criterion_interacting} presents a criterion for when a broken theory flows to an interacting SCFT in the IR.

\paragraph{{\tt Mathematica} code} As anciliary material, we provide a {\tt Mathematica} file that implements said electric dualisation algorithm on generic bad star-shaped quivers associated to theories of class $\cS$ of type $\mf{su}(N)$.

\clearpage

\section{Review of theories of class \texorpdfstring{$\mathcal S$}{S}}
\label{sec:classS_review}

The theories of class $\cS$ are four-dimensional $\mathcal N=2$ SCFTs that are realised as the low energy/infrared (IR) limit of twisted compactifications of six-dimensional $\cN=(2,0)$ theories on punctured Riemann surfaces \cite{Gaiotto:2009we,Gaiotto:2009hg}. The punctures, that we will refer to as {\it regular punctures}, correspond to the locations of certain half-BPS codimension-two defects of the 6d theory that are labelled by $\rho$, an embedding of $\mf{su}(2)$ into $\mf g$.\footnote{We do not consider more general 6d compactifications involving outer automorphism twists and higher order poles for the Higgs field here \cite{Witten2008gauge}. These typically lead to twisted theories of class $\cS$ \cite{Tachikawa:2009rb,Tachikawa:2010vg} and Argyres--Douglas theories (see, for instance, \cite{Bonelli:2011aa,Xie:2012hs}), respectively.} The four-dimensional physics of these theories is closely linked to a Hitchin system on the punctured Riemann surface of compactification. At a given puncture (placed at $z=0$ on a local chart), the Higgs field of the Hitchin system admits a simple pole with the residue lying in the (Spaltenstein) dual nilpotent orbit \cite{Chacaltana:2012zy}, labelled by the embedding $\Lambda^\vee$,\footnote{For $\mf g=\mf{su}(N)$, the embedding $\rho$ can be equivalently described by an integer partition of $N$. The embedding/partition, $\rho^\vee$ labelling the Spaltenstein dual nilpotent orbit, is then given by the transpose of the partition corresponding to $\rho$.} corresponding to the codimension-two defect,
\begin{equation}\label{eq:simp_pole}
    \Phi = \frac{e_{\rho^\vee}}{z}\,{\rm d}z + \cdots ~,
\end{equation}
where $e_{\rho^\vee}$ is the nilpotent element in the $\mf{su}(2)$ triplet corresponding to the embedding $\rho^\vee$.

Therefore, a (regular) theory of class $\cS$ is specified by the following set of data,
\begin{itemize}
    \item a simply-laced Lie algebra $\mf g$, taken to be $\mf{su}(N)$ throughout this work,
    \item a Riemann surface, $\CC_{g,s}$, with genus $g$ and $s$ punctures, referred to as the {\it UV curve}. Throughout this work, we will only focus on genus-zero surfaces, i.e.~spheres, and
    \item a homomorphism, $\rho_i:\mf{su}(2)\hookrightarrow\mf{su}(N)$, or equivalently an integer partition of $N$, for every puncture, $i=1,\dots,s$.
\end{itemize}
We will refer to these theories of class $\cS$ as $\TcS{N}{g}{s}{\rho_1,\rho_2,\dots,\rho_s}$.

\subsection{Regular punctures and trinions}
\label{subsec:regpunc_trinions}

The flavour symmetry of these theories contains an independent contribution from each puncture. For a puncture labelled by the partition $\rho\equiv[n_1^{l_1},n_2^{l_2},\dots]$, such that $\sum_i n_i l_i=N$, this contribution is the centraliser of the embedding\footnote{Indeed, it is possible that the full global symmetry is a larger group due to potential enhancements. Moreover, the faithful global symmetry could have a distinct and non-trivial global form. Throughout this paper, we will not discuss these details unless explicitly stated.}
\begin{equation}\label{eq:flavsymm}
     G_\rho = \Bigl(\oplus_i U(l_i)\Bigr)/U(1) ~.
\end{equation}
The trivial embedding, corresponding to the partition $\maxp{N}$, is a distinguished embedding that leads to the maximal flavour symmetry $SU(N)$ and is referred to as the {\it maximal} puncture. Another commonly discussed embedding is the subregular embedding, given by the partition $[N-1,1]$, corresponding to only a $U(1)$ flavour symmetry and is referred to as the {\it minimal} puncture. Finally, the principal embedding, corresponding to $[N]$, leads to a trivial centraliser (as well as a locally regular Higgs field profile) and is thus effectively equivalent to having no puncture.

The Higgs branch of the moduli space of vacua of these theories will be described as a quiver variety in Subsection \ref{subsec:3dmirror_classS}. However, we already note here that one way to obtain theories of class $\cS$ with non-maximal punctures is by moving to particular strata on the Higgs branch of the theory associated to the same UV curve with all maximal punctures \cite{Benini:2009gi,Chacaltana:2012zy}. These Higgs branch flows act on each maximal puncture independently, and are realised as a nilpotent VEV $\left\langle\mu_{SU(N)}\right\rangle\sim \mathcal{J}_\rho$ for the moment map operator of the $SU(N)$ flavour symmetry of the maximal puncture. The decoupled Nambu--Goldstone modes associated with this Higgsing are arranged as hypermultiplets corresponding to the flat directions of the moment map \cite{Maruyoshi:2013hja,Tachikawa:2013kta,Tachikawa:2015bga}
\begin{equation}\label{eq:NGmodes_number}
    \# \text{ of Nambu--Goldstone modes} = \frac12\left(N^2-\sum_is_i^2\right) \text{ , where }\; \rho^\vee = [s_1,s_2,\dots] ~.
\end{equation}
Thus, the low energy physics in such a vacuum of $\TcS{N}{g}{s}{\maxp{N}^{\times s}}$, specified by $\rho$, is given by $\TcS{N}{g}{s}{\maxp{N}^{\times(s-1)},\rho}$, the corresponding theory of class $\cS$ with one of the maximal punctures replaced by a regular puncture of type $\rho$.

The Coulomb branch of the moduli space of vacua of the theories of class $\cS$ is a freely generated special K\"ahler variety with its complex dimension referred to as the {\it rank} of the SCFT. In terms of the Hitchin system, the Seiberg--Witten curve of the 4d SCFT is identified with the spectral curve of the Hitchin system \cite{Gaiotto:2009hg}, which for $\mf{su}(N)$ takes the form,
\begin{equation}\label{eq:SWcurve}
    {\rm det}(\lambda-\Phi) = \lambda^N + \sum_{i=2}^N \lambda^{N-i}\phi_i = 0 ~,
\end{equation}
where $\lambda$ is the Seiberg--Witten differential and $\phi_i\in H^0 \left(\CC_{g,s},K_{\CC_{g,s}}^{\otimes i}\right)$ are meromorphic $k$-differentials. Thus, the Coulomb branch,  in the absence of mass deformations, is simply a graded vector space,
\begin{equation}
    V_{{\rm CB}} = \bigotimes_{i=2}^N V_i ~,
\end{equation}
where $V_i$ is the space of $\phi_i$. The $\phi_i$ have an order $p_i^{(j)}$ pole at the $j^{{\rm th}}$ puncture, with $j=1,\dots,s$, completely determined by local profiles of the form \eqref{eq:simp_pole}, and thus the set $\{p_i^{(j)}\}_{i=2,\dots,N}$ corresponding to a fixed $j$ is referred to as the ``pole structure" \cite{Gaiotto:2009we,Chacaltana:2010ks}. For instance, the pole structure of a maximal puncture is simply $\{1,2,\dots,N-1\}$, while that of a minimal puncture is $\{1,1,\dots,1\}$. The dimensions of each homogeneous vector subspace of $V_{{\rm CB}}$ can then be expressed in terms of the pole structures as follows,
\begin{equation}\label{eq:graded_CBdim}
    d_i \coloneq {\rm dim}_{\mathbb C}(V_i) = (g-1)(2i-1) + \sum_{j=1}^s p_i^{(j)} ~,
\end{equation}
with $\sum_i d_i$ giving the rank of the SCFT.

Notice that for certain choices of punctures, these numbers may be negative. As it clearly does not make sense for the spectrum of the theory to include a negative number
of operators of a certain scaling dimension, such negativity indicates some form of pathology of the particular configuration. In fact, this defines a particular 4d notion of {\it badness} which we will get back to in the next section.\footnote{Note that there is an alternative notion of \textit{badness} for class $\cS$ theories provided in \cite{Gaiotto:2012uq} from the point of view of divergent superconformal indices (and limits thereof). The two notions of badness are expected to coincide. For concreteness, throughout the paper, we will only use the requirement $d_i \geq 0$ as a \textit{goodness} condition.}

The theories of class $\cS$ corresponding to a three-punctured sphere, also referred to as {\it trinions}, $\TcS{N}{0}{3}{\maxp{N}^{\times 3}}$, with all three maximal punctures are the fundamental building blocks for this class of theories \cite{Gaiotto:2009we}. The remaining trinions can be obtained from these by way of nilpotent Higgsings as explained around \eqref{eq:NGmodes_number}. The more general theories, {\it i.e.}, those associated with $\CC_{g,s}$, can be built from trinions by exploiting the striking feature that only the complex structure moduli of the UV curve affect the 4d physics, and are in fact identified with the exactly marginal couplings in the SCFT. This leads to a natural correspondence between the topological decomposition of $\CC_{g,s}$ into pair-of-pants, $\CC_{0,3}$, and ``gluing" trinions together to form general theories of class $\cS$.

\begin{figure}[H]
    \centering
    \includegraphics[width=0.75\hsize]{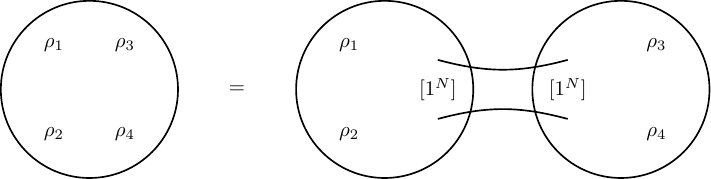}
    \caption{Pictorial representation of gluing together two maximal punctures. This can also be seen as decomposing the four-punctured sphere into two pairs-of-pants.}
    \label{fig:channel_decomp}
\end{figure}

The prescription for {\it gluing} is to gauge the diagonal subgroup of an $SU(N)\times SU(N)$ flavour symmetry, with each factor coming from the two maximal punctures that are being glued together. We pictorially depict the gluing as a tube extending between two Riemann surfaces, depicted for the case of gluing two trinions together in Figure \ref{fig:channel_decomp}.
The inverse of this operation can be understood as moving to a particular cusp on the moduli space of complex structures (or, equivalently, taking the corresponding exactly marginal coupling to infinity) that isolates a trinion from the full Riemann surface. We refer to a particular choice of doing this as a {\it channel decomposition}. Thus, one can associate multiple ``weakly-coupled" descriptions of the full SCFT by considering various ways to glue together trinions along maximal punctures.
The equivalence of different decompositions of the theory associated to $\CC_{g,s}$ with generic punctures gives rise to an instance of generalised $\cS$-duality.

As we are about to discuss in the next section, certain channel decompositions —specifically those corresponding to the ``collision"\footnote{In our language, a collision of $m$ punctures is the same as taking the channel decomposition that separates a sphere with $m$ of these punctures from the rest of the Riemann surface.} of two (or even more) regular punctures — may isolate a pathological 4d theory, a bad trinion. Such a possibility warrants extra care.

\subsection{Fixtures with irregular punctures \`a la the tinkertoys program}
\label{subsec:irregpunc}

As anticipated, an arbitrary configuration of $\CC_{g,s}$ and regular punctures $\rho_i$ does not necessarily correspond to a physical 4d $\cN=2$ SCFT. For example, for a trinion with one maximal and two minimal punctures, the dimensions of the homogeneous subspaces of its Coulomb branch using \eqref{eq:graded_CBdim} are given by
\begin{equation}\label{eq:CBdim_2minimal}
    d_i = 2-i \quad,\text{ for } i=2,\dots,N~.
\end{equation}
Since these dimensions are all negative for $i>2$, this is an ill-defined configuration when considered in isolation. However, these configurations readily arise upon channel decompositions of perfectly physical theories of class $\cS$. For instance, consider the case $\rho_{1,2}=[N-1,1]$ and $\rho_{3,4}=\maxp{N}$ in Figure \ref{fig:channel_decomp}, that corresponds to the 4d $\cN=2$ $SU(N)$ gauge theory with $N_f=2N$ \cite{Gaiotto:2009we}. In this case, the trinion with two minimal punctures arises in a particular channel decomposition of the class $\cS$ realisation of this gauge theory.

It should be immediately clear that configurations with sufficiently many punctures — or punctures that are sufficiently “large,” in the sense that the commutant of their associated flavour symmetries is small inside the maximal $SU(N)$ — correspond to positive subspace dimensions. Consequently, these configurations generically correspond to physical SCFTs. However, since apparently unphysical configurations do arise in channel decompositions of good SCFTs, there must exist a physical mechanism that cures the apparent pathologies these configurations exhibit in isolation.

In this paper, we provide an explanation for this mechanism. Our perspective is closely aligned with that of \cite{Gaiotto:2011xs}: we relate the 4d badness of a configuration to the 3d badness of the star-shaped quiver that describes its 3d mirror, as will be described in Subsection \ref{subsec:3dmirror_classS}. We then build on recent results concerning bad 3d $\cN=4$ theories to support our analysis.

Before delving into our approach, let us briefly review some alternative strategies. In principle, one can start from compactifications involving only maximal punctures. As already discussed around \eqref{eq:NGmodes_number}, by turning on vacuum expectation values (VEVs) for moment map operators associated with the flavour symmetries at these punctures, one can trigger Higgs branch flows that partially close the punctures, thereby generating the desired regular punctures. The RG flow initiated by this Higgsing procedure can, in principle, be followed to determine the weakly-coupled description associated with a given channel. However, implementing this idea in practice requires a detailed understanding of the Higgs branch relations in the $T_N$ building blocks, which is highly non-trivial.

Substantial insight can instead be gained by analysing the corresponding Seiberg--Witten curves \eqref{eq:SWcurve}. It is, in fact, possible to reconstruct the effect of channel decompositions, including the collision of general regular punctures, as discussed in \cite{Gaiotto:2009we} (also see \cite{Tachikawa:2013kta} for a review) by a careful analysis of the pole structures of the SW differential. This approach was further developed in \cite{Chacaltana:2010ks}, eventually giving rise to a classification program known as the “tinkertoys program.” The central idea of this program is to classify all possible trinions (fixtures) that can appear in a channel decomposition, along with the gauge groups (tubes) that connect various pairs of punctures. A crucial input was the inclusion of an {\it irregular} puncture\footnote{It is important to note here that this notion of irregular punctures is distinct from the irregular singularity commonly used in the vast Argyres--Douglas literature (see, for instance, \cite{Bonelli:2011aa,Xie:2012hs} amongst a multitude of references). There it corresponds to a set up where the Higgs field admits a higher order pole (with branch cuts potentially) at a marked point on the Riemann surface in the construction of Argyres--Douglas theories as twisted compactifications of 6d $\cN=(2,0)$ theories.} on the trinion and its gluing to a regular puncture that served as a cure for this particular 4d notion of badness.

Here, we selectively review this method from the lens of channel decompositions as opposed to a classification scheme. Consider the channel decomposition that isolates two regular punctures as a trinion, $\CC_{0,3}$, from the remaining Riemann surface, $\CC_{g,s-2}$. If the decomposed trinion has positive subspace dimensions in isolation, then it is connected to $\CC_{g,s}$ via a maximal puncture on both ends of the gauging, as we have already reviewed. However, if the trinion has negative subspace dimensions in isolation, then the prescription is to replace the maximal puncture on $\CC_{0,3}$ with a formal object, the aforementioned irregular puncture.

The irregular puncture is defined to have the precise pole structure required to cure the negativity of the subspace dimensions by raising those that were negative to zero. The reason that this is not a regular puncture is that it does not correspond to any $\mf{su}(2)$ embedding inside $\mf{su}(N)$. Alternatively, the pole structure $p_i=i-1$ for the maximal puncture is the maximal possible value, and thus by definition the pole structure for the irregular puncture is larger. For instance, in the case of the trinion with two minimal and a maximal puncture, the maximal puncture with pole structure $\{1,2,3,\dots,N-1\}$ is replaced by an irregular puncture with pole structure $\{1,3,5,\dots,2N-3\}$, ensuring that \eqref{eq:CBdim_2minimal} gets modified to
\begin{equation}
    d_i^{\text{irregular}} = 0 \quad,\text{ for } i=2,\dots,N~.
\end{equation}
Such a replacement not only affects the trinion, but also the gauging that connects it to the rest of the Riemann surface, $\CC_{g,s-2}$, as well as the maximal puncture on $\CC_{g,s-2}$, referred to as the conjugate puncture, that is participating in the gluing. 

We do not present the exact criterion given in \cite{Chacaltana:2010ks} to determine the gauging and the conjugate puncture, for each irregular pole structure and refer to the original reference for a detailed exposition. We instead note the physical consistency condition that must be satisfied: the subspace dimensions before decomposition, $d_i^{\text{original}}$, must agree with the sum of subspace dimensions coming from the trinion with the irregular puncture, $d_i^{\text{irregular}}$, the appropriate vector multiple based on the modified gauge group, $d_i^{\text{vector}}$, and the theory associated to the remaining surface now containing the conjugate puncture, $d_i^{\text{remaining}}$,
\begin{equation}
    d_i^{\text{original}} = d_i^{\text{irregular}} + d_i^{\text{vector}} + d_i^{\text{remaining}} \quad,\text{ for } i=2,\dots,N~.
\end{equation}
Finally, the physical content of the trinion with two regular punctures and the corresponding irregular puncture is a bunch of free hypermultiplets (if $d_i=0$ for all $i$) or an interacting SCFT (if $d_i\neq0$ for some $i$).

More concretely, reconsider the case $\rho_{1,2}=[N-1,1]$ and $\rho_{3,4}=\maxp{N}$ in Figure \ref{fig:channel_decomp} that manifestly corresponds to the 4d $\cN=2$ $SU(N)$ gauge theory with $2N$ hypermultiplets in one channel decomposition. In the other one, we decompose a trinion with two minimal and a maximal puncture. Upon replacing the maximal puncture by the corresponding irregular puncture, we modify the gauging from $SU(N)$ to $SU(2)$ and then replace the maximal puncture on the other trinion with the puncture labelled by the $\left[N-2,1^2\right]$ embedding. Thus, this provides a ``weakly-coupled" description of the original SCFT as an $SU(2)$ gauge theory coupled to a single hypermultiplet and an interacting SCFT described by the trinion $\TcS{N}{0}{3}{\maxp{N}^{\times2},\left[N-2,1^2\right]}$. This is a form of Argyres--Seiberg duality \cite{Argyres:2007cn} where the $SU(2)$ gauge theory description emerges in a strongly coupled region of the $SU(N)$ gauge theory. This is indeed the original Argyres--Seiberg duality for $N=3$, where the interacting SCFT is the rank-1 $E_6$ Minahan--Nemeschansky SCFT.

We discuss another example in lieu of the exact criterion to determine the various components in this description involving the irregular puncture. Consider the theory of class $\cS$ associated to a sphere with five punctures, $\TcS{6}{0}{5}{\maxp{6}^{\times3},[3^2],[5,1]}$. This is an interacting SCFT with positive Coulomb branch subspace dimensions as computed using \eqref{eq:graded_CBdim}. Consider the channel decomposition into the trinion with the $[3^2],[5,1]$ punctures. With the maximal puncture, the Coulomb branch subspace dimensions for this case are
\begin{equation}
    d_i = \left\lfloor\frac{i}{2}\right\rfloor-i+1 \quad,\text{ for } i=2,3,4,5,6~.
\end{equation}
The pole structure of the irregular puncture that cures the illness in this case is $\{1,3,4,6,7\}$ as opposed to $\{1,2,3,4,5\}$ corresponding to the maximal puncture. The gauge group accordingly changes to $SU(2)$ instead of $SU(6)$ and the conjugate puncture is the one labelled by the $[2^2,1^2]$ embedding. Thus the weakly-coupled description in this channel is an $SU(2)$ gauging coupled to the the interacting SCFT $\TcS{6}{0}{4}{\maxp{6}^{\times3},[2^2,1^2]}$.

As will be demonstrated, our approach will not only be able to reproduce these results, but also provide a completely physical derivation. In particular, the analysis of the putative bad star-shaped 3d mirror from the lens of its $S^3_b$ partition function exactly reproduces the physics of the pathological trinion, the associated gauging, and the conjugate puncture, instead of relying on an apriori {\it ad hoc} series of algorithmic manipulations. We now move on to discuss these star-shaped quivers as they arise from the 4d theories of class $\cS$.

\subsection{Star-shaped quivers as 3d mirrors}
\label{subsec:3dmirror_classS}

In the low energy limit of $S^1$ compactifications of the 4d $\cN=2$ theories of class $\cS$, one obtains a 3d $\cN=4$ SCFT, which typically has no Lagrangian description as a gauge theory with hypermultiplet matter.\footnote{It is however possible to obtain a {\it quasi-Lagrangian} description in the sense of a gauge theory obtained from the gauging of emergent symmetries of SCFTs with Lagrangian descriptions.} However, the 3d mirror dual of the circle reduction can be described as a 3d $\cN=4$ gauge theory that can be neatly expressed as star-shaped quiver gauge theories \cite{Benini:2010uu}.

Indeed, the 4d Higgs branch of a theory of class $\cS$ is identified with the 3d Coulomb branch of the corresponding star-shaped quiver. We refer to the star-shaped quiver corresponding to a theory of class $\cS$ as the ``3d mirror." Each regular puncture, labelled by an embedding $\rho$ corresponding to a 6d codimension-two defect, in a theory of class $\cS$ associated to the algebra $\mf{su}(N)$ corresponds to a special class of 3d $\cN=4$ SCFTs referred to as $T_\rho[U(N)]$ \cite{Gaiotto:2008ak}. We review 3d $T[U(N)]$ and $T_\rho[U(N)]$ theories in Appendix \ref{app:TSUN_TRHOSIGMA}.\footnote{A generic $T^\sigma_\rho[U(N)]$ theory differs from the $T^\sigma_\rho[SU(N)]$ theory only by a BF coupling. We prefer to use the former throughout the paper since, from the 3d point of view, we prefer to gauge $U(N)$ symmetries instead of $SU(N)$. This sometimes gives rise to $U(1)$ redundant gauge symmetries, and we will deal with this problem in Section \ref{sec:channel_decompSSQ}.}

These are unitary linear quiver gauge theories that have a $\mf g_\rho$ symmetry, given by \eqref{eq:flavsymm}, on the Coulomb branch and a manifest $\mf{su}(N)$ Higgs branch symmetry. These can be obtained from the self-dual $T[U(N)]$ theory, associated to a maximal puncture, by triggering a nilpotent VEV for the Coulomb branch moment map via a superpotential deformation \cite{Hwang:2020wpd} (as reviewed in Appendix \ref{app:TSUN_TRHOSIGMA}).
The 3d mirror of a theory of class $\cS$ associated to a sphere with $s$ punctures, $\TcS{N}{0}{s}{[\rho_1],[\rho_2],\dots,[\rho_s]}$ can then be assembled by gauging a diagonal combination of the $U(N)$ symmetry on the Higgs branch of every $T_{\rho_i}[U(N)]$ tail together.

\begin{figure}
    \centering
    \includegraphics[width=0.33\hsize]{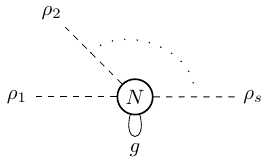}
    \caption{Three-dimensional mirror dual to the circle compactification of $\TcS{N}{g}{s}{[\rho_1],[\rho_2],\dots,[\rho_s]}$. Here the dotted line connected to a $\rho_i$ indicates the $T_{\rho_i}[SU(N)]$ tail.}
    \label{fig:gen_starshaped}
\end{figure}

Finally, the 3d mirrors of theories associated with a genus $g$ surface and $s$ punctures have the same structure with an additional $g$ adjoint hypermultiplets on the central $U(N)$ node. This results in a star-shaped quiver as depicted schematically in Figure \ref{fig:gen_starshaped}.
An important feature of $T_\rho[U(N)]$ theories is that these do not contain any unitarity-violating operators in the spectrum, and are referred to as ``good" 3d $\cN=4$ theories \cite{Gaiotto:2008ak}.  We will discuss $3d$ $\mathcal{N}=4$ bad theories in more detail in  Section \ref{subsec:badSQCD}.

Here we just note that the ``goodness" of a 3d $\cN=4$ quiver gauge theory is equivalent to having no underbalanced gauge node,
that is, each $U(N_i)$ gauge node sees at least $2N_i$ fundamental hypermultiplets. This condition ensures that each local monopole operator, {\it i.e.}, those with non-zero magnetic flux under the gauge node and some of the neighbouring ones, satisfies the unitarity bound on the conformal dimension.
Since $T_\rho[U(N)]$ theories are good, it's easy to notice that 
a star-shaped quiver can only be bad if the central node
is underbalanced.\footnote{The condition that each gauge node has positive balancing is necessary but not sufficient for a quiver gauge theory to be good. For a generic quiver gauge theory, there can be ``global" monopole operators whose conformal dimensions fall below the unitarity bound. These are global in the sense that they carry non-zero magnetic fluxes under a sequence of gauge groups that can not be localised around a single gauge node. In fact, this situation happens in star-shaped quivers. Any quiver with the shape of an affine Dynkin diagram exhibits this notion of global badness. See \cite{Gaiotto:2012uq} for a discussion on this point from the 4d point of view.} 

As already pointed out in \cite{Gaiotto:2011xs}, there is in fact a relation between the particular 4d notion of badness related to the irregular puncture \cite{Chacaltana:2010ks} and the 3d notion of badness \cite{Gaiotto:2008ak}. We review this explicitly in Appendix \ref{subsec: 4d3d_badness}.

\section{3d star-shaped quivers and channel decomposition}
\label{sec:channel_decompSSQ}

In this section, we transition to studying the star-shaped quivers as 3d $\cN=4$ gauge theories. The primary tool for our analysis is the $S^3_b$ partition function, which we review in Appendix \ref{app:S3b_partfn}. We begin by setting up the notation for star-shaped quivers obtained as 3d mirrors of theories of class $\cS$. We then discuss the 3d counterpart of the 4d channel decomposition that is crucial to the analysis of theories of class $\cS$. In a purely 3d setup, this is achieved by the insertion of an identity operator that corresponds to the CB fusion of two $T[U(N)]$ tails.

\subsection{Partition functions of star-shaped quivers}
\label{subsec:partfun_SSQ}

The 3d mirror of a theory of class $\cS$ associated to a sphere with $M$ generic regular punctures can be expressed as:
\begin{equation}\label{eq:starsh_param}
    \includegraphics[width=.9\textwidth]{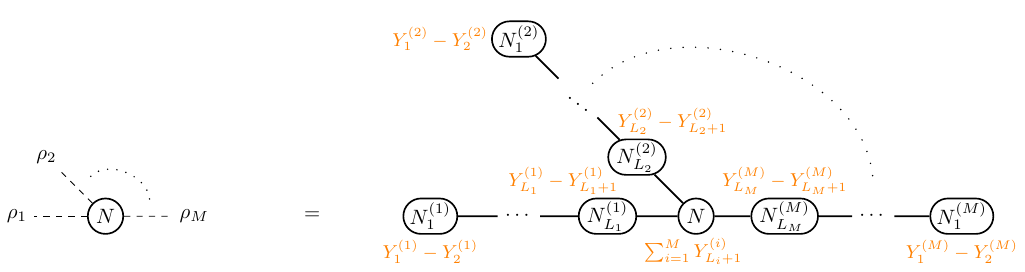}
\end{equation}
As in Figure \ref{fig:gen_starshaped}, the $T_\rho[U(N)]$ tail on the left is denoted as a dashed line ending on a $\rho$. Meanwhile, on the right, we express the same quiver with the explicit gauge nodes $N_i^{(a)}$  obtained from the partitions $\rho_a$ as described in Appendix \ref{app:TSUN_TRHOSIGMA}. We also specify our conventions for labelling the FI parameters associated with each gauge node in the star-shaped quiver on the right.

Using the prescription of \cite{Benini:2010uu} as explained in Subsection \ref{subsec:3dmirror_classS}, we can associate the following partition function in terms of $T_\rho[U(N)]$ theories:
\begin{align}\label{eq:starsh_parfun}
    & \PFS{S}\left(\vec{X}^{(1)}_{\rho_1}, \vec{X}^{(2)}_{\rho_2}, \ldots, \vec{X}^{(M)}_{\rho_M}; m\right) = \nn \\
    & = \int {\rm d}\vec{U}_N \D^{\rm HB}_N(\vec{U};m) \PFT{\rho_1}{N}( \vec{U},\vec{X}^{(1)};m)\PFT{\rho_2}{N}(\vec{U},\vec{X}^{(2)};m) \ldots \PFT{\rho_M}{N}(\vec{U},\vec{X}^{(M)};m) ~,
\end{align}
where $\vec{X}^{(i)}$ are the set of parameters associated to the $i^{\rm th}$ puncture, parameterised as in figure \eqref{eq:starsh_param}. The explicit formula for the partition function of $T_\rho[U(N)]$ is given in \eqref{eq:tsunPF}. Furthermore, it is always implied that the first entry in the partition function of $T_\rho[U(N)]$ corresponds to the HB parameters, while the second one corresponds to the CB ones. The integration measure, $\D_N^{\rm HB}(\vec{U};m)$, defined in \eqref{HBmeasure}, is the Higgs branch (HB) gluing measure, that captures the (inverse of the) contribution of a 3d $\mathcal{N}=4$ vector multiplet of the $U(N)$ gauge symmetry with associated parameters $\vec{U}$. The parameter $m$ is defined in \eqref{eq:massm} and its real part corresponds to a real mass parameter for the axial symmetry as explained in Appendix \ref{app:S3b_partfn}.

Notice that according to the definitions in Figure \ref{eq:starsh_param} and \eqref{eq:starsh_parfun}, there is always a redundant $U(1)$ gauge group, which corresponds to the diagonal $U(1)$ among all the $U(k)$ gauge symmetries. The monopole operators associated with this $U(1)$ factor always violate the unitarity bound, and are thus removed to correctly define the theory. An effective trick that can be used to decouple any $U(1)$ is to gauge the corresponding topological symmetry. Thus, we gauge the topological symmetry associated with the diagonal $U(1)$, which at the level of the partition function corresponds to an integration over the corresponding FI parameter, which we name $C$.

To determine the combination of FI parameters corresponding to $C$ we can think of redefining all the gauge Cartans $Z_i$
as  $Z_i \to \tilde Z_i + Z$, requiring then $\sum_i \tilde Z_i=0$
(the sum over $i$ then runs over all the components of all the gauge groups). This redefinition produces a term in the partition function of the form $e^{2\pi i Z C}$ which indeed represents the contribution of the  FI coupling for the diagonal $U(1)$
with 
\begin{align}\label{eqtomodU1}
    C = \sum_{\text{gauge nodes}} \text{rank} \times \text{FI} ~,
\end{align}
where the sum is taken to be over all the gauge nodes. This is thus a sum over all gauge nodes with the associated FI parameters weighted by the rank of the gauge nodes. 
We  can then decouple the diagonal $U(1)$ by performing the integration over $C$:
\begin{align}\label{prescription to decouple U(1)}
\widehat{\PFS{S}}\left(\vec{X}^{(1)}_{\rho_1}, \vec{X}^{(2)}_{\rho_2}, \ldots, \vec{X}^{(M)}_{\rho_M}; m\right) \coloneqq    \int {\rm d}C \; \D_1^{\rm CB}(C;m) \; \PFS{S}\left(\vec{X}^{(1)}_{\rho_1}, \vec{X}^{(2)}_{\rho_2}, \ldots, \vec{X}^{(M)}_{\rho_M}; m\right) ~.
\end{align}
since  $\int d C \; e^{2\pi i Z C}=  \delta(Z)$.
Note that this gauging is performed with a $U(1)$ CB measure, which is defined in \eqref{eq:CBmeasure}. Throughout the paper, we use the notation where $\widehat{\mathcal{Z}_S}$ is the partition function of the theory $S$ without the redundant gauge $U(1)$.

\subsection{Decomposing star-shaped quivers by Identity insertion}
\label{subsec:SSQdecomp_identity}

\begin{figure}
    \centering
    \includegraphics[width=\textwidth]{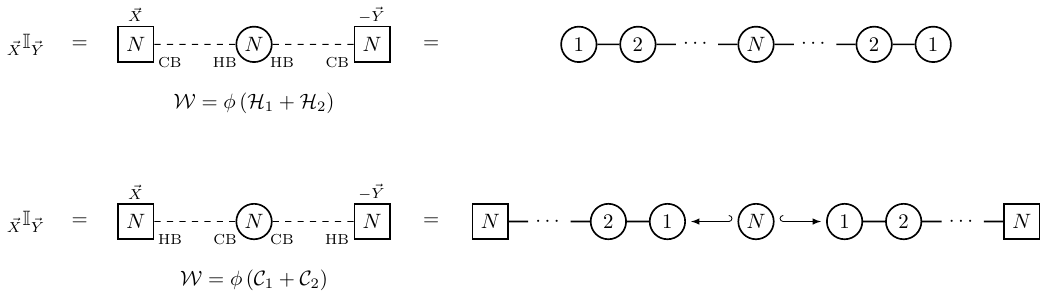}
    \caption{HB and CB fusion to Identity. Notice that the gluing prescription involves coupling the adjoint chiral $\phi$ of the $U(N)$ gauge nodes to the HB/CB moment maps $\CH_i/\CC_i$ of the first and the second quiver tails. We use $_{\vec{X}}\mathbb{I}_{\vec{Y}}$ as the shorthand for the distribution on the r.h.s.\ of \eqref{eq:tsun_delta}. Meanwhile, on the r.h.s.\ of the figure, we depict the explicit Lagrangian depiction of the two gluings. The top right depicts the linear quiver obtained by diagonal gauging of two $T[U(N)]$ theories, whereas, on the bottom right, we depict the gauging of the emergent symmetries on the Coulomb branches of two $T[U(N)]$ theories using hook arrows.}
    \label{fig:IDgluings}
\end{figure}

The three-dimensional counterpart of the pairs-of-pants decomposition of a sphere with punctures, as reviewed in Subsection \ref{subsec:regpunc_trinions}, is the channel decomposition of the corresponding star-shaped quiver into star-shaped quivers with three tails. At the level of three-dimensional physics, this operation can be performed by inserting Identity operators given by the fusion of two $T[U(N)]$ tails as given in Figure \ref{fig:IDgluings}.

At the level of the partition function, it has been shown in \cite{Bottini:2021vms} that the fusion of two $T[U(N)]$ theories yields a delta function identifying the Cartan subalgebras of the surviving $U(N)$ symmetries.\footnote{The $T[U(N)]$ theory was identified in \cite{Gaiotto:2008ak} as the S-duality wall. The fusion of two $T[U(N)]$ theories can then be interpreted as the relation $S S^{\pm 1}=\mathbb I$. This and the other $SL(2,\mathbb{Z})$ relations involving the  $T$-wall generator have been rigorously proven in field theory in \cite{Bottini:2021vms} by iterative applications of Seiberg-like dualities.} More precisely, depending on whether we gauge the CB or the HB symmetries of the two tails we have:\footnote{It is also possible to consider ``mixed" HB/CB fusion with $\mathcal{W}=\mathcal{C}_1\mathcal{H}_2$, however we will not be concerned with such a gluing in this paper.}
\begin{equation}\label{eq:tsun_delta}
\begin{split}
    & \text{HB gluing:}
    \int {\rm d}\vec{U}_{N} \;\D^{\rm HB}_{N}(\vec{U};m) \PFT{}{N}(\vec{U},\vec{X};m) \PFT{}{N}(\vec{U},-\vec{Y};m) = \sum_{\sigma \in S_N} \frac{ \prod_{j,k=1}^N \d( X_j - Y_{\sigma(k)} )}{\D^{\rm CB}_N(\vec{X};m)}
    \\
    & \text{CB gluing:}
    \int {\rm d}\vec{U}_{N} \;\D^{\rm CB}_{N}(\vec{U};m)  \PFT{}{N}(\vec{X},\vec{U};m) \PFT{}{N}(-\vec{Y},\vec{U};m) = \sum_{\sigma \in S_N} \frac{   \prod_{j,k=1}^N \d( X_j - Y_{\sigma(k)} )}{\D^{\rm HB}_N(\vec{X};m)} 
\end{split}
\end{equation}

We refer to Appendix \ref{app:S3b_partfn} for conventions on partition functions. We will come back to this property of fusion to Identity in Subsection \ref{subsec: twoTSUN_gluing} to re-interpret it from the point of view of 3d $\mathcal{N}=4$ bad theories. In the first line of \eqref{eq:tsun_delta} we are gluing together the two $T[U(N)]$ theories via the identification of their Coulomb branches, while in the second line we are instead gluing along their Higgs branches. These two identities are not independent; in fact, each follows from the other by means of the self-mirror property of the $T[U(N)]$ theory. 

We can represent the HB and CB gauging as a quiver operation, as shown in Figure \ref{fig:IDgluings}. Note that on the bottom right of Figure \ref{fig:IDgluings} we introduce a notation to indicate the gauging of an emergent Coulomb branch symmetry of a theory, which will often appear in our results. In fact, the gauge node, with gauge group $G$, with a hook arrow, means that the group $G$ on the CB of the theory pointed by the hook is being gauged. A gauge node with 2 hook arrows instead indicates the gauging of a diagonal combination of (in general, a subgroup of) the CB symmetries of the 2 theories. In general, the group $G$ being gauged is only a subgroup of the full CB symmetry, indicated by the hook arrow usually representing an inclusion.

\begin{figure}[ ]
    \centering
    \includegraphics[width=0.8\linewidth]{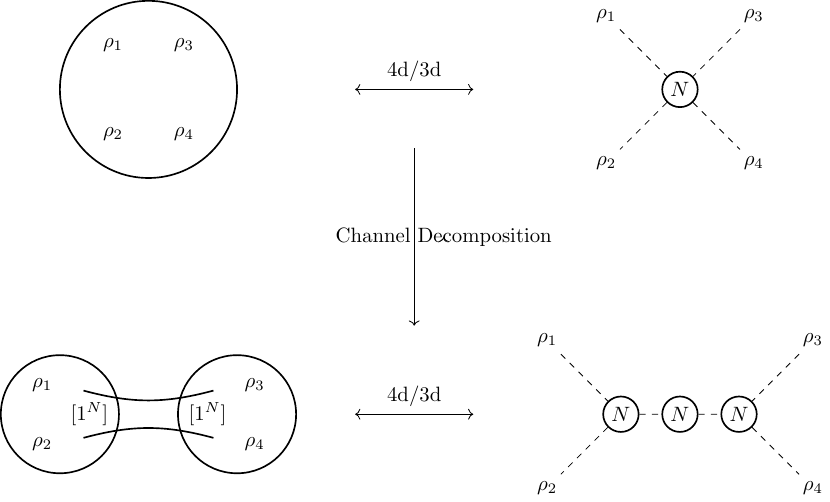}
    \caption{Parallel between the channel decomposition in 3d and in 4d. The procedure at the level of the 3d $S^3_b$ partition function is reported in \eqref{eq:chaneldecomp}. We recall that dashed lines connecting two nodes labelled by $N$ represent $T[U(N)]$ theories, while those that connect the label $\rho_i$ represent $T_{\rho_i}[U(N)].$}
    \label{fig:25062}
\end{figure}

Let us now go back to the problem of the channel decomposition of a star-shaped quiver. For simplicity, let us consider the case of a four-punctured sphere or, in terms of the star-shaped theory, a theory with four  $T_\rho[U(N)]$ tails that are diagonally gauged together as shown in Figure \ref{fig:25062}.

At the level of partition function, using \eqref{eq:tsun_delta}, one can perform the following sequence of manipulations:
\begin{equation}\label{eq:chaneldecomp}
\begin{split}
    & \mathcal{Z}_S(\vec{X}^{(1)}_{\rho_1}, \vec{X}^{(2)}_{\rho_2}, \vec{X}^{(3)}_{\rho_3}, \vec{X}^{(4)}_{\rho_4}; m) = \\
    & \quad = 
    \int {\rm d}\vec{U}_{N} \;\D^{\rm HB}_{N}(\vec{U};m) \PFT{\r_1}{N}(\vec{U},\vec{X}^{(1)};m) \PFT{\r_2}{N}(\vec{U},\vec{X}^{(2)};m) 
    \PFT{\r_3}{N}(\vec{U},\vec{X}^{(3)};m)
    \PFT{\r_4}{N}(\vec{U},\vec{X}^{(4)};m)
     \\
    & \quad = 
    \int {\rm d}\vec{U}_{N} \;\D^{\rm HB}_{N}(\vec{U};m)
    \int {\rm d}\vec{V}_{N} \;\D^{\rm HB}_{N}(\vec{V};m)
    \PFT{\r_1}{N}(\vec{U},\vec{X}^{(1)};m) \PFT{\r_2}{N}(\vec{U},\vec{X}^{(2)};m) 
     \\
    & \quad\qquad \times
    \frac{  \sum_{\sigma \in S_N} \prod_{i,j=1}^N \d( U_i - V_{\sigma(j)} )}{\D^{\rm HB}_N(\vec{V};m)}
    \PFT{\r_3}{N}(\vec{V},\vec{X}^{(3)};m) 
    \PFT{\r_4}{N}(\vec{V},\vec{X}^{(4)};m)
     \\
    & \quad =
    \int {\rm d}\vec{Y}_{N} \; \D_{N}^{\rm CB}(\vec{Y};m) 
    \int d\vec{U}_{N} \D^{\rm HB}_{N}(\vec{U};m)  
    \PFT{\r_1}{N}(\vec{U},\vec{X}^{(1)};m) \PFT{\r_2}{N}(\vec{U},\vec{X}^{(2)};m) 
    \PFT{}{N}(\vec{U},\vec{Y};m)
     \\
    & \quad \qquad \times
    \int {\rm d}\vec{V}_{N} \;\D^{\rm HB}_{N}(\vec{V};m)
    \PFT{\r_3}{N}(\vec{X}^{(3)},\vec{V};m)
    \PFT{\r_4}{N}(\vec{X}^{(4)},\vec{V};m) 
    \PFT{}{N}(-\vec{Y},\vec{V};m)
     \\
    & \quad =
    \int {\rm d}\vec{Y}_{N} \;\D^{\rm CB}_{N}(\vec{Y};m)
    \mathcal{Z}_{S_1}(\vec{X}^{(1)}_{\rho_1}, \vec{X}^{(2)}_{\rho_2}, \vec{Y}; m)
    \mathcal{Z}_{S_2}(\vec{X}^{(3)}_{\rho_3}, \vec{X}^{(4)}_{\rho_4}, -\vec{Y}; m) \,.
\end{split}
\end{equation}
Let us now explain these manipulations. In the first step, we simply use the definition of the partition function of the starting theory $S$ using \eqref{eq:starsh_parfun}, as the gauging of four $T_\rho[U(N)]$ theories. In the second step, we insert an Identity operator using the second equation of \eqref{eq:tsun_delta} to double the gauge parameters. In the third step, we then replace the Identity operator with the CB gluing of two $T[U(N)]$ theories using the first equation of \eqref{eq:tsun_delta}. In the final step, we then combine one of the two $T[U(N)]$ tails, forming the Identity wall with the regular tails $T_{\rho_{1,2}}[U(N)]$, to reconstruct the star-shaped theory $S_1$ and the other $T[U(N)]$ tail of the Identity wall with the two regular tails $T_{\rho_{3,4}}[U(N)]$ to reconstruct $S_2$.

As anticipated, the star-shaped quivers $S_1$ and $S_2$ emerging from a given channel decomposition may be bad in the three-dimensional (and hence also in the four-dimensional) sense. Thus, we now pause our discussion regarding class $\mathcal{S}$ theories to review some recent results regarding 3d $\cN=4$ bad theories, and introduce a new subclass: broken theories. We will then leverage these results in Sections \ref{sec:goodSSQ_twofull} and \ref{sec:further_results} to extract the physics emerging from the channel decomposition that isolates two generic regular punctures.

\section{Broken theories}
\label{sec:brokenbad}

This section is dedicated to the physics of three-dimensional $\cN=4$ bad theories. In particular, we review and discuss some of their features discussed in \cite{Giacomelli:2023zkk,Giacomelli:2024laq}. We then identify a distinguished family of bad theories, which we term {\it broken theories}, which will play a key role in the rest of this paper.

\subsection{Bad 3d $\cN=4$ SQCD}
\label{subsec:badSQCD}

Three-dimensional $\mathcal{N}=4$ theories are often classified by their infrared behaviour \cite{Gaiotto:2008ak}. A natural question is whether the ultraviolet (UV) $R$ symmetry algebra of a given 3d $\cN=4$ gauge theory is identified with its infrared (IR) counterpart? Such an identification is spoiled whenever a (chiral primary) operator violates the unitarity bound $\Delta \geq \tfrac{1}{2}$, where $\Delta$ is the scaling dimension. This bound is automatically satisfied by operators built from hypermultiplets, and is in fact saturated by a single free hypermultiplet. However, depending on the gauge and matter content of the theory, it is possible that some monopole operators violate this bound, leading to a mismatch between the UV and IR $R$-symmetries.

This leads to a trichotomy of 3d $\cN=4$ theories based on their IR behaviour as proposed in \cite{Gaiotto:2008ak}. A theory is termed {\it good} if it flows to an IR SCFT whose $R$ symmetry algebra is the same as the UV one. It is termed \textit{ugly} when the dimension of a monopole operator saturates the unitarity bound, in which case the theory flows to an SCFT, with an identical IR $R$ symmetry as the UV one, along with a decoupled free sector. Finally, a theory is termed \textit{bad} if some monopole operator violates the unitarity bound, as computed using the UV $R$-symmetry in a standard fashion. Due to these operators decoupling from the interacting part of the theory, the IR dynamics of bad theories are quite intricate.

It was argued in \cite{Giacomelli:2023zkk} that partition functions of bad theories are distributions rather than ordinary functions. More precisely, their partition functions are given by a sum of contributions, called \textit{frames}, that consist of a Dirac delta distribution, enforcing a particular constraint on the  FI parameters, which multiplies the
partition function of an interacting SCFT, and a collection of singlets describing the free sector. Possibly, there can also be a special frame that does not come with a Dirac delta distribution.

Let us now consider concretely the $U(N)$ SQCD case, which is good for $F\geq 2N$.
In the bad/ugly case $N \leq F \leq 2N-1$ the partition function schematically becomes \cite{Giacomelli:2023zkk}:
\begin{equation}
    \includegraphics[]{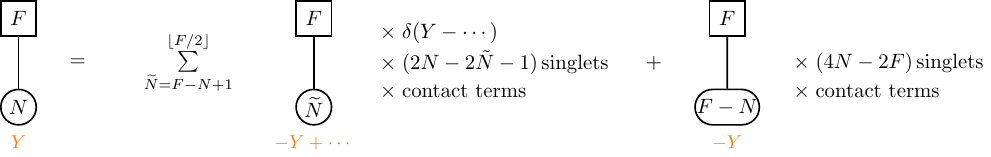} \nn
\end{equation}
Notice in particular that for $F=2N-1$, when the theory is ugly, there are no frames carrying delta-functions. In this case the partition function is exactly equal to that of a good $U(N-1)$ SQCD with one free hyper \cite{Gaiotto:2008ak,Kapustin:2010mh}.
The bad SQCD for $F \leq N-1$ is said to be \textit{evil} and the partition function is schematically equal to:
\begin{equation}
    \includegraphics[]{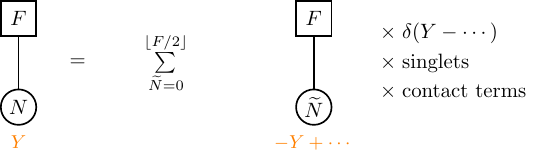} \nn
\end{equation}
Notice that every frame carries a delta function.
 
In general, the Dirac delta constrains the FI parameters to a specific value which was interpreted in \cite{Giacomelli:2023zkk} as an indication that a monopole operator of the original bad SQCD is acquiring a VEV, driving the bad theory to a vacuum where the effective description of the IR is provided by a good $U(\widetilde N)$ SQCD. Since this special vacuum is reached due to a monopole operator acquiring a VEV, the topological symmetry in this vacuum can be considered to be spontaneously broken.

Each frame additionally comes with a total of $(2N-2\widetilde{N}-1)$ chiral singlets. These can be interpreted as monopole operators in the UV theory that become free in the IR, corresponding to the particular frame. Interpreting the delta function as a chiral multiplet with vanishing $R$-charge, it combines with the $(2N-2\widetilde{N}-1)$ free chirals to a total of $(N-\widetilde{N})$ free twisted hypermultiplets with the canonical $R$-charge in the IR. The correct assignment is thanks to a mixing between the UV $R$-symmetry and an emergent global symmetry that rotates the chiral multiplets. Each frame also contains a contact term in the form of BF couplings between the topological symmetry and the flavour symmetry.

For $N \leq F \leq 2N-1$, there is always a distinguished frame corresponding to the good $U(F-N)$ SQCD with $F$ flavours that does not contain a delta function, and thus this frame is not reached via a monopole VEV. This frame coincides with the one found in \cite{Kim:2012uz,Yaakov:2013fza}. For $F<N$, such a frame without the Dirac delta is not present, and hence all the frames in this case are reached by a monopole VEV.

These results are consistent with the findings of \cite{Assel:2017jgo}, where the full quantum moduli space of the bad $U(N)$ SQCD was analysed using the techniques developed in \cite{Bullimore:2015lsa}. It was shown that at the most singular locus of the moduli space, there exists a good $U (\lfloor F/2 \rfloor)$SQCD with $ N-\lfloor F/2 \rfloor$ twisted hypermultiplets. Notably, this singular locus does not have the maximal possible codimension, as the gauge group rank is reduced—an expected feature of bad theories. Upon turning on a generic FI parameter, this locus—as well as other loci corresponding to the other frames where we find lower rank good SQCD theories—is lifted. As a result, only the frame with a $U(F-N)$ SQCD gauge group, as found in \cite{Kim:2012uz,Yaakov:2013fza}, remains.

We now present this result at the level of  the $S^3_b$ partition function, for the cases with $N-F < \widetilde N \leq \lfloor F/2 \rfloor$ we have:
\begin{equation}\label{eq:bad_sqcd}
    \begin{split}
        &\CZ_{{\rm SQCD}_U(N,N\leq F<2N)}\left(\vec X,Y_1-Y_2;m\right) \times e^{-\pi i(Y_1-Y_2)\sum_{i=1}^F X_i} = \\
        &\; = \sum_{r=0}^{\left\lfloor\frac{F}{2}\right\rfloor-(F-N)-1} \sum_{\substack{\beta=1 \text{ if } n-\epsilon=0, \\ \beta=\pm1 \text{ otherwise}}} \Bigg[ \delta\Big( Y_1-Y_2+\beta(n-\epsilon)(iQ-2m) \Big) \;\times \\
        &\qquad \times\; e^{-\pi i(W_2-W_1)\sum_{j=1}^F X_j} \; \frac{\prod_{j=1}^{N-F+\left\lfloor\frac{F}{2}\right\rfloor-1-r} s_b\left(\frac{iQ}{2}+j(iQ-2m)\right)}{\prod_{j=0}^{N-F+\left\lfloor\frac{F}{2}\right\rfloor-1-r} s_b\left(\frac{iQ}{2}-(j-(2N-F-1))(iQ-2m)\right)} \\
        &\qquad \times\; \prod_{j=1}^{N-\left\lfloor\frac{F}{2}\right\rfloor+r} s_b\left(\frac{iQ}{2}-\left(j-N+\frac{F}{2}\right)(iQ-2m)\pm(W_1-W_2)\right) \\
        &\qquad \times\; \left.\CZ_{{\rm SQCD}_U\left(\left\lfloor\frac{F}{2}\right\rfloor-r,F\right)} \left(\vec X,W_2-W_1;m\right) \right\vert_{W_{1,2}=Y_{1,2}\mp\frac12\beta\left(N-F+\left\lfloor\frac{F}{2}\right\rfloor-r\right)(iQ-2m)} \Bigg] \\
        &\quad\; +\; e^{-\pi i(Y_2-Y_1)\sum_{j=1}^F X_j} \prod_{j=1}^{2N-F} s_b\left(\frac{iQ}{2}-\left(j-N+\frac{F}{2}\right)(iQ-2m)\pm(Y_1-Y_2)\right) \\
        &\qquad \times\; \CZ_{{\rm SQCD}_U\left(F-N,F\right)} \left(\vec X,Y_2-Y_1;m\right) ~,
    \end{split}
\end{equation}
where,
\begin{equation}\label{eq:epsilon_exp}
    \epsilon = 
    \begin{cases}
        \quad0 & \text{for } F \text{ even}, \\
        -\frac12 &\text{for } F \text{ odd}.
    \end{cases}
\end{equation}

Notice that delta functions freeze the FI parameters to particular multiples of $(iQ-2m)$.\footnote{Recall that the $SU(2)_H \times SU(2)_C$ $R$-symmetry in 3d $\cN=4$ theories has two real parameters. The parameter associated to the $U(1)_R = U(1)_{H-C}$ R-symmetry subgroup is $\tfrac{iQ}{2}$, whereas the real part of the parameter $m$ is the real mass of the $U(1)_A$ axial symmetry, which is the commutant of $U(1)_R$ in $SU(2)_H \times SU(2)_C$.} The specific monopole operator acquiring a VEV can be read off from the argument of the delta function. One should read the argument as a polynomial with real masses associated with the symmetries as the variables and respective charges as their coefficients. In particular, from \eqref{eq:bad_sqcd} we can see that the operators acquiring a VEV are fundamental monopole operators with different levels of dressing. 

Also note that the contribution of the chiral singlets in each frame needs to be simplified by using the constraint implied by the Dirac deltas to match the expected number of chirals. Each frame also contains a contact term (appearing as an exponential prefactor) in the form of BF couplings between the topological symmetry and the flavour symmetry.

In the evil case, $F<N$, we find instead:
\begin{equation}\label{eq:evil_sqcd}
    \begin{split}
        &\CZ_{{\rm SQCD}_U(N,F2N)}\left(\vec X,Y_1-Y_2;m\right) \times e^{-\pi i(Y_1-Y_2)\sum_{i=1}^F X_i} = \\
        &\; = \sum_{r=0}^{\left\lfloor\frac{F}{2}\right\rfloor} \sum_{\substack{\beta=1 \text{ if } n-\epsilon=0, \\ \beta=\pm1 \text{ otherwise}}} \Bigg[ \delta\Big( Y_1-Y_2+\beta(n-\epsilon)(iQ-2m) \Big) \;\times \\
        &\qquad \times\; e^{-\pi i(W_2-W_1)\sum_{j=1}^F X_j} \; \frac{\prod_{j=1}^{N-F+\left\lfloor\frac{F}{2}\right\rfloor-1-r} s_b\left(\frac{iQ}{2}+j(iQ-2m)\right)}{\prod_{j=0}^{N-F+\left\lfloor\frac{F}{2}\right\rfloor-1-r} s_b\left(\frac{iQ}{2}-(j-(2N-F-1))(iQ-2m)\right)} \\
        &\qquad \times\; \prod_{j=1}^{N-\left\lfloor\frac{F}{2}\right\rfloor+r} s_b\left(\frac{iQ}{2}-\left(j-N+\frac{F}{2}\right)(iQ-2m)\pm(W_1-W_2)\right) \\
        &\qquad \times\; \left.\CZ_{{\rm SQCD}_U\left(\left\lfloor\frac{F}{2}\right\rfloor-r,F\right)} \left(\vec X,W_2-W_1;m\right) \right\vert_{W_{1,2}=Y_{1,2}\mp\frac12\beta\left(N-F+\left\lfloor\frac{F}{2}\right\rfloor-r\right)(iQ-2m)} \Bigg] ~,
    \end{split}
\end{equation}
where $\epsilon$ can be determined by using \eqref{eq:epsilon_exp}. As stated previously, notice that each frame in this case necessarily comes along with a Dirac delta. 

Note that our discussion so far in this section has been at the level of the $S^3_b$ partition function. We naturally expect similar results to hold for the superconformal index (involving periodic Dirac delta functions), or for other 3-manifolds on which we can perform supersymmetric localisation.

However, we actually believe that the full path integral of the theory will take the form of a functional distribution. This, in a sense, generalises the fact that the path integral of a pure $U(1)$ gauge theory coupled to a background vector multiplet $\widehat V$ for
the topological symmetry is a functional delta function. Indeed, $\widehat V$ is coupled to the dynamical vector multiplet $V$ via a BF term
\begin{equation}
    S_{\rm BF} (V,\widehat V )=\int {\rm d}^3 x\, {\rm d}\theta \,{\rm d}\bar\theta  \; \widehat V \Sigma -\int {\rm d}^3 x\, {\rm d}^2\theta \left(\Phi \widehat \Phi+cc\right) ~,
\end{equation}
where $V/\widehat V,\Sigma/\widehat \Sigma$, and $\Phi/\widehat \Phi$ are the dynamical/background vector, linear, and adjoint scalar multiplets, respectively. In the IR (where we can drop the kinetic term $1/g^2 F_{\mu\nu}^2$) performing the path integral of $V$ leads to a functional delta function setting $\widehat V=0$.\footnote{The interpretation of this BF coupling term as a kernel for a functional Fourier transform was the key idea for the piecewise derivation of abelian mirror symmetry of \cite{Kapustin:1999ha}.}

We will see a direct generalisation of this idea in Section \ref{subsec: twoTSUN_gluing} when considering the bad theory resulting from the diagonal gauging of two $T[U(N)]$ theories. In general, our interpretation is that the path integral of bad theory is a functional distribution with support at special points given by the solutions of the equations of motion. 

Indeed, as shown in \cite{Giacomelli:2023zkk},
from the analysis of the $\mathcal{N}=2^*$ EOM  of the $U(N)$ SQCD, in the good case, one finds the familiar situation in which at a generic point on the HB there are no residual CB directions (since $\Phi$ and $\Sigma$ cannot take a VEV), and the gauge group is unbroken and equal to $U(N)$ at the origin of the HB. On the other hand, in the bad case, the analysis of the EOM shows that there is a always residual CB at each point of the HB, and the gauge group is still broken at the origin of the HB, so there is no point in the moduli space where the full $U(N)$ gauge group is preserved. In particular, there are extra solutions to the equations of motion in the bad case that correspond to the multiple frames: SQCD with gauge group $U(\widetilde N)$ with $F-N \leq\widetilde N\leq \left\lfloor\frac{F}{2}\right\rfloor$ and a free sector describing the residual CB. These extra solutions appear only for specific values of the FI parameter, corresponding to the conditions imposed by the delta functions.

\subsection{Bad quiver theories and the electric dualisation algorithm} \label{subsec: elec_algorithm}
\label{subsec:electric_algo}

While we only considered bad SQCDs in the previous subsection, the discussion can be extended to generic 3d $\cN=4$ unitary quiver gauge theories with bifundamental matter. A quiver theory could be bad if it contains one or more gauge nodes that are underbalanced. There could also be quiver theories that are bad, that do not contain any gauge node. For the purpose of this paper, we will focus only on the former case, while the analysis of the latter is beyond the scope of this paper. One may then expect that bad quiver gauge theories enjoy properties similar to those of the SQCD, in the sense that their $S^3_b$ partition functions can be written as a sum over frames, each corresponding to a good theory multiplied by a Dirac delta, and (possibly) a frame with the lowest rank with no Dirac delta. From the partition function perspective, it is then possible to extend the classification also to quiver theories. An \textit{ugly} quiver theory has a partition function that is equal to that of a good theory times a free sector. When more than one frame is present, we say that the quiver theory is \textit{bad}, and if each frame carries at least one Dirac delta distribution, we say that it is \textit{evil}.

An algorithmic way to derive such a sum-over-frames expression was proposed in \cite{Giacomelli:2024laq}, and is commonly referred to as the \textit{electric (dualisation) algorithm}. While its validity was tested at first only for linear quiver gauge theories, the algorithm can be applied to generic quivers. The algorithm consists of the following steps:
\begin{enumerate}
    \item Choose a bad gauge node in the initial quiver gauge theory, with no preferred choice.
    \item Un-gauge the neighbouring gauge nodes, {\it i.e.}, freeze the gauge interactions between the chosen bad node and those connected to it by bifundamental hypermultiplets. This amounts to locally extracting a bad SQCD.
    \item Use \eqref{eq:bad_sqcd} or \eqref{eq:evil_sqcd}, as appropriate, to {\it dualize} the bad SQCD by replacing it with a sum of frames.
    \item Each frame must then be glued back to the remaining quiver by re-introducing the gauge interactions that were frozen in step 2. This will generically produce a sum of frames corresponding to the original bad quiver, where each frame represents a quiver where the bad node chosen in Step 1 has been replaced by a good node.
    \item Starting from the quiver gauge theories in each frame, locate and choose a bad node and repeat steps 2 to 4 to obtain a set of new frames, for each frame at the end of step 4. Repeat this until every frame generated by repeated dualisations contains no bad node.\footnote{Note that the algorithm may produce additional ``unphysical” frames that must be discarded. A physical frame is required to satisfy two consistency conditions: 1) The constraints from the set of delta functions associated with a frame must admit a solution for generic values of $m$ and $Q$, 2) the singlets appearing in the frame must remain non-singular upon imposing the delta function constraints.}
\end{enumerate}

\begin{figure}
    \centering
    \includegraphics[width=\linewidth]{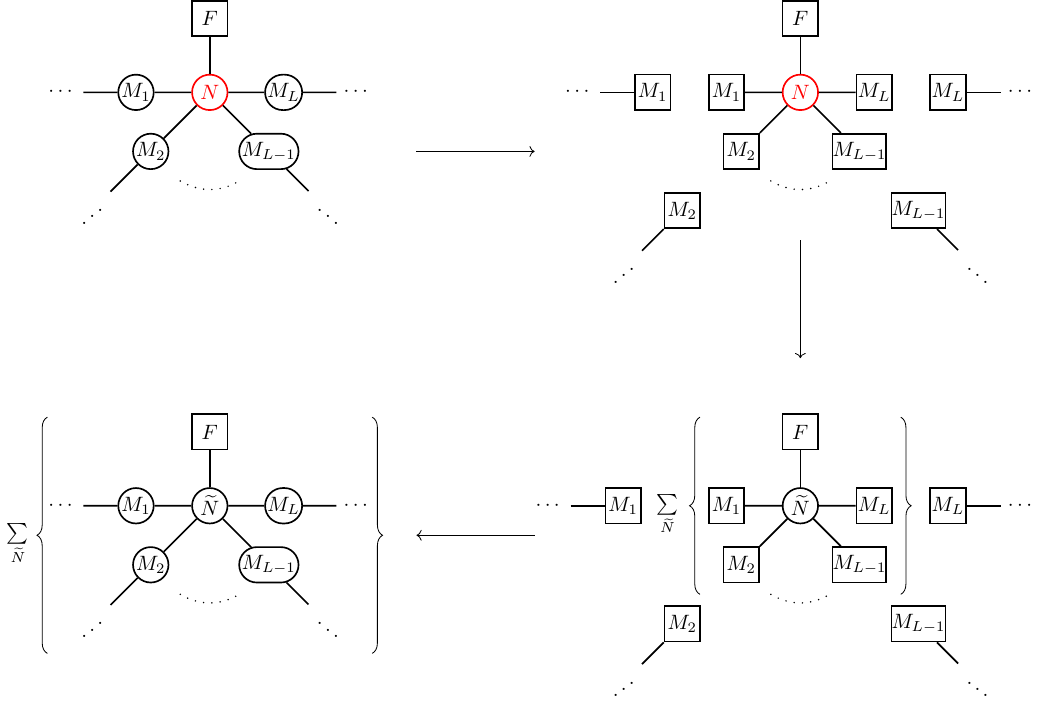}
    \caption{Schematic representation of the electric algorithm for bad theories. This represents the steps 1 through 4 of the algorithm that first isolate a local bad SQCD (we consider $\sum_{j=1}^L M_j+F<N$ so that the $U(N)$ gauge node is bad) by ungauging the neighbours, followed by a dualisation using \eqref{eq:bad_sqcd} or \eqref{eq:evil_sqcd} and then re-gauging. Since the replacement $N\rightarrow\widetilde N$ is a reduction in the gauge rank, some of the neighbours may become bad. Thus, the badness generically propagates, and these four steps need to be iterated as explained in step 5 until there are no bad nodes remaining in any frame.}
    \label{fig: elec_algorithm}
\end{figure}

Steps 1 to 4 of this algorithm are summarised schematically in Figure \ref{fig: elec_algorithm}.
Starting from the $S^3_b$ partition function of the initial quiver gauge theory, the electric algorithm along with formulae \eqref{eq:bad_sqcd} and \eqref{eq:evil_sqcd} furnish for an exact partition function identity. By carefully keeping track of all the details during the algorithm, one obtains a sum of frames with all the relevant physical information, such as the delta functions, charges of the free chirals, and any possible BF couplings, explicitly.

While \cite{Giacomelli:2024laq} developed the electric algorithm to study bad linear quivers, we instead focus on the more general bad {\it star-shaped} quivers. In particular, we are interested in those that arise from as the 3d mirrors of theories of class $\cS$, crucially including those that correspond to 4d configurations that arise in certain ``bad" channel decompositions. As already mentioned at the end of Subsection \ref{subsec:3dmirror_classS} and shown in Appendix \ref{app:roadmap_4dto3d}, these cases correspond to 3d mirrors with a bad central node and $T_\rho[U(N)]$ tails.

We have implemented the electric algorithm for star-shaped quivers with a bad central node in \texttt{Mathematica}, which is available as an ancillary file. While the result of the dualisations using the electric algorithm does not depend on the order in which the bad nodes are chosen and dualised (see Step 1 of the algorithm), we fix an order in the code for concreteness. Since only the central node is bad in the initial quiver, we first dualise the central node. We then dualise the bad nodes that arise in the tails due to this and go tail-wise. Once all the bad nodes have been dualised, we may have a bad central node once again. This process is then repeated until there are no bad nodes remaining in the full quiver. We refer to the actual code for more details.

\subsection{Broken theories and spontaneous symmetry breaking}
\label{subsec:brokenbad_SSB}

We now present a subclass of 3d $\mathcal{N}=4$ \textit{bad} theories that we coin \textit{broken} theories. These are bad theories characterised by the fact that, upon running the electric algorithm (see Section \ref{subsec:electric_algo}), their $S^3_b$ partition function consists of a collection of frames that are all identical. This implies that the frames are all equal up to the action of a global symmetry transformation that reshuffles the real parameters. As an immediate consequence, a \textit{broken} theory must be an \textit{evil} theory, since the presence of a frame without Dirac deltas immediately implies the existence of a distinct frame. 

We now attempt to interpret the physics of these theories. As already noted in Subsection \ref{subsec:badSQCD}, evil theories are characterised by the fact that they do not possess vacua for generic values of the parameters, but instead only for their special values. Each individual frame corresponds to a collection of monopole operators acquiring a VEV, implying that the topological symmetry must be partially broken, as indicated by the Dirac deltas that freeze the FI parameters. The distinguishing property of broken theories is that the set of monopole operators that acquire a VEV and thus trigger the flow to each frame is equivalent -- implying that they are the same up to global symmetry transformations.

This is in contrast to a generic evil theory where the set of monopoles driving the theory to a good frame is inequivalent among the frames. For example, in the evil SQCD case, the various frames are reached by giving a VEV to monopoles with different levels of dressing, making them inequivalent operators. We can then conclude that \textit{broken} theories only possess a single vacuum, up to global symmetry transformations; therefore, the global symmetry of the original bad theory is spontaneously broken\footnote{Hence the name {\it broken} theories.} in this vacuum.

In what follows, we provide a couple of concrete examples of theories that exhibit this behaviour. The presented examples were first discussed in \cite{Gaiotto:2008sa} and then later related to class $\cS$ theories in \cite{Gaiotto:2011xs}. In the following section, and then the remaining bulk of the paper, we will discuss the key role of broken theories in the study of channel decomposition of theories of class $\cS$, providing a novel infinite class of examples of broken theories.

\subsubsection*{Example 1: Gluing two full quiver tails}\label{subsec: twoTSUN_gluing}

Let us consider the theory obtained by fusing together two $T[U(N)]$ tails along the Higgs branch. This corresponds to a sphere with two maximal punctures and was studied from a similar perspective in \cite{Gaiotto:2011xs}. Below, we present the quiver theory along with a compact notation to represent it:
\begin{align}\label{fig: gluing_2tsun}
    \includegraphics[]{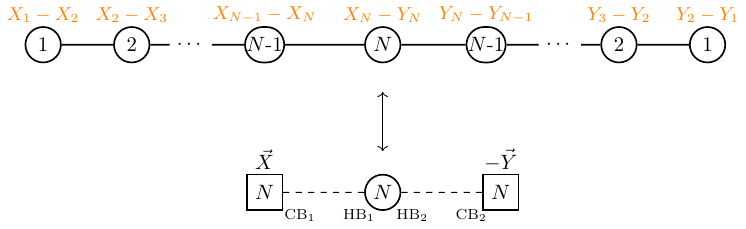}
\end{align}

As discussed in Section \ref{subsec:SSQdecomp_identity}, this corresponds to the HB gluing of two full tails yielding, at the level of the $S^3_b$ partition function, the Dirac delta function that identifies the Cartans of the two $U(N)$ global symmetries in \eqref{eq:tsun_delta}. However, as observed in \cite{Giacomelli:2024laq}, we can also regard the quiver resulting from the gluing of two $T[U(N)]$ theories as a bad theory. Indeed, the central $U(N)$ node sees only $N-2$ fundamental hypermultiplets; therefore, it is bad and can be dualised into a sum of two frames using \eqref{eq:bad_sqcd}. The rank of the central node becomes at most $N-1$ after the dualisation, leading to the two $U(N-1)$ nodes becoming bad, and the procedure thus propagates into the two tails.

By running the dualisation algorithm, we recover \eqref{eq:tsun_delta} that we repeat here for convenience:
\begin{equation}\label{eq:tsun_delta2}
    \quad 
    \int {\rm d}\vec{U}_{N} \;\D^{\rm HB}_{N}(\vec{U};m)  \PFT{}{N}(\vec{U},\vec{X};m) \PFT{}{N}(\vec{U},-\vec{Y};m) = \sum_{\sigma \in S_N} \frac{ \prod_{j=1}^N \d( X_j - Y_{\sigma(j)} )}{\D^{\rm CB}_N(\vec{X};m)} ~.
\end{equation}
The result is a sum over $N!$ identical frames, each corresponding to a collection of Dirac delta functions and chiral multiplets that can be repackaged to reconstruct the (inverse of) the contribution of a $\mathcal{N}=4$ vector multiplet of a $U(N)$ symmetry $\D_N^{\rm HB}(\vec{X};m)$. We once again refer the reader to Appendix \ref{app:S3b_partfn} for the $S^3_b$ partition function convention and the definition of the partition function of the $T[U(N)]$ theory.

This theory then has characteristic features of a broken theory, being a sum of identical frames such that each frame contains the same good theory, the same number of delta functions expressing the same constraints, and the same chiral singlets. The original bad UV theory in \eqref{fig: gluing_2tsun} has a naive global symmetry which is $S[U(N)\times U(N)]$, where each $U(N)$ factor arises on the CB from the balanced nodes of a single tail. However, the theory is bad, and the global symmetry is spontaneously broken in the IR to the diagonal $U(N)$ subgroup. In fact, the frames are rotated by an $S_N$ symmetry, which is the Weyl group of $U(N) \times U(N)$ modded by Weyl transformations of the unbroken $U(N)$ symmetry.

The operator acquiring a VEV and inducing the spontaneous symmetry breaking is a collection of monopole operators in \eqref{fig: gluing_2tsun} whose scaling dimension is 0, when computed using the UV R-symmetry group. These monopoles are  those with non-zero magnetic flux through a sequence of adjacent gauge nodes—with no gaps—that includes the central
$U(N)$ gauge group. There are a total of $N^2$ of these, and they can be assembled in a bifundamental representation of the naive global symmetry $U(N) \times U(N)$. Let us call this operator  $\Pi_{i,j}$ with $i,j=1,..,N$.

Recall, as commented below \eqref{eq:bad_sqcd}, the set of operators acquiring a VEV can be identified by looking at the argument of the delta functions. The set of delta functions in \eqref{eq:tsun_delta2}, in the frame where $\sigma$ is the identity element of $S_N$, is $\prod_{j=1}^N \delta( X_j - Y_j )$, and the argument corresponds to the charges of the diagonal components of the bifundamental, $\Pi_{jj}$. This is a monopole that carries magnetic flux $+1$ under all the gauge nodes from the $U(j)$ node of the left tail to the  $U(j)$ node of the right tail. Indeed, an equal VEV for all the $\Pi_{jj}$ causes the breaking of the $U(N) \times U(N)$ global symmetry to the diagonal $U(N)$. In the sum-over-frames, we are simply summing over the many possible ways to identify together two $U(N)$ gauge groups. When part of the $U(N) \times U(N)$ global symmetry is gauged, then the VEV causes a partial Higgsing of the gauge group.

\subsubsection*{Example 2: Gluing a full and a regular quiver tail}

Let us now consider the case where we glue a full tail to a regular tail, a.k.a.\ the $T_\rho[U(N)]$ quiver, by gauging the diagonal combination of their $U(N)$ HB symmetries. We depict this in Figure \ref{fig: tsun&trho_delta} along with the corresponding compact notation:
\begin{align}\label{fig: tsun&trho_delta}
    \includegraphics[width=0.8\linewidth]{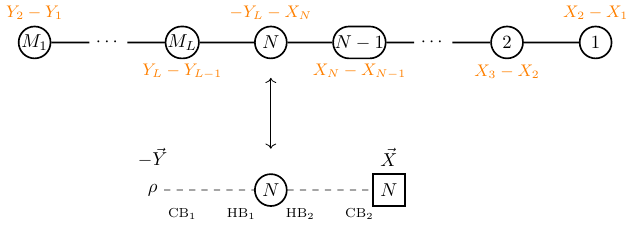}
\end{align}

The central node is underbalanced, and by running the electric dualisation algorithm, we find that for any regular tail, this theory has the features of a broken bad theory: it is a sum of identical frames, each with the same good theory, the same number of delta functions expressing the same constraints, and the same number of singlets.

In the case of $\rho = \left[N-M,1^M\right]$ (corresponding to ranks $M_i=i$ and $L=M$), the result was already computed in \cite{Bottini:2021vms} and reads:
\begin{equation}\label{eq:tsuntrho_delta}
\begin{split}
    & \int {\rm d}\vec{U}_{N} \;\D^{\rm HB}_{N}(\vec{U};m) \PFT{[N-M,1^M]}{N}(\vec{U},-\vec{Y};m)
     \PFT{}{N}(\vec{U},\vec{X};m)
    = \\
    & = \sum_{\sigma \in S_N} \underbrace{\prod_{j=1}^M \d( X_{\sigma(j)} - Y_j )}_{\dd{\text{ex}}} 
    \underbrace{\prod_{j=M+1}^N \d\left( X_{\sigma(j)} - Y_{M+1} + \frac{N-M+1-2j}{2}(iQ - 2m )\right)}_{\dd{\text{Nil}}} 
    \times \\
    & \qquad \times 
    \frac{1}{\D^{\rm CB}_N(\vec{X};m)\;\text{Ch}_{\text{Nil}}^{\left[N-M,1^M\right]}}\,.
\end{split}
\end{equation}
where $Y_{1,\ldots,M}$ and $Y_{M+1}$ parametrise the $U(M)_{\vec{Y}} \times U(1)_{Y_{M+1}}$ CB 
symmetry of the $\rho=\left[N-M, 1^M\right]$ quiver tail.

The term $\text{Ch}_{\text{Nil}}^{\left[N-M,1^M\right]}$ is a set of chiral multiplets whose role will be clarified around \eqref{eq: tsun&trho_chirvev}. The delta functions identify the Cartans of the $U(N)_{\vec{X}}$ global symmetry associated to the CB of the full tail $T[U(N)]$ with the Cartans of the the $U(M)_{\vec{Y}} \times U(1)_{Y_{M+1}}$ global symmetry carried by the $T_{\left[N-M,1^M\right]}[U(N)]$ theory. As in the previous example corresponding to the fusion of two full tails, we could read the set of monopole operators that acquire a VEV from the argument of the delta functions. Such a VEV causes the spontaneous breaking of the  $U(N)_{\vec{X}}\times U(M)_{\vec{Y}} \times U(1)_{Y_{M+1}}$ global symmetry down to a diagonal $U(M) \times U(1)$.

Furthermore, we like to distinguish the contributions of the Dirac deltas in two sets: $\dd{\text{Ex}}$ and $\dd{\text{Nil}}$. Firstly, we notice that the argument of the Dirac delta functions in $\dd{\text{Nil}}$ can be linearly combined to imply the following constraints (WLOG consider $\sigma = \mathbb I$):
\begin{equation}\label{eq: tsun&trho_constrnil}
\begin{split}
    &X_{j} - X_{j+1} = iQ - 2m  \qquad\qquad\quad,\; j=M+1,\ldots,N-1 ~, \nn \\
    &X_{N} = Y_{M+1} + \frac{M-1}{2}(iQ-2m) ~.
\end{split}
\end{equation}
The first line can be interpreted as the constraint imposed by a nilpotent VEV
\begin{equation}\label{eq:C2nilvev}
    \left\langle \, \mathcal{C}_{2} \, \right\rangle \sim \mathcal{J}_{\left[N-M,1^M\right]} ~,
\end{equation}
where $\mathcal{C}_{2}$ is the moment map associated to the $U(N)_{\vec{X}}$ global symmetry on the CB of the full tail. This breaks $U(N)_{\vec{X}}\longrightarrow U(M)_{\vec{X}} \times U(1)_{X_N}$ by identifying the Cartans of $U(N-M)_{\vec{X}} \subset U(N)_{\vec{X}}$ with the one of $U(1)_{X_N}$. The second line of \eqref{eq: tsun&trho_constrnil} instead corresponds to the constraint imposed by a VEV acquired by the monopole operator in \eqref{fig: tsun&trho_delta} with non-zero magnetic charge only under the central node. This has the effect of further identifying $U(1)_{X_N}$ with $U(1)_{Y_{M+1}}$. On the other hand, the set $\dd{\text{Ex}}$ simply plays the role of identifying the Cartans of the surviving $U(M)_{\vec{X}}$ symmetry of the full tail to the $U(M)_{\vec{Y}}$ symmetry of the regular tail. Overall, the monopole VEVs are responsible for the spontaneous breaking of the UV symmetry $U(N)_{\vec{X}} \times U(M)_{\vec{Y}} \times U(1)_{Y_{M+1}} \longrightarrow U(M)_{\vec{Y}} \times U(1)_{Y_{M+1}}$.

\begin{figure}
    \centering
    \includegraphics[width=0.85\hsize]{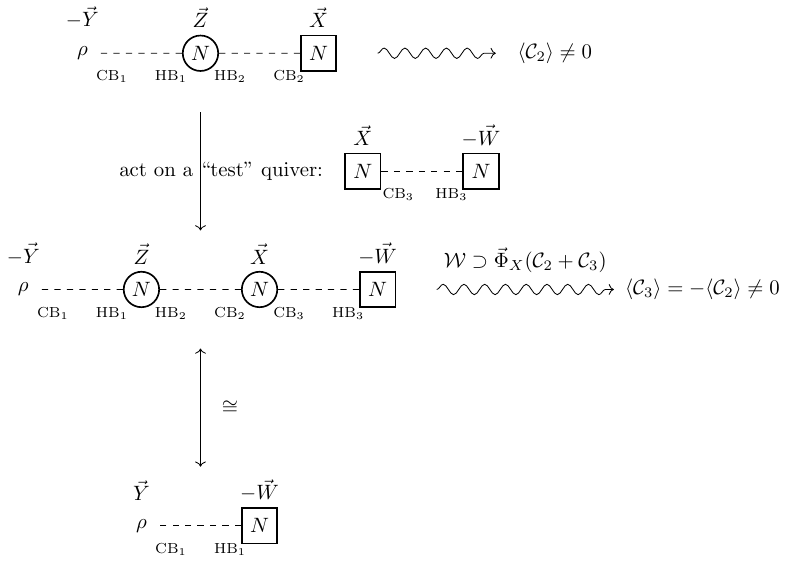}
    \caption{The first theory, the regular quiver tail, is glued to the second theory, the full tail, via HB gauging (this is the theory in \eqref{fig: tsun&trho_delta}). We then gauge the CB moment map of the second quiver theory with another (test) full quiver tail.}
    \label{action}
\end{figure}

We now want to understand what happens when we gauge the $U(N)_{\vec X}$ CB symmetry of the full tail in \eqref{fig: tsun&trho_delta}. As we have seen already, the partition function of this theory is a distribution, and we need to consider its action on a test function to understand its effect. We choose another full quiver as test quiver theory, {\it i.e.} consider the case in which the quiver \eqref{fig: tsun&trho_delta} is glued to another full tail. We consider the gluing corresponding to gauging the diagonal CB $U(N)_{\vec{X}}$ symmetry as depicted in Figure \ref{action}.

We refer to the moment maps of the first, second, and third theories as $\mathcal{C}_{1,2,3}/\mathcal{H}_{1,2,3}$. As explained above, the theory in the first line in Figure \ref{action}, obtained via the $U(N)_{\vec{Z}}$ gauging of the regular and the full tail, is bad and there is a non vanishing VEV for the CB $U(N)_{\vec X}$ moment map $\mathcal{C}_2$. We now let this theory ``act," via the gauging of the $U(N)_{\vec X}$ node on the ``test" full tail, as in the second line in Figure \ref{action}. Due to the gluing superpotential, $\mathcal{W} = \phi (\mathcal{C}_2 + \mathcal{C}_3)$, the HB moment map $\mathcal{C}_2$ of the test theory also acquires a non-zero VEV: $\left\langle\mathcal{C}_2\right\rangle=-\left\langle\mathcal{C}_3\right\rangle\neq 0.$

This implies that even in the test $T[U(N)]$ quiver theory, there is a nilpotent VEV that breaks its CB $U(N)_{\vec{X}}$ symmetry to $U(M)_{\vec{Y}} \times U(1)_{Y_{M+1}}$. Therefore, from the quiver theory in the second line in Figure \ref{action}, depicting the concatenation of a regular and two full tails, one obtains the regular tail, $T_{\left[N-M,1^M\right]}[U(N)]$. In terms of the partition function, we have the identity:
\begin{equation}
\begin{split}
    & \int {\rm d}\vec{Z}_N \;\Delta^{\rm HB}_N (\vec{Z};m) 
    \int {\rm d}\vec{X}_N \;\Delta^{\rm CB}_N (\vec{X};m) 
\PFS{T^N_{\left[N-M,1^M\right]}}(\vec{Z},-\vec{Y};m) 
    \PFS{T^N} (\vec{Z},\vec{X};m)
    \PFS{T^N}(-\vec{W},\vec{X};m)  \\
    & \qquad = 
    \int {\rm d}\vec{X}_N \;\Delta^{\rm CB}_N (\vec{X};m) \PFS{T^N}(-\vec{W},\vec{X};m) 
    \sum_{\sigma \in S_N} \prod_{j=1}^M \d( X_{\sigma(j)} - Y_j ) \times \\
    & \qquad \qquad \times 
    \prod_{j=M+1}^N \d\left( X_{\sigma(j)} - Y_{M+1} + \frac{N-M+1-2j}{2}(iQ - 2m )\right)
    \frac{1}{\D^{\rm CB}_N(\vec{X};m)\text{Ch}_{\text{Nil}}^{\left[N-M,1^M\right]}} \\
    & \qquad = 
    \PFS{T^N}(-\vec{W},\vec{X};m) \vert_{\dd{\text{Nil}},\dd{\text{Ex}}} \frac{1}{\text{Ch}_{\text{Nil}}^{\left[N-M,1^M\right]}} \\
    & \qquad = 
    \PFS{T_{\left[N-M,1^M\right]}^N}(-\vec{W},\vec{Y};m) ~,
\end{split}
\end{equation}
where the steps are self-explanatory, and crucially, we have used that,
\begin{equation}\label{eq: tsun&trho_chirvev}
    \PFT{}{N}(-\vec{W},\vec{X};m)\vert_{\dd{\rm{Nil}}} =
    \PFT{\left[N-M,1^M\right]}{N}(-\vec{W},\vec{Y};m) \times \text{Ch}^{\left[N-M,1^M\right]}_{\text{Nil}} ~.
\end{equation}
This equality implies that the $T[U(N)]$ quiver evaluated on the constraint set by $\dd{\text{Nil}}$, corresponding to the nilpotent VEV given in \eqref{eq:C2nilvev} for the CB moment map $\mathcal{C}_3$, flows to the quiver tail $T_\rho[U(N)]$ with $\rho=\left[N-M,1^M\right]$ together with a free sector which exactly cancels with the chirals produced in \eqref{eq:tsuntrho_delta}. We refer to Appendices \ref{app:TSUN_TRHOSIGMA} and \ref{app:higgsing_TSUN} for further details.

Evidently, we can also regard all of this as a consistency check. Indeed, if we first imagine the fusion of the two tails along $U(N)_{\vec X}$ and then use \eqref{eq:tsun_delta}, we get an Identity wall glued to a regular tail, which clearly coincides with the regular tail itself.

\subsubsection*{Broken theories in class $\mathcal{S}$ }

Going beyond these two examples, we return to the star-shaped quivers obtained as 3d mirrors of theories of class $\cS$. We observe that the various bad star-shaped quivers that emerge upon the channel decomposition of these good star-shaped quivers are broken theories. As such, based on the collection of examples that we have analysed, we propose the following conjecture.
\begin{conjecture}\label{conj:broken}
    Every 3d $\mathcal{N}=4$ star-shaped quiver with a $T[SU(N)]$ tail and any number of generic $T_\rho[SU(N)]$ tails, such that the central node is underbalanced, is a broken theory. Moreover, the number of frames is given by the following ratio of dimensions of the Weyl groups:
    \begin{equation}
        \# \text{ of frames } =\; \frac{\lvert W_{U(N)} \rvert}{\lvert W_{H\subset U(N)} \rvert} ~,
    \end{equation}
    where $H\subset U(N)$ is the subgroup of the global symmetry $U(N)$, corresponding to the full tail, that is preserved by the monopole VEVs.
\end{conjecture}
The various frames are all rotated by a discrete permutation symmetry that reflects the number of ways (related by $G$-conjugation) a fixed embedding of $H$ into $U(N)$ can be realised.  We are now finally ready to study the channel decompositions that give rise to bad star-shaped quivers and exemplify our discussion on broken theories, using the electric dualisation algorithm.

Apart from these ones that naturally emerge in the class $\cS$ setting, we expect many other broken theories to exist. For example, in an upcoming work \cite{badusp}, more broken theories will be analysed, extending and generalising the class described in this paper. We also expect them to play an important role in the problem of understanding the gauging of emergent Coulomb branch symmetries of 3d $\mathcal{N}=4$ theories \cite{cbgauge}. 

\section{Star-shaped quivers with two full  and two regular tails}
\label{sec:goodSSQ_twofull}

In this section, we bring together all the discussions so far to study the channel decompositions of star-shaped quivers. In particular, we describe a procedure to analyse the cases where the quivers obtained from channel decompositions are bad quiver gauge theories. For concreteness, we restrict the presentation in this section to good star-shaped quivers with four tails: two full $T[U(N)]$ tails, and two generic $T_{\rho_{1,2}}[U(N)]$ tails.

\begin{figure}[]
    \centering
    \includegraphics[width=0.8\linewidth]{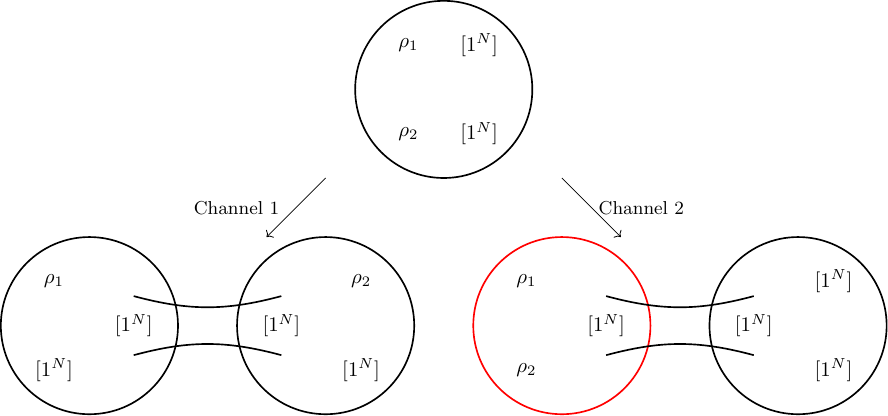}
    \caption{Schematic depiction of the channel decomposition of a four-punctured sphere, with two maximal and two generic regular punctures $\rho_{1,2}$. In the first channel, we isolate a maximal and a regular puncture, whereas in the second channel, we isolate the two regular punctures. For specific choices of $\rho_1,\rho_2$, the trinion, highlighted in red, in the second channel may be a pathological 4d theory and thus correspond to a bad 3d star-shaped quiver.}
    \label{fig: 2f2s_generic}
\end{figure}

As shown in Figure \ref{fig: 2f2s_generic}, there are two distinct channel decompositions for this setup. The focus of this section is the bad trinion emerging in the second channel decomposition that contains the two regular and a maximal puncture. In the case where this trinion is a pathological 4d configuration, as discussed in Subsection \ref{subsec:irregpunc}, we will observe that the corresponding 3d mirror is a broken theory, in accordance with Conjecture \ref{conj:broken}. We will then carefully examine the effect of the renormalisation group flow triggered by the monopole VEVs present in the broken trinion, and examine how it changes the gauge symmetry and the other trinion that contains the three maximal punctures.

We implement this procedure in several examples and analyse the theories that emerge at the endpoints of the flows. In all cases, we find perfect consistency with the three-dimensional counterparts of the results obtained via the ``atomic" tinkertoys program \cite{Chacaltana:2010ks}. Our approach offers a systematic framework allowing us to precisely determine the matter sector in each weakly-coupled channel. In particular, we can exactly determine the interacting part in $\widetilde S_1$, $\widetilde S_2$ and the representation of the twisted hypers. Notably, our approach provides the exact map of the fugacities for the global symmetries across each channel. 

In the next section, we will see that our method streamlines the analysis of the theories with many punctures, as it allows us to bypass the channel decomposition into trinions.

\subsection{Generic features under channel decomposition}
\label{subsec:channel_decomposition}

We will now discuss some general features of the possible channel decomposition of star-shaped quivers. We restrict the presentation to the example of two full and two generic regular tails for concreteness. We label the 3d mirror of the four-punctured sphere as $S$, and its partition function as $\PFS{S}\left(\vec{X}^{(1)}_{\rho_1}, \vec{X}^{(2)}_{\rho_2}, \vec{X}^{(3)}, \vec X^{(4)}; m\right)$, which depends on the real mass parameters for the various topological symmetries as described in Subsection \ref{subsec:partfun_SSQ}.

Recall that the star-shaped theory, $S$, contains a redundant $U(1)$ gauge transformation that can be removed by performing an integration over a suitably chosen combination of the FI parameters, as in \eqref{prescription to decouple U(1)}. For the chosen case, this combination is given by:
\begin{equation}\label{eq: Ccomb_initS}
    C = \sum_{j=1}^N (X^{(3)}_j + X^{(4)}_j) + f_{\rho_1}(\vec{X}^{(1)}) + f_{\rho_2}(\vec{X}^{(2)}) 
\end{equation}
where $f_{\rho}(\vec{X})$ is a linear combination of the $\vec{X}$ parameters that depends on the choice of partition $\rho$. The partition function without the redundant $U(1)$ can then be obtained as:
\begin{equation}\label{eq: pf_hatS}
    \widehat{\PFS{S}}\left(\vec{X}^{(1)}_{\rho_1}, \vec{X}^{(2)}_{\rho_2}, \vec{X}^{(3)}, \vec X^{(4)}; m\right) = \int {\rm d}C \;\DCB{1} (C;m) \;
    \PFS{S}\left(\vec{X}^{(1)}_{\rho_1}, \vec{X}^{(2)}_{\rho_2}, \vec{X}^{(3)}, \vec X^{(4)}; m\right) \,.
\end{equation}

We study the two equivalent decompositions that the chosen theory $S$ admits: the one where the full tail is paired with a regular, and the other one where the two regular punctures are paired together. These channel decompositions are schematically depicted in Figure \ref{fig: 2f2s_generic}, and we now discuss them in detail.

\paragraph{Channel I:}
In the decomposition where the regular punctures are split among the two trinions, as on the bottom left side of Figure \ref{fig: 2f2s_generic}, based on Subsection \ref{subsec:channel_decomposition}, we have:
\begin{equation}\label{boring}
\begin{split}
    & \PFS{S}\left(\vec{X}^{(1)}_{\rho_1}, \vec{X}^{(2)}_{\rho_2}, \vec{X}^{(3)}, \vec X^{(4)}; m\right) = \\
    &=\int d\vec{U}_{N} \;\D_{N}^{\rm CB}(\vec{U};m)
    \PFS{S_1}\left(\vec{X}^{(1)}_{ \rho_1 }, \vec X^{(3)} ,\vec{U}; m\right) \PFS{S_2}\left(-\vec{U}, \vec{X}^{(2)}_{ \rho_2 }, \vec X^{(4)}; m\right)\,.
\end{split}
\end{equation}
The trinions $S_1$ and $S_2$ isolated in this channel decomposition carry two maximal and a regular puncture and are thus always good 3d $\cN=4$ theories.

\begin{figure}[]
    \centering
    \includegraphics[width=.9\linewidth]{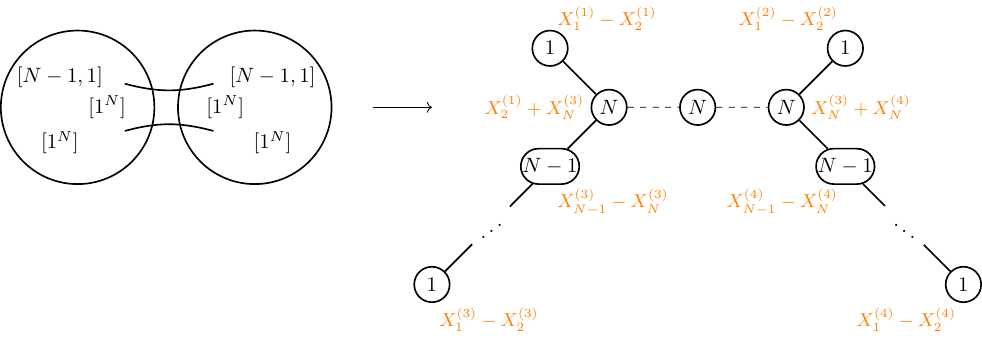}
    \caption{3d mirror quiver representation of the Channel I decomposition in the special case $\rho_{1,2}=[N-1,1]$. }
    \label{fig:2minimalgeneralNstart}
\end{figure}

\begin{figure}[]
    \centering
    \includegraphics[width=.8\linewidth]{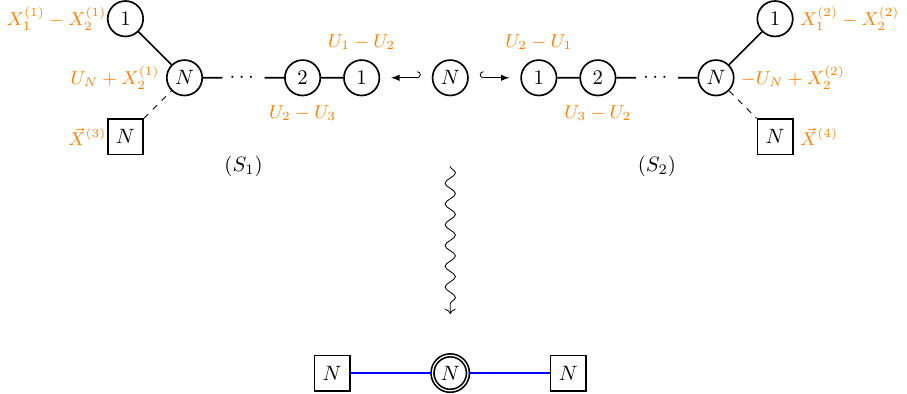}
    \caption{Channel decomposition isolating two ugly star-shaped quivers $S_1$, $S_2$. Iterative dualizations of the ugly nodes yield two sets of (twisted) hypers which are then gauged together to realise the $U(N)$ SQCD with $2N$ flavours. Note that the hypers shown in blue are twisted with respect to those in the original theory.
  }
    \label{fig:2minimalfig10}
\end{figure}

In the particular case depicted in Figure \ref{fig:2minimalgeneralNstart} in which $\rho_1$ and $\rho_2$ are minimal punctures, i.e. $\r_{1,2}=[N-1,1]$, the trinions isolated by the channel decomposition are free theories consisting of a set of $N^2$ free twisted hypers. Correspondingly, the 3d star-shaped quivers $S_1$ and $S_2$, emerging from this channel decomposition, which we report at the top of Figure \ref{fig:2minimalfig10}, are {\it ugly} theories. Indeed, the central node in $S_{1,2}$ is ugly, {\it i.e.} it sees $2N-1$ fundamental hypers. As pointed out in \cite{Gaiotto:2008ak}, the ugly $U(N)$ SQCD with $2N-1$ hypers is dual to a $U(N-1)$ SQCD and a free twisted hypermultiplet.

After using this duality to reduce the rank of the central node, the $U(N-1)$ nodes adjacent to it see one less flavour. Since they were previously balanced, this duality makes them ugly instead. Re-applying the duality once again on each $U(N-1)$ replaces them with a $U(N-2)$ node along with the contribution of a free twisted hypermultiplet. As before, this causes the neighbouring nodes to become ugly, and hence we use the duality for the ugly SQCD once again. This procedure iterates repeatedly and can easily be seen to terminate once the remaining theory has no gauge groups and thus only consists of a collection of free twisted hypermultiplets. Evidently, we could have also run the \textit{electric algorithm} for bad theories, explained in Subsection \ref{subsec: elec_algorithm}. This way, each dualisation step uses only the duality for the ugly SQCD, and we obtain the same result.

For theory $S_1$, at the level of the partition function, we obtain
\begin{equation}\label{eq: 2full1simp_genn}
\begin{split}
    \PFS{S_1}\left(\vec{X}^{(1)}_{[N-1,1]}, \vec{X}^{(3)},\vec U;m\right) =
    & \prod_{j,k=1}^N s_b \left(m \pm (X^{(1)}_2 + U_k + X^{(3)}_j) \right) \times \\
    & \qquad\times \frac{\delta\left( \sum_{j=1}^N (U_j + X^{(3)}_j) + X^{(1)}_1 + (N-1)X^{(1)}_2\right)}{ \D_1^{\rm CB}(\vec{U};m) } \,.
\end{split}
\end{equation}

The collection of double-sine functions encodes the $N^2$ free hypermultiplets\footnote{Note that these are twisted hypermultiplets from the perspective of the star-shaped quiver. However, since the star-shaped quivers are themselves the 3d mirrors of the 4d theories of class $\cS$, the free hypermultiplets obtained here are the direct circle reductions of those in four dimensions.} that are in a bifundamental representation of the $U(N)\times U(N)$ global symmetries of the original theory, where each $U(N)$ is the symmetry associated to each maximal puncture\footnote{Indeed, since the theory is free, the free hypers have a bigger global symmetry $USp(2N^2)$. However, since we will study the theory only while it is gauged to some other matter, it is important to know how the free hypers are charged under the naive global symmetry -- as read from the punctures or by looking at the balancing of the nodes in the star-shaped quiver.} The Dirac delta and the term $\Delta_1^{\rm CB}(\vec{U},m)$ in \eqref{eq: 2full1simp_genn} are present due to the redundant $U(1)$ gauge symmetry present in the original $S_1$ theory. These two factors in \eqref{eq: 2full1simp_genn} are indeed removed once we decouple the redundant $U(1)$ from $S_1$, thus leaving only $N^2$ free hypermultiplets.

Plugging this result, along with the analogue for $S_2$, in \eqref{boring}, we obtain the following expression:
\begin{equation}
\begin{split}
    \PFS{S} & \left(\vec{X}^{(1)}_{[N-1,1]}, \vec{X}^{(2)}_{[N-1,1]}, \vec{X}^{(3)}, \vec X^{(4)}; m\right) = \\
    & = \int {\rm d}\vec{U}_{N} \;\D_{N}^{\rm CB}(\vec{U};m) \prod_{j,k=1}^N s_b \left(m \pm (X^{(1)}_2 + U_k + X^{(3)}_j)  \right) s_b \left(m \pm (X^{(2)}_2 - U_k + X^{(4)}_j)  \right) \\
    & \qquad\times \frac{\delta\left( \sum_{j=1}^N (U_j + X^{(3)}_j) + X^{(1)}_1 + (N-1)X^{(1)}_2\right)}{ \D_1^{\rm CB}(\vec{U};m) } \\
    & \qquad\times \frac{\delta\left( \sum_{j=1}^N (-U_j + X^{(4)}_j) + X^{(2)}_1 + (N-1)X^{(2)}_2\right)}{ \D_1^{\rm CB}(\vec{U};m) } \,.
\end{split}
\end{equation}
In the case at hand, where both $\rho_{1,2} = [N-1,1]$, the condition in \eqref{eq: Ccomb_initS} reads,
\begin{equation}
    C = \sum_{j=1}^N (\vec{X}^{(3)}_j + \vec{X}^{(4)}_j) + X^{(1)}_1 + X^{(2)}_1 + (N-1) (X^{(1)}_2 + X^{(2)}_2) \,.
\end{equation}
Thus, to remove the decoupled redundant $U(1)$, we perform the $C$ integration in \eqref{eq: pf_hatS} by implementing one of the Dirac deltas to obtain:\footnote{To perform the $C$ integration, we use that if $C=A+B$ then $\delta(U+A) \delta(-U+B)= \delta(U+A) \delta(A+B)$.}
\begin{equation}
\begin{split}
    \widehat{\PFS{S}} & \left(\vec{X}^{(1)}_{\rho_1}, \vec{X}^{(2)}_{\rho_2}, \vec{X}^{(3)}, X^{(4)}; m\right) = \\
    & = \int {\rm d}\vec{U}_{N} \;\D_{N}^{\rm CB}(\vec{U};m) \prod_{j,k=1}^N s_b \left(m \pm (X^{(1)}_2 + U_k + X^{(3)}_j)  \right) s_b \left(m \pm (X^{(2)}_2 - U_k + X^{(4)}_j)  \right) \times \\
    & \qquad \times 
    \frac{\delta\left( \sum_{j=1}^N U_j - (\sum_{j=1}^NX^{(4)}_j + X^{(2)}_1 + (N-1)X^{(2)}_2 \right)}{ \D_1^{\rm CB}(\vec{U};m) } \,.
\end{split}
\end{equation}
Lastly, we use the remaining Dirac deltas to constrain $\sum_{j=1}^N U_j$, which implies that the $U(1)$ part of the $U(N)$ gauge group is frozen, leaving only an $SU(N)$ gauge symmetry.

The result we obtain is the partition function of the $SU(N)$ SQCD with $2N$ fundamental flavours, where each of the two spheres contributed half of the flavours, as shown at the bottom of Figure \ref{fig:2minimalfig10}. Notice that the hypers (in blue) in the SQCD are twisted w.r.t.~to those in the star-shaped quiver; indeed, this is the direct (non-mirror) 3d reduction of the 4d theory associated with this channel.

\begin{figure}
    \centering
    \includegraphics[width=0.9\linewidth]{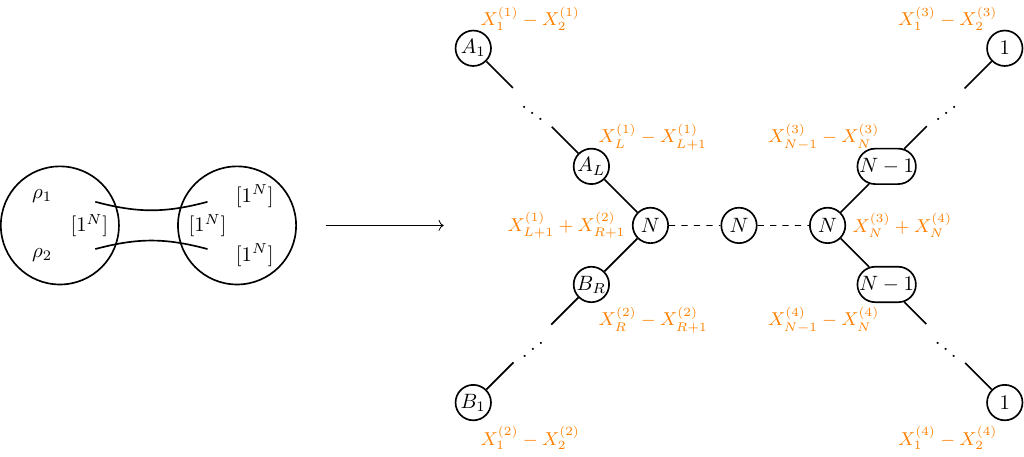}
    \caption{Quiver representation of the channel decomposition corresponding to Channel II. The two tails on the left correspond to the regular punctures, where the ranks $A_i,B_i$ are determined from $\rho_1,\rho_2$, respectively. The two tails on the right are $T[U(N)]$ theories and correspond to the maximal punctures. We recall that the dashed lines represent $T[U(N)]$ theories, so that in the middle, we are gluing two of them, corresponding to the tube in the geometric representation.}
    \label{fig: decomp_genn}
\end{figure}

\begin{figure}[ ]
    \centering
    \includegraphics[width=0.8\linewidth]{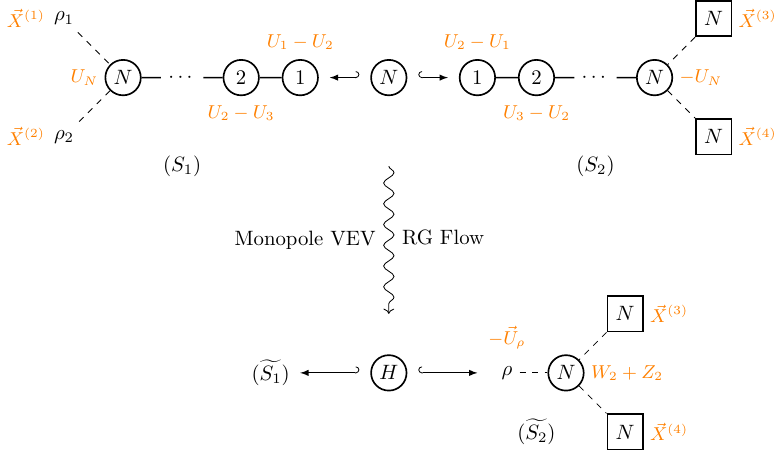}
    \caption{Channel decomposition isolating a broken star-shaped quiver $S_1$ and good quiver $S_2$. By running the electric algorithm on the broken theory and following the RG flow generated by the monopole VEV, we find the diagonal gauging of an $H\subset U(N)$ symmetry of $\widetilde{S}_1$ and $T_\rho[U(N)]$ tail in $\widetilde S_2$. Here $\tilde S_1$ is a possibly interacting theory, while $\tilde S_2$ is the original $S_2$ theory with the full tail Higgsed down to a regular tail $T_\rho[U(N)]$.} 
    \label{fig:enter-label}
\end{figure}

\paragraph{Channel II:}
Let us now consider the case in which the two regular punctures $\rho_1,\rho_2$ are clubbed together in a single trinion as shown in Figure \ref{fig: decomp_genn}. At the level of the partition function, we have:

\begin{align}\label{eq: 2reg2fullGenericFeature}
    & \PFS{S}\left(\vec{X}^{(1)}_{\rho_1}, \vec{X}^{(2)}_{\rho_2}, \vec{X}^{(3)}, \vec X^{(4)}; m\right) = \nn \\
    &=\int {\rm d}\vec{U}_{N} \;\D_{N}^{\rm CB}(\vec{U};m)
    \PFS{S_1}\left(\vec{X}^{(1)}_{ \rho_1 }, \vec{X}^{(2)}_{ \rho_2 }, \vec{U}; m\right) \PFS{S_2}\left(-\vec{U}, \vec{X}^{(3)}, \vec{X}^{(4)}; m\right) \,,
\end{align}
where $S_1$ and $S_2$ are the star-shaped trinions resulting from this channel decomposition, which we report at the top of Figure \ref{fig:enter-label}.

Depending on $\rho_1$ and $\rho_2$, the star-shaped theory $S_1$ can be bad. In this case, by running the \textit{electric dualisation algorithm} we find that its partition function is a distribution that generically takes the form:
\begin{equation}\label{eq: stsh_distrib_gen}
    \PFS{S_1}\left(\vec{X}^{(1)}_{\rho_1}, \vec{X}^{(2)}_{\rho_2},\vec U;m\right)=
    \sum_{\sigma \in W_{U(N)}/W_{H \subset U(N)}} \PFS{\widetilde{S}_1}(\ldots;m)\times \frac{\D_{H \subset U(N)} (\ldots;m)}{\D_N(\vec{U};m) } \times \bf \frac{\dd{\text{Nil}}}{\text{Ch}^\r_\text{Nil}} \times \dd{\text{Ex}} \,.
\end{equation} 
As anticipated from Conjecture \ref{conj:broken}, $S_1$  is a \textit{broken} theory. Thus, the partition function of $S_1$ is a sum of frames that are all equal and correspond to a theory $\widetilde{S}_1$, which can possibly be an interacting theory. Since the partition function of $\widetilde{S}_1$ depends in a complicated fashion on combinations of the 
parameters $\vec{X}^{(1)}_{\rho_1}, \vec{X}^{(2)}_{\rho_2},\vec U$, we avoid giving this general dependence explicitly. As remarked earlier, the various frames are identical and are rotated by a permutation symmetry given by the Weyl group of $U(N)$ modulo that of $H$.

As explained in Subsection \ref{subsec:badSQCD}, the Dirac deltas encode information about the set of monopole operators that acquire a VEV\footnote{Recall that the star-shaped quivers are the 3d mirror dual of the circle reduction of the 4d theories of class $\mathcal{S}$ theory. Therefore, monopole operators of the 3d star-shaped quivers are associated with the Higgs branch operators of the 4d theory.}. In particular, as in the second example in Subsection \ref{subsec:brokenbad_SSB}, the set $\dd{\text{Nil}}$ induces a VEV for some of the components of the moment map of the $U(N)$ global symmetry associated to the full tail of $S_1$. This VEV alone causes the breaking of the $U(N)$ global symmetry down to  $U(M)\subset U(N)$ group. The rest of the Dirac deltas $\dd{\text{Ex}}$ are other monopole operators that acquire a VEV, and are typically charged under the global symmetry of the regular punctures. This second VEV further breaks $U(M)$ to a subgroup $H\subset U(M)$ which we identify to be either an $SU(L)$ or $USp(2L)$ group.\footnote{This is in agreement with the prediction that for theories of class $\cS$ of type $\mf{su}(N)$, the only existing tubes correspond to unitary or symplectic gauge group \cite{Chacaltana:2010ks}.} Indeed, the total number of delta functions satisfies the identity,\footnote{This is true only when $\widetilde{S}_1$ is a free theory. When $\widetilde{S}_1$ is interacting, as in the example in Subsection \ref{subsubsec:su5_2hooks}, one less delta is obtained. This is because the interacting $\widetilde{S}_1$ theory still has a redundant $U(1)$ gauge group left to decouple, which provides the extra Dirac delta.}
\begin{align}
    \# \text{ delta} = N - \text{rank}(H) \,.
\end{align}
Thus, if the $U(N)$ global symmetry associated to the full tail of $S_1$ is gauged, which is the case under consideration, then the VEV causes the Higgsing of $U(N)$ gauge group to $H$.

Let us now review this phenomenon from the point of view of the $S^3_b$ partition function. We first rewrite \eqref{eq: 2reg2fullGenericFeature} as follows:
\begin{equation}
\begin{split}
    & \int {\rm d}\vec{U}_{N} \;\D_{N}^{\rm CB}(\vec{U};m)
    \PFS{S_1}\left(\vec{X}^{(1)}_{ \rho_1 }, \vec{X}^{(2)}_{ \rho_2 }, \vec{U}; m\right) \PFS{S_2}(-\vec{U}, \vec{X}, \vec{Y}; m)= \\
    & = \int {\rm d}\vec{U}_{N} \;\D_{N}^{\rm CB}(\vec{U};m)
    \int {\rm d}\vec{V}_{N} \;\D^{\rm HB}_{N}(\vec{V};m) \times \\
    & \qquad \times\;
    \PFS{S_1}\left(\vec{X}^{(1)}_{ \rho_1 }, \vec{X}^{(2)}_{ \rho_2 }, \vec{U};m\right)
    \PFT{}{N}(-\vec{U},\vec{V};m)
    \PFT{}{N}(\vec{X},\vec{V};m)
    \PFT{}{N}(\vec{Y},\vec{V};m)\,.
\end{split}
\end{equation}
Next, we regard the $U(N)_{\vec{U}}$ integration as the action of the $S_1$ theory \textit{distribution} \eqref{eq: stsh_distrib_gen} on the $T[U(N)]$ tail of the $S_2$ theory:
\begin{align}\label{eq: NilVEV}
    &\int {\rm d}\vec{U}_{N} \;\D_{N}^{\rm CB}(\vec{U};m)
    \PFS{S_1}\left(\vec{X}^{(1)}_{ \rho_1 }, \vec{X}^{(2)}_{ \rho_2 }, \vec{U}; m\right)
    \PFT{}{N}(-\vec{U},\vec{V};m) = \nonumber \\
    &=\int {\rm d}\vec{U}_{N} \;\D_{N}^{\rm CB}(\vec{U};m) \sum_{\sigma \in W_{U(N)}/W_H} \PFS{\widetilde{S}_1}(\ldots;m) \, \frac{\D_{H \subset U(N)} (\ldots;m)}{\D_N(\vec{U};m) } \,  \dd{\rm{Ex}} \, \dd{\rm{Nil}} \, \PFT{}{N}(-\vec{U},\vec{V};m)
\end{align}
As stated earlier, we can deduce the VEV acquired by the CB moment map $\mathcal{C}$ of the maximal puncture of the theory $S_1$ from $\dd{\text{Nil}}$. However, notice that $\mathcal{C}$ is coupled to the moment map $\mathcal{C}'$ of the $T[U(N)]$ tail of the theory $S_2$ via the gluing superpotential, $\Phi_U (\mathcal{C}+\mathcal{C}')$, where $\Phi_U$ is the adjoint chiral in the 3d $\mathcal{N}=4$ vector multiplet. The equations of motion of $\Phi_U$ then imply:
\begin{equation}
    \langle \, \mathcal{C} \, \rangle \sim
    \langle \, \mathcal{C}' \, \rangle \sim
    \mathcal{J}_{\rho} \,,
\end{equation}
where $\mathcal{J}_\rho$ is the Jordan normal matrix associated to the partition $\rho$ of $N$. Since $\mathcal{C}'$ acquires a nilpotent VEV as well, the $T[U(N)]$ tail of the $S_2$ theory is Higgsed to a $T_\rho[U(N)]$ (see Appendix \ref{app:TSUN_TRHOSIGMA}):
\begin{equation}\label{neq: TtoTr_ch}
    \PFT{}{N}(-\vec{U},\vec{V};m)\vert_{\dd{\rm{Nil}}}=
    \PFT{\r}{N}(-\vec{U}_\rho,\vec{V};m) \times \text{Ch}_{\text{Nil}}^\rho \,.
\end{equation}

All in all, starting from \eqref{eq: NilVEV}, we obtain,
\begin{equation}
\begin{split}
  & \int {\rm d}\vec{U}_{N} \;\D_N(\vec{U},m) \sum_{\sigma \in W_{U(N)}/W_H} \PFS{\widetilde{S}_1}(\ldots;m) \frac{\D_{H \subset U(N)} (\ldots;m)}{\D_N(\vec{U},m) } \times \frac{\dd{\rm{Nil}}}{\text{Ch}^\r_{\text{Nil}}} \times \dd{\rm{Ex}} \times \PFT{}{N}(-\vec{U}_\rho,\vec{V};m) \\
  & = \int {\rm d}\vec{U}_{N} \;\D_N(\vec{U},m) \sum_{\sigma \in W_{U(N)}/W_H} \PFS{\widetilde{S}_1}(\ldots;m) \frac{\D_{H \subset U(N)} (\ldots;m)}{\D_N(\vec{U};m) } \times \dd{\rm{Nil}} \times \dd{\rm{Ex}} \times \PFT{\r}{N}(-\vec{U}_\rho,\vec{V};m) \\
  & = \int {\rm d}\vec{U}_{H\subset U(N)} \;\D^{\rm CB}_{H\subset U(N)} (\vec{U};m) \PFS{\widetilde{S}_1}(\ldots;m) \times  \PFT{\r}{N}(-\vec{U}_\rho,\vec{V};m)\vert_{\dd{\rm{Ex}}} ~.
\end{split}
\end{equation}
In the first equality, we have used \eqref{neq: TtoTr_ch}, noticing that the factor ``$\text{Ch}^\rho_\text{Nil}$" from \eqref{eq: stsh_distrib_gen} and from \eqref{neq: TtoTr_ch} cancel out, indicating that the two set of chiral fields give mass to each other and decouple. In the last equality, we have implemented all the Dirac deltas, that effectively break the $U(N)$ gauge group down to $H$, while the correct measure for $H$ is reconstructed thanks to the chirals in the term $\tfrac{\D_{H \subset U(N)} (\ldots;m)}{\D_N(\vec{U};m) }$. Further notice that the sum over the permutation group is absent in the last step, as this was used to take into account the change of the factor $|W_{U(N)}|$ into $|W_H|$ in the integration measure.

Thus, the end result is the gauging of the subgroup $H\subset U(N)$ of the Coulomb branch global symmetry of the theory $T_\rho[U(N)]$ as shown at the bottom of Figure \ref{fig:enter-label}. Through various examples, we will explicitly show how all these details combine together delicately to correctly give rise to this Higgsing, {\it i.e.} the result is simply written as the diagonal gauging of an $H$ symmetry of $\widetilde{S}_1$ and $T_\rho[U(N)]$ tail in $\widetilde S_2$.

If $\tilde S_1$ and/or $\tilde S_2$ are interacting star-shaped quivers, we can further take their mirror duals to obtain the 3d electric reduction of the theory associated with a sphere containing two full and two regular punctures, in the channel where the two regular punctures are clubbed together. If the algorithm also produces a free sector, this consists of a set of free twisted hypermultiplets from the perspective of the original star-shaped quiver (see footnote 23). These hypers are then the direct dimensional reduction of the 4d electric hypers and do not require further mirror dualization.

As mentioned in the introduction, by starting from the theory shown at the bottom of Figure \ref{fig:enter-label} and studying the effect of CB gauging of a subgroup $H \subset U(N)$, we can actually reconstruct the original star-shaped quiver. We will demonstrate this explicitly in the next example.
Although we will not repeat the analysis for all subsequent examples, the technique is easily generalizable. More results on CB gauging will appear in \cite{cbgauge}.

\paragraph{Agreement with the tinkertoys program} 

We close this section by comparing our results with the analysis of \cite{Chacaltana:2010ks}. Starting from a bad theory, we run the electric dualisation algorithm to find an expression of the form \eqref{eq: 2reg2fullGenericFeature}, which can be compared to \cite{Chacaltana:2010ks} as follows:
\begin{itemize}
    \item The $\widetilde{S}_1$ theory corresponds to the sphere with two regular and the corresponding irregular puncture, in the language of \cite{Chacaltana:2010ks}. 
    \item From the collection of delta functions and chirals, we deduce the unbroken gauge group and, in particular, read off the conjugate regular puncture from $\dd{\text{Nil}}$. Thus, we fully characterise the gauge group that connects the irregular puncture to the conjugate puncture (referred to as the tube), in the language of \cite{Chacaltana:2010ks}.
\end{itemize}

An advantage of our proposed method is that the theory resulting from a channel decomposition, which produces a bad trinion, is generated dynamically. Moreover, our method can be directly applied to spheres with any number of punctures without relying on a pair-of-pants decomposition. An example of such an application is discussed in Subsection \ref{subsec:su4_1max3minimal}. We can also apply our method to more complicated situations, such as when the channel decomposition of a four-punctured sphere produces two bad trinions. We report examples of such a decomposition in Subsection \ref{subsec:su4_1minimal2box1ntmax}, obtaining results that are in agreement with those provided in \cite{Chacaltana:2010ks}.

\subsection{\texorpdfstring{$N=3$}{N=3} example: \texorpdfstring{$\rho_{1,2}=[2,1]$}{rho12=[2,1]}}
\label{subsec:su3_example}

\begin{figure}
    \centering
    \includegraphics[width=0.8\linewidth]{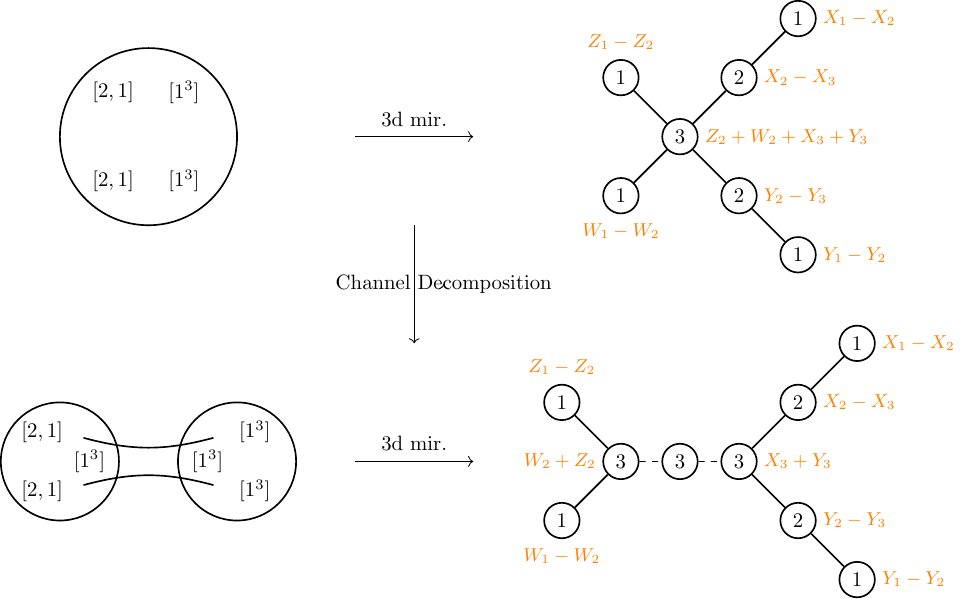}
    \caption{The $N=3$ example with channel decompositions, expressed at the level of the Riemann surfaces associated to the corresponding theory of class $\cS$, and the star-shaped quivers corresponding to their 3d mirror.}
    \label{fig: su3ex_start}
\end{figure}

In this subsection, we study the theories of class $\cS$ of type $\mf{su}(3)$ associated to a sphere with two maximal and two minimal punctures. As discussed in the previous subsection, there are two possible decomposition channels. The first channel, which splits the minimal punctures among the two trinions, corresponds to an $SU(3)$ SQCD with six fundamental hypermultiplets. The other channel, where we group the two regular punctures into a single trinion, is depicted in Figure \ref{fig: su3ex_start}.

\begin{figure}
    \centering
    \includegraphics[width=0.9\linewidth]{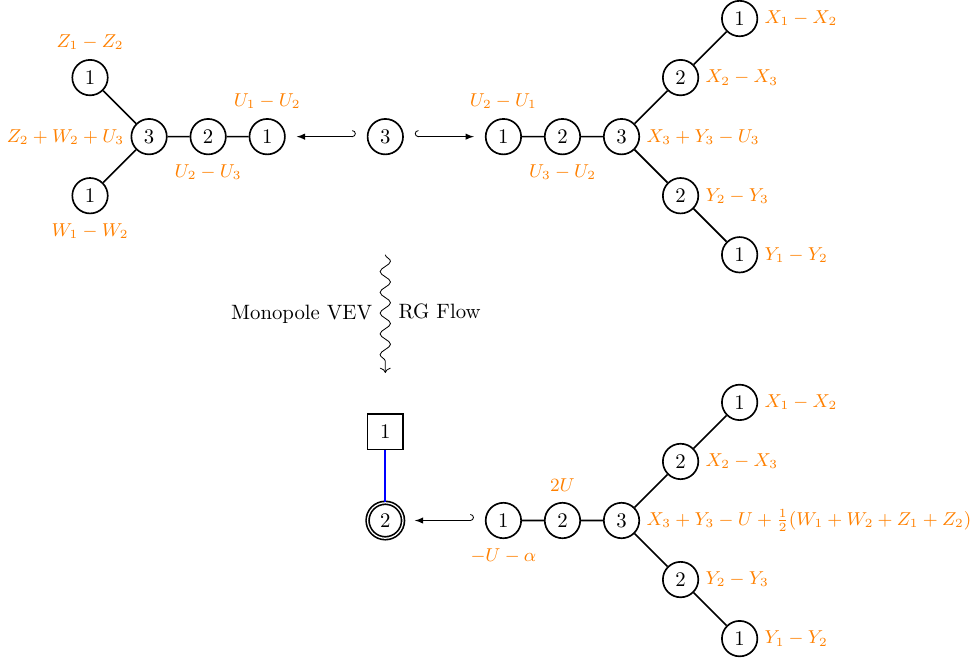}
    \caption{At the top, we depict the quiver corresponding to the 3d mirror of the chosen channel decomposition. Notice that we use a gauge node with hook arrows to indicate the gauging of a $SU(2)$ symmetry on the Coulomb branch of the two connected full tails. As pointed out in the main text, $S_1$ is a bad theory, and there is an RG flow driven by monopole VEVs leading to the quiver theory depicted on the bottom. In the interest of brevity, we denote the following combination of FI parameters as $\alpha=\tfrac12(W_1+Z_1-W_2-Z_2)$. Note that the hypers shown in blue are twisted with respect to those in the original theory.}
    \label{fig: su3ex_finish}
\end{figure}

The starting theory $S$ is depicted on the top-right corner of Figure \ref{fig: su3ex_start} as the star-shaped quiver. Its $S^3_b$ partition function is
\begin{equation}\label{eq: Shat_SU3}
    \begin{split}
        \widehat{\PFS{S}}\left(\vec{X}, \vec{Y}, \vec{W}_{ [2,1] }, \vec{Z}_{ [2,1] } ; m\right) = \intd C \;\D^{\rm CB}_1 (C;m) \PFS{S}\left(\vec{X}, \vec{Y}, \vec{W}_{ [2,1] }, \vec{Z}_{ [2,1] } ; m\right) \,.
    \end{split}
\end{equation}
The combination of real mass parameters corresponding to $C$ is
\begin{equation}\label{eq: conditmodU1_N=3}
    C \equiv \sum_{i=1}^3(X_i+Y_i)+Z_1+2Z_2+W_1+2W_2 \,,
\end{equation}
recalling that the integral over $C$ has the effect of ungauging the diagonal decoupled $U(1)$ gauge symmetry in the star-shaped quiver, as explained in Section \ref{sec:channel_decompSSQ}. The channel decomposition that separates the two minimal punctures from the two maximal corresponds to the following partition function identity:
\begin{equation}\label{eqN=3channel2}
\begin{split}
    \PFS{S} & \left(\vec{X}, \vec{Y}, \vec{W}_{ [2,1] }, \vec{Z}_{ [2,1] } ; m \right) =\\
    & = \intd\vec{U}_{3} \;\D^{\rm CB}_{3}(\vec{U};m) \PFS{S_1}\left(\vec{Z}_{[2,1]}, \vec{W}_{ [2,1] }, \vec{U}; m\right) \PFS{S_2}\left(-\vec{U}, \vec{Y}, \vec{X}; m\right) \,.
\end{split}
\end{equation}
For convenience, we present the quiver corresponding to the 3d mirror of the channel decomposition again at the top of Figure \ref{fig: su3ex_finish}.

Notice that the theory $S_1$, corresponding to the trinion with one maximal and two minimal punctures, is a bad theory since the central node is a $U(3)$ gauge node which sees only 4 fundamental hypers. Thus, we analyse the partition function of $S_1$ using the electric algorithm for bad theories (see Section \ref{subsec: elec_algorithm}), to obtain
\begin{equation}\label{Nequal3specificacalculation}
\begin{split}
    & \PFS{S_1}\left(\vec{Z}_{ [2,1] }, \vec{W}_{ [2,1] }, \vec{U}; m\right) = \\[1mm]
    & \qquad\; = 
    \d( U_1 + W_2 + Z_2) \d( U_2 + U_3 + W_1 + W_2 + Z_1 + Z_2 ) \times \\[1mm]
    & \qquad\;\quad \times 
    \prod_{j=1}^2 \left[s_b\left(\frac{iQ}{2}\pm(U_3+W_j+Z_j)\right)s_b\left(2m-\frac{iQ}{2}\pm(U_3+W_j+Z_j)\right)\right] \times \\ 
    & \qquad\;\quad \times \;
    s_b\left(2m-\frac{iQ}{2}\right)^2s_b(m\pm(U_3+W_1+Z_2)) s_b(m\pm(U_3+W_2+Z_1)) + \\[1mm]
    & \qquad\quad\quad + \;
    ( \, U_1 \leftrightarrow U_2 \, ) + ( \, U_1 \leftrightarrow U_3 \, ) \,.     
\end{split}
\end{equation}
This partition function consists of a sum of 3 frames, where each of them is a Wess--Zumino model, {\it i.e.} a collection of chirals/hypermultiplets, and two delta functions. The two frames in the last line of \eqref{Nequal3specificacalculation} are obtained from the first by permuting the $U_{1,2,3}$ variables, as indicated in the last line.

Upon examination, we find that the theory $S_1$ admits a VEV that spontaneously Higgses the naive $U(3)$ global symmetry, supported by the full tail, into $SU(2)$. This can be extracted from the arguments of the Dirac deltas: the first delta constraints one Cartan of $U(3)$, thus breaking it to $U(2)$, and then the second delta constraints the sum of the remaining variables, thus breaking $U(2)$ to $SU(2)$. This can be also confirmed by observing that the permutation acting on the $U_j$ to generate the two extra frame are indeed elements of the $U(3)$ Weyl group, $S_3$, that acts by reshuffling all the $U_i$ variables, modulo the action of the Weyl group of the unbroken $SU(2)$ group, $\mathbb{Z}_2$, that acts by swapping two of the $U_j$.

Indeed, we can further confirm the picture by rewriting \eqref{Nequal3specificacalculation} in terms of various pieces identified in and around \eqref{eq: stsh_distrib_gen}:
\begin{equation}\label{eq: su3ex_ch}
\left.
\begin{aligned}
    &\PFS{\widetilde{S}_1} =\; s_b(m\pm(U_3+W_1+Z_2)) \; s_b\left(m\pm(U_3+W_2+Z_1)\right) \\[0.69mm]
    &\dd{\rm{Nil}} =\; 1  \\
    &\dd{\rm{Ex}} =\; \d( U_1 + W_2 + Z_2) \; \d( U_2 + U_3 + W_1 + W_2 + Z_1 + Z_2 ) \\[2.1mm]
    &\left.\frac{\D_{SU(2)}\left(U_3+\tfrac{W_1+W_2+Z_1+Z_2}{2};m\right)}{\D_{3}(\vec{U};m)} \right\vert_{\dd{\text{Ex}}} = s_b\left(2m-\frac{iQ}{2}\right)^2 \times \\
    &\quad\quad\times\; \prod_{j=1}^2  \left[s_b\left(2m-\frac{iQ}{2}\pm(U_3+W_j+Z_j)\right) \right. 
    \left. s_b\left(\frac{iQ}{2}\pm(U_3+W_j+Z_j)\right) \right]  
\end{aligned}
\right\vert\;\text{\small Frame 1}
\end{equation}
Let us use this equation to identify the monopole operators in $S_1$ that are acquiring a VEV. The argument of the first Dirac delta in $\dd{\text{Ex}}$ in \eqref{eq: su3ex_ch} corresponds to a monopole with $+1$ magnetic charge under all the gauge groups in the full tail, including the central gauge group. This monopole lives in a triplet of the emergent $U(3)$ global symmetry and has a UV $R$-charge $0$, thus violating the unitarity bound. A single component of an operator in the fundamental representation of $U(3)$, indeed, breaks the symmetry down to $U(2)$, where the two additional frames correspond to the permutation of the component in the triplet that is acquiring the VEV.

We then consider the argument of the second Dirac delta, which corresponds to a monopole with $+1$ magnetic charge under all the gauge groups in $S_1$. This monopole is associated with the diagonal $U(1)$ gauge symmetry, which is trivial in $S_1$. A VEV for this monopole corresponds to quotienting by the diagonal $U(1)$ global symmetry on the Coulomb branch of $S_1$, which then reflects the Higgsing of $U(2)$ to $SU(2)$.

Having analysed the partition function of $S_1$ in isolation, we now substitute \eqref{Nequal3specificacalculation} into \eqref{eqN=3channel2}. The two Dirac deltas in $\dd{\text{Ex}}$ now freeze two out of three integration parameters, leaving only one corresponding to the unbroken $SU(2)$ gauge group. The extra chirals serve to correctly convert the $U(3)$ integration measure to that of $SU(2)$. Indeed, the chirals evaluated on the constraints implied by the Dirac deltas give the ratio between the Vandermonde determinant of $SU(2)$, the new group obtained after the Higgsing, and that of $U(3)$, the starting group (modulo a shift of variables). Thus, the partition function of $S$ can be expressed as
\begin{equation}\label{channel2Nequal32minimalPFS}
\begin{split}
     &\PFS{S} \left(\vec{X}, \vec{Y}, \vec{W}_{ [2,1] }, \vec{Z}_{ [2,1] } ; m \right) \\
     & \quad =  \intd U_{SU(2)} \;\D_{SU(2)}\left(U+\tfrac{W_1+W_2+Z_1+Z_2}{2};m\right) \; \times \\
     & \qquad \times \; s_b(m \pm ( U + W_1 + Z_2)) s_b(m \pm ( U + W_2 + Z_1)) \;\PFS{S_2}(-\vec{U}^{\rm spec}, \vec{X}, \vec{Y}; m) \\[2mm]
     & \quad = \intd U_{SU(2)} \;\D_{SU(2)}(U;m) s_b\left(m \pm U \pm \frac{W_1-W_2-Z_1+Z_2}{2}\right) \PFS{S_2}(-\vec{U}^{\rm spec}, \vec{X}, \vec{Y}; m) ~,
\end{split} 
\end{equation}
where $U$ (equal to any of the $U_j$) is the only remaining gauge parameter. To go from the first to the second expression, we shifted $U \longrightarrow U - \frac{W_1+W_2+Z_1+Z_2}{2}$. After implementing the constraints implied by the Dirac deltas and performing the $U$ redefinition, the partition function of $S_2$ depends on $\vec{U}^{\rm spec}$, which can be expressed as
\begin{equation}\label{FIPARAMETRIZATIONNequal3channel2}
\begin{split}
    \vec{U}^{\rm spec}& =(U_1,U_2,U_3)|_{\text{$\dd{\rm{Ex}}$+shift}} \\[1.5mm]
    & = \left(-W_2-Z_2,\;-U-\frac{W_1+W_2+Z_1+Z_2}{2},\;U-\frac{W_1+W_2+Z_1+Z_2}{2}\right) \,.
\end{split} 
\end{equation}

Thus, the 3d theory we obtain in this channel is given by an $SU(2)$ gauging of a single hypermultiplet in the fundamental of $SU(2)$, and of an $SU(2)$ subgroup of the $E_6$ symmetry of the 3d mirror of the $T_3$ theory. We report a drawing of the resulting theory in the second line of Figure \ref{fig: su3ex_finish}. Notice that the hypers (in blue) gauged to the $SU(2)$ node are twisted with respect to the hypers in the quivers shown in the first line of Figure \ref{fig: su3ex_finish}, indeed if we mirror back $S_2$ we obtain the direct (non-mirror) 3d reduction of the 4d theory associated to this channel.
Note that $S_2$ is still a star-shaped quiver with a redundant $U(1)$ gauge symmetry that needs to be decoupled. On the other hand, upon decoupling the $U(1)$ from the initial theory $S$, by using the definition \eqref{eq: Shat_SU3}, we obtain
\begin{equation}
    \begin{split}
        \widehat{\PFS{S}} & \left(\vec{X}, \vec{Y}, \vec{W}_{ [2,1] }, \vec{Z}_{ [2,1] } ; m\right) = \intd C \;\D^{\rm CB}_1 (C;m) \PFS{S}\left(\vec{X}, \vec{Y}, \vec{W}_{ [2,1] }, \vec{Z}_{ [2,1] } ; m\right) \\
        = & \intd C \D^{\rm CB}_1 (C;m) \intd U_{SU(2)} \;\D_{SU(2)}(U;m) \; s_b\left(m \pm U \pm \frac{W_1-W_2-Z_1+Z_2}{2}\right) \\
        & \hspace{6.5cm} \times \; \PFS{S_2}(-\vec{U}^{\rm spec}, \vec{X}, \vec{Y}; m) \\[2mm]
        = & \intd U_{SU(2)} \;\D_{SU(2)}(U;m) \; s_b\left(m \pm U \pm \frac{W_1-W_2-Z_1+Z_2}{2}\right) \widehat{\PFS{S_2}}(-\vec{U}^{\rm spec}, \vec{X}, \vec{Y}; m) \,,
    \end{split}
\end{equation}
where to go from the second to the third line, we observe that $C$ is precisely the combination of FI parameters in $S_2$ that we need to integrate over to decouple the redundant $U(1)$ gauge symmetry.

The equivalence of the descriptions in the two different channels corresponds to the (3d reduction of the) Argyres--Seiberg duality \cite{Argyres:2007cn}, and is indeed a direct analogue of the class $\cS$ result \cite{Gaiotto:2009hg}. We further note that our calculations precisely reproduce the data associated with the irregular puncture involved in the second channel in \cite{Chacaltana:2010ks} in this particular channel. In their notation, the maximal puncture in the trinion containing the two minimal punctures is replaced by an irregular puncture with pole structure $\{1,3\}$, along with an $SU(2)$ gauging. Moreover, our theory labelled $\widetilde S_1$ containing a free hyper in the fundamental of $SU(2)$ matches their prediction regarding the trinion containing two minimal and the $\{1,3\}$ irregular puncture.

\subsubsection*{Going back}

In this section, we outline the procedure for reconstructing the original star-shaped quiver shown in the top right corner of Figure~\ref{fig: su3ex_start}, starting from the final theory depicted in the second line of Figure~\ref{fig: su3ex_finish}. We begin by ungaging the \( U(3) \) gauge group at the central node of the \( S_2 \) theory as illustrated below and focus on the  CB gauging of the $T[U(3)]$ tail.
\begin{equation}\label{fig:su3ex_ungauging}
    \includegraphics[width=0.5\linewidth]{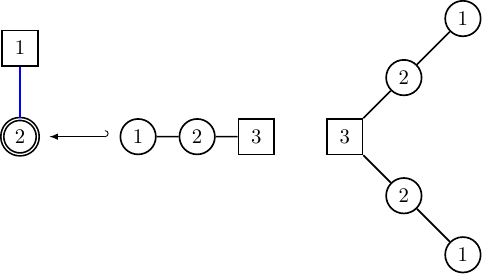}
\end{equation}
Next, as shown in the first step of \eqref{fig:su3ex_mirror_steps}, we take the mirror dual of the  \( T[U(3)] \). The gauging becomes Lagrangian on the HB, and we obtain a quiver in which all hypermultiplets are twisted relative to those in the original star-shaped quiver (they are shown in blue), as depicted in the final step of \eqref{fig:su3ex_mirror_steps}.
\begin{equation}\label{fig:su3ex_mirror_steps}
    \includegraphics[width=0.9\linewidth]{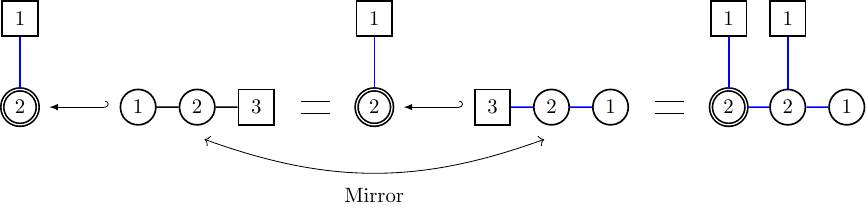}
\end{equation}
To obtain a more convenient description of the final quiver, we proceed as follows. We begin with the first quiver in \eqref{fig:ugly_sqcd_steps}, which differs from the last quiver in \eqref{fig:su3ex_mirror_steps} only at the first node—now a \( U(2) \) gauge node.
This \( U(2) \) node is ugly, so we dualise it. The second node then becomes ugly as well; dualising it leads to the second quiver in \eqref{fig:ugly_sqcd_steps}. In the third step, we take the mirror of the resulting abelian quiver and obtain an SQED with three flavours.
\begin{equation}\label{fig:ugly_sqcd_steps}
    \includegraphics[width=0.75\linewidth]{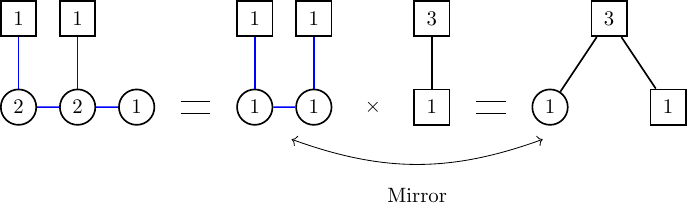}
\end{equation}
If we now gauge the \( U(1) \) topological symmetry of the first \( U(2) \) node in the first theory of \eqref{fig:ugly_sqcd_steps}—effectively turning it into an \( SU(2) \) node—and simultaneously gauge the corresponding \( U(1) \) flavor symmetry in the final theory of the same figure, we obtain a new dual description of the quiver in the last step of \eqref{fig:su3ex_mirror_steps}. This dual theory consists of two decoupled copies of SQED with three flavours. Finally, by reinstating the two \( T[U(3)] \)  tails which were previously omitted during the computation and gauging the \( U(3) \) flavour symmetry, then clearly reproduces the theory displayed in the top right corner of \eqref{fig: su3ex_start}.

\subsection{$N=4$ examples}
\label{subsec:su4_examples}

In this subsection, we study the theories of class $\cS$ of type $\mf{su}(4)$ associated to a sphere with two maximal and two generic regular punctures. The aim is to study the channel decomposition in which the two regular punctures are paired together in a single trinion. The set of regular punctures for $\mf{su}(4)$ that can possibly give rise to a bad trinion is $[3,1],[2^2],[2,1^2]$. Among these, if both the regular punctures are of the type $[2,1^2]$, or one is of type $[2,1^2]$ while the other is $[2^2]$, then the resulting trinion is not bad, but instead corresponds to a free theory consisting of free hypers in certain representations of $SU(4)$. We focus on all the remaining possible combinations of these three $\mf{su}(4)$ punctures that give rise to a bad trinion.

In each case, we match the prediction of \cite{Chacaltana:2010ks}, regarding both matter content of the trinions and the residual gauge group under the channel decomposition, from the point of view of their 3d mirrors.

\subsubsection{\texorpdfstring{$\rho_{1,2}=[3,1]$}{rho12=[3,1]}}\label{subsubsec:su4_2minimal}

\begin{figure}[]
    \centering
    \includegraphics[width=0.9\linewidth]{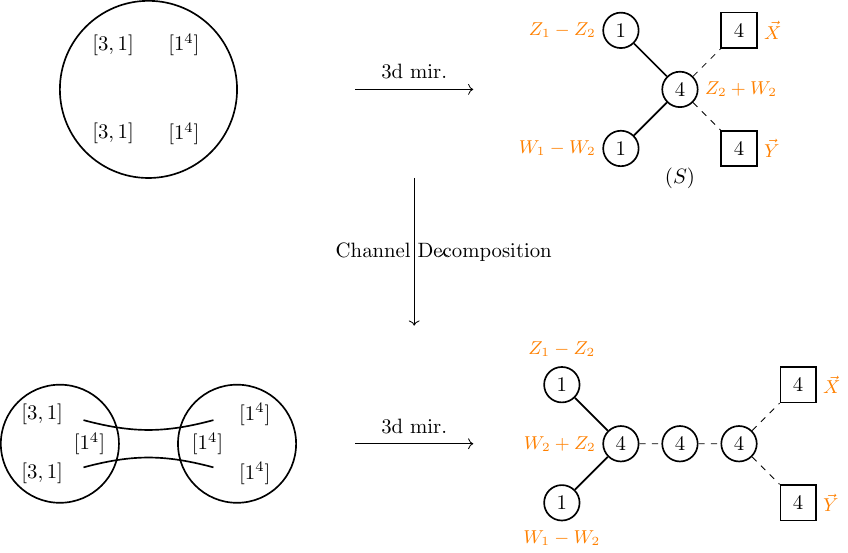}
    \caption{$N=4$ example with two minimal punctures and the associated channel decompositions. These are expressed at the level of the Riemann surfaces associated to the corresponding theory of class $\cS$, and the star-shaped quivers corresponding to their 3d mirror. The original theory $S$ is labelled on the top right corner, while the dashed lines denote the full $T[U(4)]$ tails.}
    \label{eq:n=42minimal}
\end{figure}

We first consider the case with two minimal punctures, $\rho_{1,2} = [3,1]$. As usual, there are two possible decomposition channels, and in the one that splits the two minimal punctures among different trinions, we obtain an $SU(4)$ SQCD with 8 fundamental hypermultiplets, as discussed in Subsection \ref{subsec:genresult_2minimal}. We instead focus on the second channel decomposition, in which we club the two minimal punctures together. We depict the Riemann surface, the chosen channel decomposition, and the associated 3d mirror star-shaped quivers in Figure \ref{eq:n=42minimal}.

\begin{figure}[]
    \centering
    \includegraphics[width=.8\linewidth]{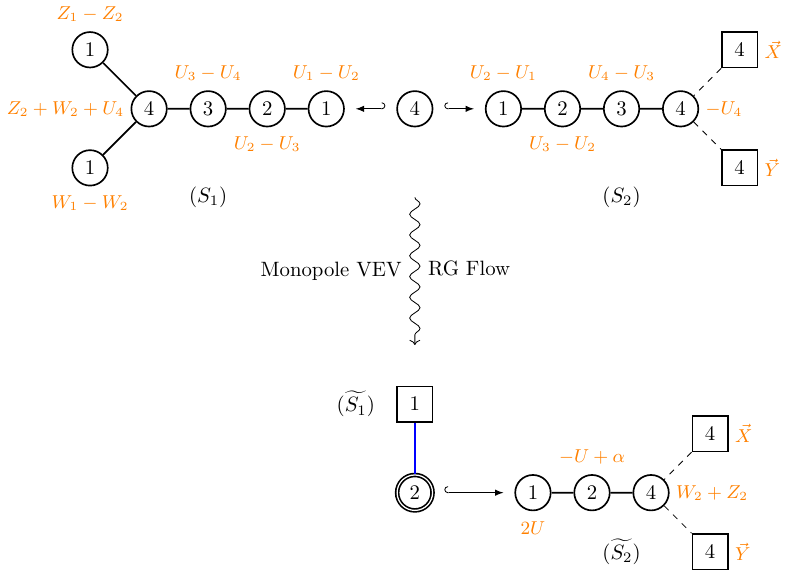}
    \caption{At the top, we depict the quiver corresponding to the 3d mirror of the chosen channel decomposition. Notice that we use a gauge node with hook arrows to indicate the gauging of a $U(4)$ symmetry on the Coulomb branch of the two connected full tails. As pointed out in the main text, $S_1$ is a bad theory, and there is an RG flow driven by monopole VEVs leading to the quiver theory depicted on the bottom. In the interest of brevity, we denote the following combination of FI parameters as $\alpha=\tfrac12(W_1+Z_1-W_2-Z_2)$. Note that the hypers shown in blue are twisted with respect to those in the original theory.}
    \label{fig:26061}
\end{figure}

The partition function of the original theory $S$ can be expressed as
\begin{equation}\label{eq:su42simp_zhat}
    \begin{split}
        \widehat{\PFS{S}}\left(\vec{X}, \vec{Y}, \vec{W}_{ [3,1] }, \vec{Z}_{ [3,1] } ; m\right) = \intd C \, \DCB{1}(C;m) \;\PFS{S}\left(\vec{X}, \vec{Y}, \vec{W}_{ [3,1] }, \vec{Z}_{ [3,1] } ; m\right) \,,
    \end{split}
\end{equation}
where the combination $C$, that is integrated over in order to decouple the redundant $U(1)$ gauge group, is
\begin{equation}\label{conditmodU1N=32full2minimal}  
C=\sum_{i=1}^4(X_i+Y_i)+Z_1+3Z_2+W_1+3W_2 \,.
\end{equation}
The partition function associated with this channel decomposition is:
\begin{align}\label{eqN=42min2fullchannel2}
    \PFS{S}&\left(\vec{X}, \vec{Y}, \vec{W}_{ [3,1] }, \vec{Z}_{ [3,1] } ; m \right) = \intd\vec{U}_{4} \;\D^{\rm CB}_{4}(\vec{U};m) \PFS{S_1}\left(\vec{Z}_{[3,1]}, \vec{W}_{ [3,1] }, \vec{U}; m\right) \PFS{S_2}(-\vec{U}, \vec{Y}, \vec{X}; m) \,.
\end{align}
For convenience, we report the 3d quiver corresponding to the channel decomposition again at the top of Figure \ref{fig:26061}.

By design, the trinion containing the two minimal and a maximal puncture corresponds to a bad theory since the central $U(4)$ node only sees $5$ fundamental flavours. Thus, upon running the electric dualisation algorithm on the theory $S_1$, we obtain
\begin{equation}\label{T12minimalNequal4}
\begin{split}
   & \PFS{S_1}\left(\vec{Z}_{ [3,1] }, \vec{W}_{ [3,1] }, \vec{U}; m\right)= \\
   & \quad =
   \delta \left(U_3+U_4+W_1+W_2+Z_1+Z_2\right) \delta \left(m-\frac{i Q}{2}+U_1+W_2+Z_2\right)  \\
   & \qquad \times \; \delta \left(-m+\frac{i Q}{2}+U_2+W_2+Z_2\right)s_b(m\pm(U_4+W_2+Z_1))s_b(m\pm(U_4+W_1+Z_2)) \\
    & \qquad \times \; \prod_{j=1}^2 \bigg[ s_b(m\pm(W_j+Z_j+U_4))s_b(iQ-m\pm(W_j+Z_j+U_4)) \\
    & \qquad \times \; s_b(3m-iQ\pm(W_j+Z_j+U_4)) \bigg] s_b\left(2m-\frac{iQ}{2}\right)^3s_b\left(4m-\frac{3iQ}{2}\right)s_b\left(-2m+\frac{3iQ}{2}\right) \\
    & \qquad + \text{ 11 permutations}
\end{split}
\end{equation}
From the arguments of the Dirac deltas, we deduce that the $U(4)$ global symmetry is spontaneously broken to $SU(2)$. The sum over frames is in accordance with this prediction, since the 11 other frames are related to the first one only by permutations of the $U_j$ parameters corresponding to the action of the quotient of the two Weyl groups, $S_4/\mathbb Z_2$. Making contact with \eqref{eq: stsh_distrib_gen}, we can group the terms of the first frame as follows:
\begin{equation}\label{eq: su4ex_ch}
\left.
\begin{aligned}
    &\PFS{\widetilde{S}_1} =\; s_b(m\pm(U_4+W_2+Z_1))s_b(m\pm(U_4+W_1+Z_2)) \\[0.69mm]
    &\dd{\rm{Nil}} =\; \delta \left(m-\frac{i Q}{2}+U_1+W_2+Z_2\right)\delta \left(-m+\frac{i Q}{2}+U_2+W_2+Z_2\right)  \\[1.1mm]
    &\dd{\rm{Ex}} =\; \delta \left(U_3+U_4+W_1+W_2+Z_1+Z_2\right) \\[2.1mm]
    &\left.\frac{1}{\text{Ch}^{[2,1^2]}_{\text{Nil}}}\frac{\DCB{SU(2)}\left(U_4+\tfrac{W_1+W_2+Z_1+Z_2}{2};m\right)}{\DCB{4}(\vec{U};m)} \right\vert_{\dd{\text{Ex}},\dd{\text{Nil}}} = \prod_{j=1}^{2} \bigg[(s_b(iQ-m\pm(W_j+Z_j+U_4))\\[2mm]
    & \qquad \qquad \times
    s_b(3m-iQ\pm(W_j+Z_j+U_4))s_b(m\pm(W_j+Z_j+U_4)) \bigg] \\[1mm]
    & \qquad \qquad \times
    s_b\left(2m-\frac{iQ}{2}\right)^3s_b\left(4m-\frac{3iQ}{2}\right)s_b\left(-2m+\frac{3iQ}{2}\right) ~. 
    \\[3mm]
\end{aligned}
\right\vert\;\text{\small Frame 1}
\end{equation}

Having analysed $S_1$ in isolation, let us study the partition function when $S_1$ is glued to $S_2$. Since the $S_1$ theory is bad, it generates a VEV that spontaneously breaks the $U(4)$ gauge symmetry to $SU(2)$. The two Dirac deltas in $\dd{\text{Nil}}$ in \eqref{eq: su4ex_ch} correspond to a nilpotent VEV that Higgses the $U(4)$ gauge symmetry down to $U(2)$. Additionally, the $T[U(4)]$ theory associated to the full tail of $S_2$ is Higgsed to a $T_{[2,1^2]}[U(4)]$ theory due to the gluing superpotential imposing the same nilpotent VEV for the CB moment map of $T[U(4)]$ in $S_2$. Based on the general discussion in Subsection \ref{subsec:channel_decomposition} (see also Appendices \ref{app:TSUN_TRHOSIGMA} and \ref{app:higgsing_TSUN}), we have that:
\begin{equation}\label{eq: su4_TtoTr_ch}
    \PFT{}{4}(-\vec{U},\vec{V};m)\vert_{\dd{\rm{Nil}}}= 
    \PFT{[2,1^2]}{4}(-\vec{U}_{[2,1^2]},\vec{V};m) \times \text{Ch}^{[2,1^2]}_{\text{Nil}} \,,
\end{equation}
where $\text{Ch}_{\text{Nil}}^{[2,1^2]}$ is precisely the contribution of chirals appearing in the fourth line of \eqref{eq: su4ex_ch}. Finally, the single Dirac delta corresponding to $\dd{\text{Ex}}$ simply has the effect of freezing the diagonal $U(1)$ of the unbroken $U(2)$ gauge group, thus breaking it to $SU(2)$.

In summation, by substituting \eqref{T12minimalNequal4} back in \eqref{eqN=42min2fullchannel2}, and \eqref{eq: su4_TtoTr_ch}, we obtain
\begin{equation}\label{eqriferimentoultimasezione}
\begin{split}
\PFS{S}&(\vec{X}, \vec{Y}, \vec{W}_{ [3,1] }, \vec{Z}_{ [3,1] } ; m ) = \intd \vec{U}_{SU(2)} \;\DCB{SU(2)}(U;m) \times \\[1mm]
& \qquad\qquad \times \; s_b\left(m \pm U \pm \frac{W_1-W_2-Z_1+Z_2}{2}\right) \PFS{\widetilde{S}_2}\left( -\vec{U}^{\text{spec}}_{[2,1^2]}, \vec{X}, \vec{Y} \right) ~,
\end{split}
\end{equation}
where $U$ (equal to any of the starting $U_j$) is the only remaining gauge parameter and is associated with the unbroken $SU(2)$ gauge group. Additionally, the partition function of $\widetilde S_2$, $\PFS{\widetilde{S}_2}( -\vec{U}^{\text{spec}}_{[2,1^2]}, \vec{X}, \vec{Y})$ after the nilpotent VEV, corresponds to that of a star-shaped quiver with two full and a $[2,1^2]$ regular tail. Note that \eqref{eqriferimentoultimasezione} is obtained by the following shifting of the unbroken Cartan of the $SU(2)$: $U\rightarrow U-\frac{W_1+W_2+Z_1+Z_2}{2}$, along with the definition,
\begin{equation}\label{FIPARAMETRIZATIONNequal4channel2}
\begin{split}
\vec{U}^{\text{spec}}_{[2,1^2]} & = 
\left( U_3,U_4,W_2+Z_2 \right)|_{\text{$\dd{\rm{Ex}}$+shift}} \\[1.5mm]
& = \left( -U-\frac{W_1+W_2+Z_1+Z_2}{2},\; U-\frac{W_1+W_2+Z_1+Z_2}{2},\; W_2+Z_2 \right) \,.  
\end{split} 
\end{equation}
The redundant $U(1)$ gauge symmetry still present in $\widetilde S_2$ is appropriately integrated out when considering $\widehat Z_S$ via the integration over $C$, {\it c.f.} \eqref{eq:su42simp_zhat}. Once again, the quantity $C$ for the final theory is the same as the one for the initial theory, as in \eqref{conditmodU1N=32full2minimal}, as expected.

Note that $\widetilde{S}_2$ has a redundant $U(1)$ gauge symmetry, indeed in $\widehat{\PFS{S}}$ we still have the integral over $C$ that removes this overall $U(1)$ transformation. By performing this integration, we mod out correctly the redundant gauge $U(1)$ from the resulting theory $\widetilde{S}_2$. Notice that the quantity $C$ for the final theory is the same as that of \eqref{conditmodU1N=32full2minimal}, as expected.

The final result in this channel is the 3d mirror of the generalised Argyres--Seiberg duality for the 4d $\mathcal{N}=2$ $SU(4)$ SQCD with 8 flavours, as expected from the class $\cS$ description \cite{Gaiotto:2009hg}. We depict the resulting quiver theory in Figure \ref{fig:26061}. Notice that the hypers (in blue) gauged to the $SU(2)$ node are twisted with respect to the hypers in the quivers shown in the first line of Figure \ref{fig:26061}, indeed if we mirror back $\widetilde{S}_2$ we obtain the direct (non-mirror) 3d reduction of the 4d theory associated to this channel. Once again, we match the prediction under this channel decomposition in \cite{Chacaltana:2010ks}. In their language, this corresponds to a $\{1,3,5\}$ irregular puncture, that is glued to a $[2,1^2]$ puncture on the other trinion via an $SU(2)$ gauging. Moreover, our theory labelled $\widetilde S_1$ containing a free hyper in the fundamental of $SU(2)$ matches their prediction regarding the trinion containing two minimal and the $\{1,3,5\}$ irregular puncture. This is precisely what we derive at the level of the 3d mirror via a structured calculation.

\subsubsection{\texorpdfstring{$\rho_{1,2}=[2^2]$}{rho12=[2,2]}} 
\label{subsubsec:su4_2boxes}

\begin{figure}[]
    \centering
    \includegraphics[width=0.8\linewidth]{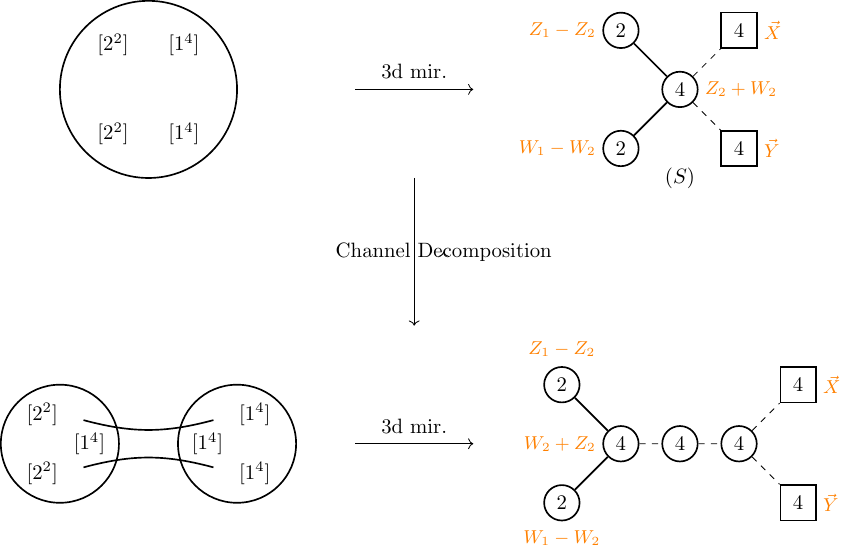}
    \caption{$N=4$ example with two $[2^2]$ punctures and the associated channel decompositions. These are expressed at the level of the Riemann surfaces associated to the corresponding theory of class $\cS$, and the star-shaped quivers corresponding to their 3d mirror. The original theory $S$ is labelled on the top right corner, while the dashed lines denote the full $T[U(4)]$ tails.}
    \label{fig:su4ex2start}
\end{figure}

Next, we consider the case with two identical regular punctures of the type $\rho_{1,2} = [2^2]$. Once again, we focus on the channel decomposition wherein the two regular punctures are clubbed together in a single trinion. We will not be interested in the other channel decompositions as the two trinions there are both good. We depict the Riemann surface, the chosen channel decomposition, and the associated 3d mirror star-shaped quiver in Figure \ref{fig:su4ex2start}.

\begin{figure}[]
    \centering
    \includegraphics[width=.8\linewidth]{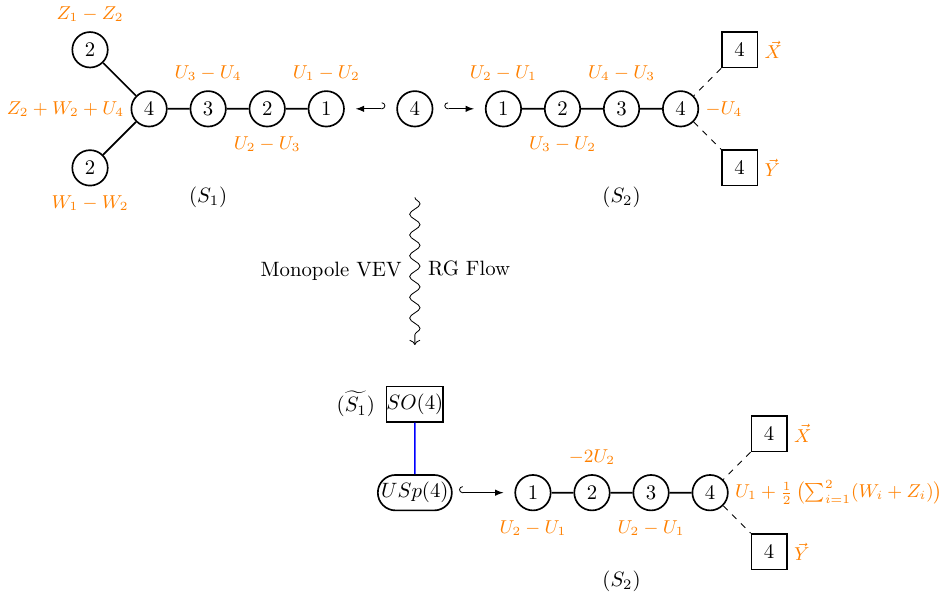}
    \caption{
    At the top, we report the quiver corresponding to the 3d mirror of the chosen channel decomposition. Notice that we use a gauge node with hook-arrows to indicate the gauging of a $U(4)$ symmetry on the Coulomb branch of the two connected full tails. As pointed out in the main text, $S_1$ is a bad theory, and there is an RG flow driven by monopole VEVs leading to the quiver theory depicted on the bottom. Note that the hypers shown in blue are twisted with respect to those in the original theory.} 
    \label{fig:ssu4ex2finish}
\end{figure}

The partition function of the original theory $S$ can be expressed as
\begin{equation}\label{eqnequal4punct22}
\begin{split}
   \widehat{\PFS{S}}&\left(\vec{X}, \vec{Y}, \vec{W}_{ [2^2] }, \vec{Z}_{ [2^2] } ; m \right)=\intd C \DCB{1}(C;m)\PFS{S}\left(\vec{X}, \vec{Y}, \vec{W}_{ [2^2] }, \vec{Z}_{ [2^2] } ; m \right) \,,
\end{split}
\end{equation}
where the combination $C$, that is integrated over in order to decouple the redundant $U(1)$ gauge group, is
\begin{equation}
    C=\sum_{i=1}^4(X_i+Y_i)+2\sum_{j=1}^2(W_j+Z_j) \,.
\end{equation}
The partition function associated with this decomposition channel is:
\begin{equation}\label{eq: 1062}
    \PFS{S}\left(\vec{X}, \vec{Y}, \vec{W}_{ [2^2] }, \vec{Z}_{ [2^2] } ; m\right) =\intd\vec{U}_{4}\;\DCB{4}(\vec{U};m)\;\PFS{S_1}\left(\vec{Z}_{ [2^2] }, \vec{W}_{ [2^2] }, \vec{U}; m\right) \PFS{S_2}(-\vec{U}, \vec{X}, \vec{Y}; m) \,.
\end{equation}
For convenience, we report the 3d quiver corresponding to the channel decomposition again at the top of Figure \ref{fig:ssu4ex2finish}.

By design, the theory $S_1$ is a bad theory and the electric algorithm yields
\begin{equation}\label{eqN=42squared2squaredbadS1}
\begin{split}
    & \PFS{S_1}(\vec{Z}_{ [2^2] }, \vec{W}_{ [2^2] }, \vec{U}; m) \\[1.5mm]
    & \quad = \delta(U_2+U_3+W_1+W_2+Z_1+Z_2)\delta(U_1+U_4+W_1+W_2+Z_1+Z_2) \\
    & \qquad \times \; \prod_{j=3}^4 \bigg[s_b(m\pm (U_j+W_2+Z_2))\;s_b(m\pm (U_j+W_1+Z_2)) \\ 
    & \qquad\qquad\qquad \times \; s_b(m\pm (U_j+W_1+Z_1))\;s_b(m\pm (U_j+W_2+Z_1)) \bigg] \\
    & \qquad \times \; s_b\left(2m-\frac{iQ}{2}\right)^2 s_b\left(\frac{iQ}{2}\pm(U_3-U_4)\right)s_b\left(2m-\frac{iQ}{2}\pm(U_3-U_4)\right) \\
    & \qquad \times \; s_b\left(\frac{iQ}{2}\pm(U_3+U_4+W_1+W_2+Z_1+Z_2)\right) \\
    & \qquad \times \; s_b\left(2m-\frac{iQ}{2}\pm(U_3+U_4+W_1+W_2+Z_1+Z_2)\right) \\
    & \qquad + \text{ 2 permutations}
\end{split}
\end{equation}

From the arguments of the Dirac deltas, we deduce that the $U(4)$ global symmetry is spontaneously broken to $USp(4)$. The sum over frames is in accordance with this prediction, since the two other frames are related to the first by permutations of the $U_j$ parameters corresponding to the action of the quotient of the two Weyl groups, $S_4/D_4$. Making contact with \eqref{eq: stsh_distrib_gen}, we can group the terms of the first frame as follows:
\begin{equation}\label{eq: su4ex2_ch}
\left.
\begin{aligned}
    &\PFS{\widetilde{S}_1} =\; \prod_{j=3}^4 \bigg[s_b(m\pm (U_j+W_2+Z_2))s_b(m\pm (U_j+W_1+Z_2)) \\
    &\qquad\qquad\qquad \times \; s_b(m\pm (U_j+W_1+Z_1))s_b(m\pm (U_j+W_2+Z_1))\bigg] \\
    &\dd{\rm{Nil}} =\; 1  \\[0.3mm]
    &\dd{\rm{Ex}} =\; \delta(U_2+U_3+W_1+W_2+Z_1+Z_2)\delta(U_1+U_4+W_1+W_2+Z_1+Z_2) \\[1mm]
    &\left.\frac{\DCB{USp(4)}\left(U_i+\tfrac{W_1+W_2+Z_1+Z_2}{2};m\right)}{\DCB{4}(\vec{U};m)} \right\vert_{\dd{\text{Ex}}} = s_b\left(2m-\frac{iQ}{2}\right)^2 s_b\left(\frac{iQ}{2}\pm(U_1-U_2)\right) \\[2mm]
    & \qquad \qquad \times 
    s_b\left(2m-\frac{iQ}{2}\pm(U_1-U_2)\right)s_b\left(\frac{iQ}{2}\pm(U_1+U_2+W_1+W_2+Z_1+Z_2)\right) \\[1mm]
    & \qquad\qquad \times 
    s_b\left(2m-\frac{iQ}{2}\pm(U_1+U_2+W_1+W_2+Z_1+Z_2)\right)  \\[1.5mm]
\end{aligned}
\right\vert\;\text{\small Frame 1}
\end{equation}
Evidently, the set $\dd{\text{Nil}}$ is empty, while the set $\dd{\text{Ex}}$ contains two Dirac deltas that impose a constraint on the $U(4)$ parameters, such it is broken to $USp(4)$. 

Having analysed $S_1$ in isolation, let us study the partition function when $S_1$ is glued to $S_2$. As in the previous examples, after substituting \eqref{eqN=42squared2squaredbadS1} back in \eqref{eq: 1062} and making the shift:
\begin{equation}
U_{i}\rightarrow U_i-\frac{W_1+W_2+Z_1+Z_2}{2} \quad , \quad i=1,2 ~,
\end{equation}
we obtain
\begin{equation}\label{eq: 1061}
    \begin{split}
        \PFS{S}&(\vec{X}, \vec{Y}, \vec{W}_{ [2^2] }, \vec{Z}_{ [2^2] } ; m )=\\&=\intd\vec{U}_{USp(4)} \; \DCB{USp(4)}(\vec{U};m)\prod_{j=1}^2\bigg[ s_b(m\pm U_j\pm(W_1+Z_1-W_2-Z_2)) \\
        & \qquad \times \; s_b(m\pm U_j\pm(W_1-Z_1-W_2+Z_2)) \bigg]\PFS{S_2}(-\vec{U}^{\text{spec}},\vec{X},\vec{Y};m) ~.
    \end{split}
\end{equation}
Here, we have used the definition
\begin{equation}
\begin{split}
    \vec{U}^{\rm spec} &= (U_1,U_2,U_3,U_4)\vert_{\text{$\dd{\rm{Ex}}$+shift}} \\[2mm]
    & = \left( U_1-\frac{W_1+W_2+Z_1+Z_2}{2}, \; U_2-\frac{W_1+W_2+Z_1+Z_2}{2}, \right. \\ 
    & \left. \qquad-U_2-\frac{W_1+W_2+Z_1+Z_2}{2}, \; -U_1-\frac{W_1+W_2+Z_1+Z_2}{2} \right) ~, 
\end{split} 
\end{equation}
where $U_1,U_2$ are the gauge parameters of $USp(4)$, and can be any pair of the starting $U_i$. Once again, the diagonal redundant $U(1)$ gauge group in $S_2$ can be decoupled by the integral over $C$ in $\widehat{\PFS{S}}$.

The final result in this channel corresponds to a $USp(4)$ gauging of two half-hypermultiplets in the fundamental of $USp(4)$ and of a $USp(4)$ subgroup of the $U(4)$ symmetry associated to the full tail in $S_2$ (which in this case is the 3d mirror of the $T_4$ theory). We depict the resulting theory in Figure \ref{fig:ssu4ex2finish}. Notice that the hypers (in blue) gauged to the $USp(4)$ node are twisted with respect to the hypers in the quivers shown in the first line of Figure \ref{fig:ssu4ex2finish}, indeed if we mirror back $S_2$ we obtain the direct (non-mirror) 3d reduction of the 4d theory associated to this channel. Once again, we match the prediction under this channel decomposition in \cite{Chacaltana:2010ks}. In their language, this corresponds to a $\{1,3,3\}$ irregular puncture, that is glued to a maximal puncture on the other trinion via a $USp(4)$ gauging. Moreover, our theory labelled $\widetilde S_1$ containing two free half-hypers of $USp(4)$ matches their prediction regarding the trinion containing two minimal and the $\{1,3,3\}$ irregular puncture. This is precisely what we derive at the level of the 3d mirror via a structured calculation.

\subsubsection{\texorpdfstring{$\rho_1=[2^2],\rho_2=[3,1]$}{rho1=[2,2],rho2=[3,1]}}
\label{subsubsec:su4_1minimal1box}

\begin{figure}[ ]
    \centering
    \includegraphics[width=0.9\linewidth]{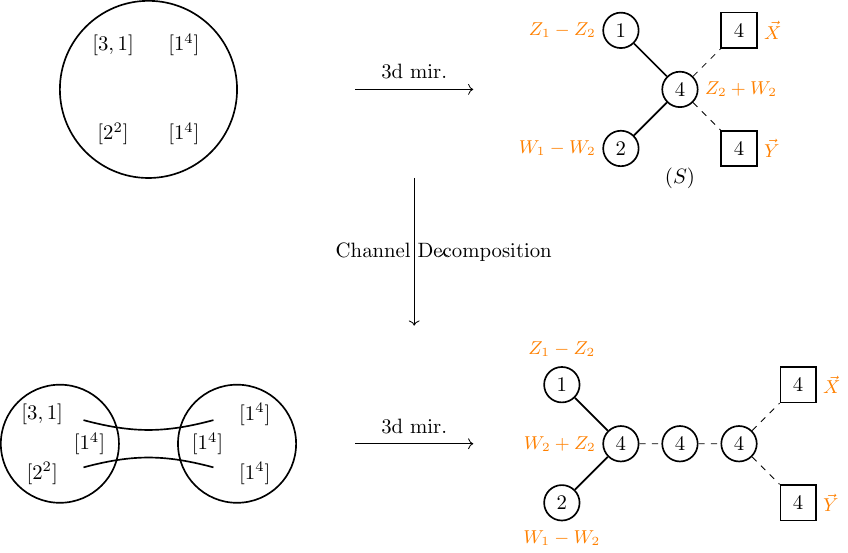}
    \caption{$N=4$ example with one $[2^2]$ and one minimal puncture and the associated channel decompositions. These are expressed at the level of the Riemann surfaces associated to the corresponding theory of class $\cS$, and the star-shaped quivers corresponding to their 3d mirror. The original theory $S$ is labelled on the top right corner, while the dashed lines denote the full $T[U(4)]$ tails.}
    \label{fig:su4ex3start}
\end{figure}

We now consider the case with $\rho_1 = [2^2]$ and $\rho_2=[3,1]$. Once again, we focus on the channel decomposition wherein the two regular punctures are clubbed together in a single trinion. We will not be interested in the other channel decompositions as the two trinions there are both good. We depict the Riemann surface, the chosen channel decomposition, and the associated 3d mirror star-shaped quiver in Figure \ref{fig:su4ex3start}.

\begin{figure}[]
    \centering
    \includegraphics[width=.9\linewidth]{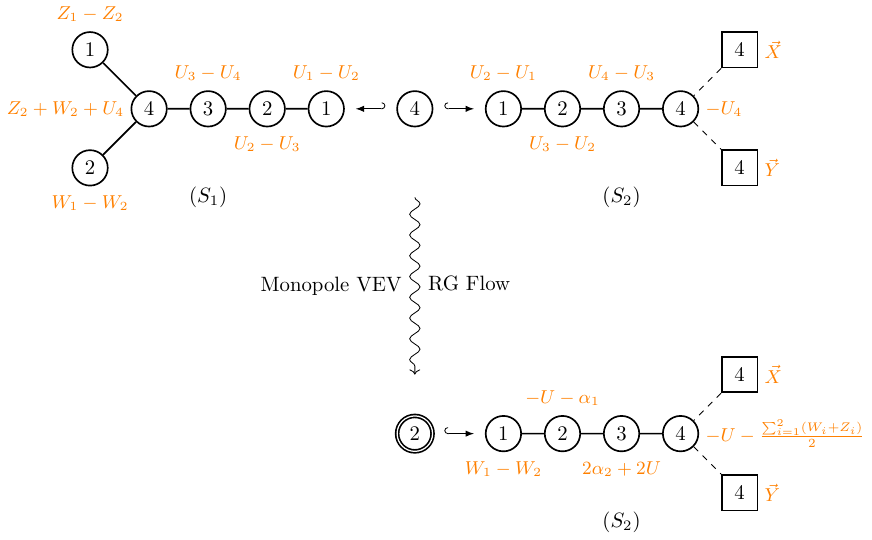}
    \caption{At the top, we depict the quiver corresponding to the 3d mirror of the chosen channel decomposition. Notice that we use a gauge node with hook arrows to indicate the gauging of a $U(4)$ symmetry on the Coulomb branch of the two connected full tails. As pointed out in the main text, $S_1$ is a bad theory, and there is an RG flow driven by monopole VEVs leading to the quiver theory depicted on the bottom. In the interest of brevity, we denote the following combinations of FI parameters as  $\alpha_1=\frac12(3W_1+3Z_1+W_2+Z_2),\;\alpha_2=W_1+W_2+Z_1+Z_2$. }
    \label{fig:su4ex3finish}
\end{figure}

The partition function of the original theory $S$ can be expressed as
\begin{equation}\label{eqnequal4punct21}
    \begin{split}
     \widehat{\PFS{S}}&\left(\vec{X}, \vec{Y}, \vec{W}_{ [2^2] }, \vec{Z}_{ [3,1] } ; m \right) = \intd C \; \DCB{1}(C;m) \; \PFS{S}\left(\vec{X}, \vec{Y}, \vec{W}_{ [2^2] }, \vec{Z}_{ [3,1] } ; m\right) \,,
    \end{split}
\end{equation}
where the combination $C$, that is integrated over in order to decouple the redundant $U(1)$ gauge group, is
\begin{equation}
    C=\sum_{i=1}^4(X_i+Y_i)+2W_1+2W_2+Z_1+3Z_2 ~.
\end{equation}
The partition function associated with this decomposition channel is:
\begin{equation}\label{eqnequal4punct212}
\begin{split}
    \PFS{S}&\left(\vec{X}, \vec{Y}, \vec{W}_{ [2^2] }, \vec{Z}_{ [3,1] } ; m \right)=\intd \vec{U}_{4} \; \DCB{4}(\vec{U};m) \; \PFS{S_1}\left(\vec{Z}_{ [3,1] }, \vec{W}_{ [2^2] }, \vec{U}; m\right) \PFS{S_2}(-\vec{U}, \vec{X}, \vec{Y}; m)
\end{split}
\end{equation}
For convenience, we report the 3d quiver corresponding to the channel decomposition again at the top of Figure \ref{fig:su4ex3finish}.

By design, the theory $S_1$ is a bad theory and the electric algorithm yields
\begin{equation}\label{eq:3061}
\begin{split}
    & \PFS{S_1}(\vec{U}, \vec{W}_{ [2^2] }, \vec{Z}_{ [3,1] }; m)
    \\[1mm]
    & \quad = \delta \left(U_2+W_1+Z_2\right) \delta \left(U_1+W_2+Z_2\right) \delta \left(U_3+U_4+W_1+W_2+Z_1+Z_2\right) \\
    & \qquad \times \prod_{j,k=1}^2\bigg[s_b\left(2m-\frac{iQ}{2}\pm (U_4+W_j+Z_k)\right)s_b\left(\frac{iQ}{2}\pm (U_4+W_j+Z_k)\right)\bigg] \\
    & \qquad \times \; s_b\left(2m-\frac{iQ}{2}\pm(W_1-W_2)\right) s_b\left(\frac{iQ}{2}\pm(W_1-W_2)\right) s_b\left(2m-\frac{iQ}{2}\right)^3 \\
    & \qquad + \; \text{ 11 permutations}
\end{split}
\end{equation}
From the arguments of the Dirac deltas, we deduce that the $U(4)$ global symmetry is spontaneously broken to $SU(2)$. The sum over frames is in accordance with this prediction, since the two other frames are related to the first by permutations of the $U_j$ parameters corresponding to the action of the quotient of the two Weyl groups, $S_4/\mathbb Z_2$. Making contact with \eqref{eq: stsh_distrib_gen}, we can group the terms of the first frame as follows:
\begin{equation}\label{eq: su4ex3_ch}
\left.
\begin{aligned}
    &\PFS{\widetilde{S}_1} =\; 1 \\[0.69mm]
    &\dd{\rm{Nil}} =\; 1  \\[1.1mm]
    &\dd{\rm{Ex}} =\; \delta \left(U_2+W_1+Z_2\right) \delta \left(U_1+W_2+Z_2\right) \delta \left(U_3+U_4+W_1+W_2+Z_1+Z_2\right) \\[2.1mm]
    &\left.\frac{\DCB{SU(2)}\left(U_4-\tfrac{W_1+W_2+Z_1+Z_2}{2};m\right)}{\DCB{4}(\vec{U};m)} \right\vert_{\dd{\text{Ex}}} = \prod_{j,k=1}^2 \bigg[s_b\left(2m-\frac{iQ}{2}\pm (U_4+W_j+Z_k)\right) \\ 
    & \qquad\qquad \times \; s_b\left(\frac{iQ}{2}\pm (U_4+W_j+Z_k)\right)\bigg] s_b\left(2m-\frac{iQ}{2}\right)^3s_b\left(2m-\frac{iQ}{2}\pm(W_1-W_2)\right) \\ 
    & \qquad\qquad \times \; \left.s_b(\frac{iQ}{2}\pm(W_1-W_2)) \; \right\vert_{\dd{\text{Ex}}} \\[3mm]
\end{aligned}
\right\vert\;\text{\small Frame 1}
\end{equation}
Evidently, the set $\dd{\text{Nil}}$ is empty; therefore, there is no nilpotent VEV for the moment map of $U(4)$, while the set $\dd{\text{Ex}}$ contains three Dirac deltas in total. While the first two freeze two Cartans, effectively breaking $U(4)$ to $SU(2)$, the third one constrains the sum of the remaining two Cartans, breaking the $U(2)$ further down to $SU(2)$. Additionally, notice that $\widetilde S_1$ is a trivial theory.

Having analysed $S_1$ in isolation, let us study the partition function when $S_1$ is glued to $S_2$. As in the previous examples, after substituting \eqref{eqN=42squared2squaredbadS1} back in \eqref{eqnequal4punct212} and making the shift: $U_4\rightarrow U_4 +\frac{W_1+W_2+Z_1+Z_2}{2}$ (and renaming $U_4 \equiv U$), we obtain
\begin{equation}\label{eq:3063}
    \begin{split}
        \PFS{S}&\left(\vec{X}, \vec{Y}, \vec{W}_{ [2^2] }, \vec{Z}_{ [3,1] } ; m \right)=\intd\vec{U}_{SU(2)} \; \DCB{SU(2)}(U)\PFS{S_2}(-\vec{U}^{\text{spec}}, \vec{X}, \vec{Y}; m) ~,
    \end{split}
\end{equation}
where
\begin{equation}
\begin{split}
\vec{U}^{\rm spec} & = 
(U_1,U_2,U_3,U_4)\vert_{\text{$\dd{\rm{Ex}}$+shift}} = \\[2mm]
& = \left(-W_1-Z_2, \; -W_2-Z_2, \; -U-\frac{3(W_1+W_2+Z_1+Z_2)}{2}, \; U +\frac{W_1+W_2+Z_1+Z_2}{2}\right) ~.    
\end{split} 
\end{equation}
Once again, the diagonal redundant $U(1)$ gauge group in $S_2$ can be decoupled by the integral over $C$ in $\widehat{\PFS{S}}$.

The final result in this channel is an $SU(2)$ gauging of the 3d mirror of the $T_4$ theory. We depict the resulting theory in Figure \ref{fig:su4ex3finish}. Once again, we match the prediction under this channel decomposition in \cite{Chacaltana:2010ks}. In their language, this corresponds to a $\{1,3,4\}$ irregular puncture, that is glued to a maximal puncture on the other trinion via an $SU(2)$ gauging. Moreover, our theory labelled $\widetilde S_1$ being an empty theory matches their prediction regarding the trinion containing a $[2^2]$ and $[3,1]$ regular punctures, and the $\{1,3,4\}$ irregular puncture. This is precisely what we derive at the level of the 3d mirror via a structured calculation.

\subsubsection{\texorpdfstring{$\rho_1=[2,1^2],\rho_2=[3,1]$}{rho1=[2,1,1],rho2=[3,1]}}
\label{subsubsec:su4_1minimal1ntmax}

\begin{figure}[]
    \centering
    \includegraphics[width=0.9\linewidth]{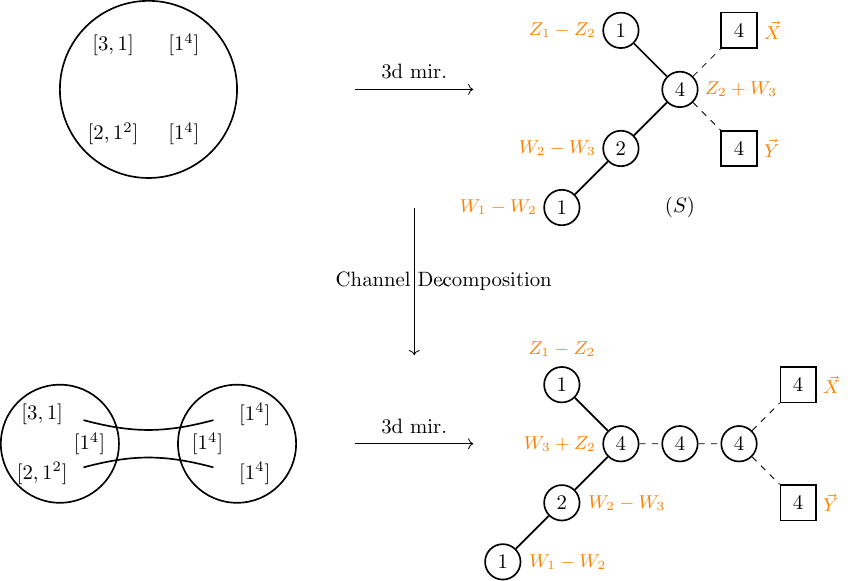}
    \caption{$N=4$ example with one $[2,1^2]$ and one minimal puncture and the associated channel decompositions. These are expressed at the level of the Riemann surfaces associated to the corresponding theory of class $\cS$, and the star-shaped quivers corresponding to their 3d mirror. The original theory $S$ is labelled on the top right corner, while the dashed lines denote the full $T[U(4)]$ tails.}
    \label{fig:su4ex4start}
\end{figure}

We now consider the case with $\rho_1 = [2,1^2]$ and $\rho_2=[3,1]$. Once again, we focus on the channel decomposition wherein the two regular punctures are clubbed together in a single trinion. We will not be interested in the other channel decompositions as the two trinions there are both good. We depict the Riemann surface, the chosen channel decomposition, and the associated 3d mirror star-shaped quiver in Figure \ref{fig:su4ex4start}.

\begin{figure}[]
    \centering
    \includegraphics[width=.9\linewidth]{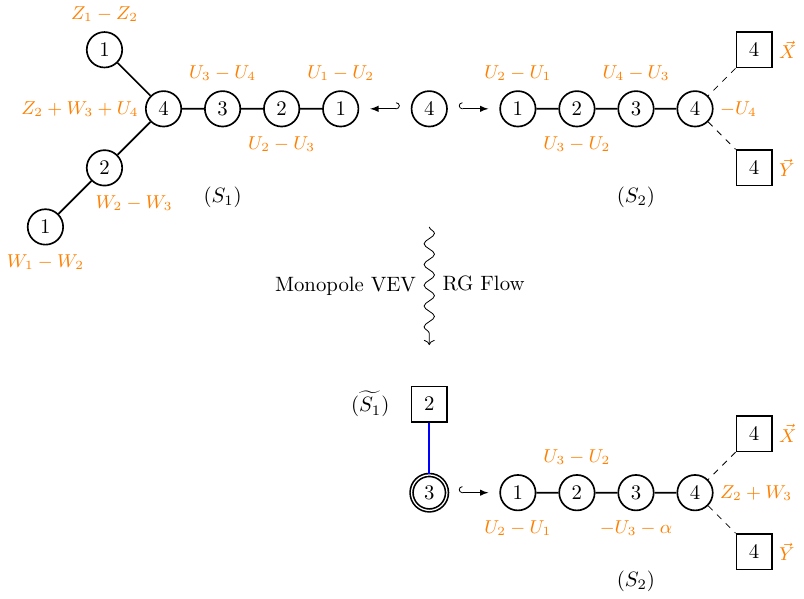}
    \caption{At the top, we depict the quiver corresponding to the 3d mirror of the chosen channel decomposition. Notice that we use a gauge node with hook arrows to indicate the gauging of a $U(4)$ symmetry on the Coulomb branch of the two connected full tails. As pointed out in the main text, $S_1$ is a bad theory, and there is an RG flow driven by monopole VEVs leading to the quiver theory depicted on the bottom. In the interest of brevity, we denote the following combinations of FI parameters as $\alpha=\tfrac13(2W_3-Z_1+Z_2-W_1-W_2)$. Note that $U_1,U_2,U_3$ are not independent, but are related by the Dirac delta appearing in \eqref{eq:05061}. Recall that the hypers shown in blue are twisted with respect to those in the original theory.}
    \label{fig:su4ex4finish}
\end{figure}

The partition function of the original theory $S$ can be expressed as
\begin{equation}\label{eqnequal4punct21}
    \begin{split}
     \widehat{\PFS{S}}&\left(\vec{X}, \vec{Y}, \vec{W}_{ [2,1^2] }, \vec{Z}_{ [3,1] } ; m \right)=\intd C \; \DCB{1}(C;m) \; \PFS{S}\left(\vec{X}, \vec{Y}, \vec{W}_{ [2,1^2] }, \vec{Z}_{ [3,1] } ; m \right) ~,
    \end{split}
\end{equation}
where the combination $C$, that is integrated over in order to decouple the redundant $U(1)$ gauge group, is
\begin{equation}
   C=\sum_{i=1}^4(Y_i+X_i)+W_1+W_2+2W_3+Z_1+3Z_2 ~.
\end{equation}
The partition function associated with this decomposition channel is:
\begin{equation}\label{eqnequal4punct212}
\begin{split}
    \PFS{S}&\left(\vec{X}, \vec{Y}, \vec{W}_{ [2,1^2] }, \vec{Z}_{ [3,1] } ; m \right)=\intd \vec{U}_{4} \; \DCB{4}(\vec{U};m) \; \PFS{S_1}\left(\vec{Z}_{ [3,1] }, \vec{W}_{ [2,1^2] }, \vec{U}; m\right) \PFS{S_2}(-\vec{U}, \vec{X}, \vec{Y}; m) \,.
\end{split}
\end{equation}
For convenience, we report the 3d quiver corresponding to the channel decomposition again at the top of Figure \ref{fig:su4ex4finish}.

By design, the theory $S_1$ is a bad theory and the electric algorithm yields
\begin{equation}\label{ZS1lastcaseN=4}
    \begin{split}
    &\PFS{S_1}(\vec{Z}_{[3,1]}, \vec{W}_{ [2,1^2] }, \vec{U}; m) \\[2mm]
    & \quad = \delta(U_4+Z_2+W_3)\delta(U_1+U_2+U_3+Z_1+2Z_2+W_1+W_2+W_3) \\
    & \qquad\qquad \times \; \prod_{j=1}^3[s_b\left(\frac{iQ}{2}\pm(U_j+Z_2+W_3)\right)s_b\left(2m-\frac{iQ}{2}\pm(U_j+Z_2+W_3)\right)] \\
    & \qquad\qquad \times \; s_b\left(2m-\frac{iQ}{2}\right)^2\prod_{k=1}^3 \bigg[ s_b\left(m\pm\left(U_k-\frac{W_1-2W_2+W_3+Z_1-Z_2}{3}\right)\right) \\
    & \qquad\qquad \times \; s_b\left(m\pm\left(U_k+\frac{2W_1-W_2-W_3-Z_1+Z_2}{3}\right)\right) \bigg] \\
    & \qquad\qquad + \text{ 3 permutations} ~.
    \end{split}
\end{equation}
From the arguments of the Dirac deltas, we deduce that the $U(4)$ global symmetry is spontaneously broken to $SU(3)$. The sum over frames is in accordance with this prediction, since the three other frames in \eqref{ZS1lastcaseN=4} are related to the first by the quotient of the two Weyl groups, $S_4/S_3$. Making contact with \eqref{eq: stsh_distrib_gen}, we can group the terms of the first frame as follows:
\begin{equation}\label{eq: su4ex4_ch}
\left.
\begin{aligned}
    &\PFS{\widetilde{S}_1} =\; \prod_{i=1}^3 \bigg[ s_b\left(m\pm\left(U_i-\frac{W_1-2W_2+W_3+Z_1-Z_2}{3}\right)\right) \times\\& \qquad \qquad \times s_b\left(m\pm\left(U_i+\frac{2W_1-W_2-W_3-Z_1+Z_2}{3}\right)\right) \bigg] \\[0.69mm]
    &\dd{\rm{Nil}} =\; 1  \\[1.1mm]
    &\dd{\rm{Ex}} =\; \delta(U_4+Z_2+W_3)\delta(U_1+U_2+U_3+Z_1+2Z_2+W_1+W_2+W_3) \\[2.1mm]    &\left.\frac{\DCB{SU(3)}\left(U_i+\frac{Z_1+2Z_2+W_1+W_2+W_3}{3};m\right)}{\DCB{4}(\vec{U};m)} \right\vert_{\dd{\text{Ex}}} = \prod_{j=1}^3\bigg[s_b\left(\frac{iQ}{2}\pm(U_j+Z_2+W_3)\right)\times\\&\qquad\qquad\qquad\qquad\qquad\qquad\qquad\qquad\qquad\quad \times \; s_b\left(2m-\frac{iQ}{2}\right)^2\\&\qquad\qquad\qquad\qquad\qquad\qquad\qquad\qquad\qquad\quad \left.\times \; s_b\left(2m-\frac{iQ}{2}\pm(U_j+Z_2+W_3)\right)\bigg]\right\vert_{\dd{\text{Ex}}}\\[3mm]
\end{aligned}
\right\vert\;\text{\small Frame 1}
\end{equation}
Evidently, the set $\dd{\text{Nil}}$ is empty; therefore, there is no nilpotent VEV for the moment map of $U(4)$, while the set $\dd{\text{Ex}}$ contains two Dirac deltas in total. While the first one freezes a single Cartan, effectively breaking $U(4)$ to $U(3)$, the second one constrains the sum of the remaining parameters, breaking the $U(3)$ further down to $SU(3)$.

Having analysed $S_1$ in isolation, let us study the partition function when $S_1$ is glued to $S_2$. As in the previous examples, after substituting \eqref{ZS1lastcaseN=4} back in \eqref{eqnequal4punct212} and making the shift:
\begin{equation}
    \begin{split}
        U_i &\rightarrow U_i -\frac{Z_1+2Z_2+W_1+W_2+W_3}{3} \quad , \quad i=1,2,3 ~,
    \end{split}
\end{equation}
we obtain
\begin{equation}\label{eq:05061}
\begin{split}
    \PFS{S}\left(\vec{X}, \vec{Y}, \vec{W}_{ [2,1^2] }, \vec{Z}_{ [3,1] } ; m \right) & = \intd \vec{U}_{U(3)} \; \DCB{U(3)}(\vec{U};m) \; \d(U_1+U_2+U_3) \\
    & \times \; \prod_{i=2}^4 \bigg[ s_b\left(m\pm\left(U_i-\frac{W_1-2W_2+W_3+Z_1-Z_2}{3}\right)\right) \\
    & \times \; s_b\left(m\pm\left(U_i+\frac{2W_1-W_2-W_3-Z_1+Z_2}{3}\right)\right) \bigg] \\
    & \times \; \PFS{S_2}(-\vec{U}^{\text{spec}}, \vec{X}, \vec{Y}; m) ~.
\end{split}
\end{equation}
Here, we have used the definition
\begin{equation}
\begin{split}       
    \vec{U}^{\rm spec} &= \vec{U}|_{\text{$\dd{\rm{Ex}}$+shift}} \\ 
    & = \left( U_1-\frac{Z_1+2Z_2+W_1+W_2+W_3}{3} , \; U_2-\frac{Z_1+2Z_2+W_1+W_2+W_3}{3} \right., \\
    & \left. \qquad U_3-\frac{Z_1+2Z_2+W_1+W_2+W_3}{3} , \; -Z_2-W_3 \right) \\
\end{split}
\end{equation}
Once again, the diagonal redundant $U(1)$ gauge group in $S_2$ can be decoupled by the integral over $C$ in $\widehat{\PFS{S}}$.

The final result in this channel corresponds to a $SU(3)$ gauging of two hypermultiplets in the fundamental of $SU(3)$ and of an $SU(3)$ subgroup of the $U(4)$ symmetry associated to the full tail in $S_2$ (which in this case is the 3d mirror of the $T_4$ theory). We depict the resulting theory in Figure \ref{fig:su4ex4finish}. Notice that the hypers (in blue) gauged to the $SU(3)$ node are twisted with respect to the hypers in the quivers shown in the first line of Figure \ref{fig:su4ex4finish}, indeed if we mirror back $S_2$ we obtain the direct (non-mirror) 3d reduction of the 4d theory associated to this channel. Once again, we match the prediction under this channel decomposition in \cite{Chacaltana:2010ks}. In their language, this corresponds to a $\{1,2,4\}$ irregular puncture, that is glued to a maximal puncture on the other trinion via an $SU(3)$ gauging. Moreover, our theory labelled $\widetilde S_1$ containing two free fundamental hypers of $SU(3)$ matches their prediction regarding the trinion containing a $[2,1^2]$ and a minimal puncture, and the $\{1,2,4\}$ irregular puncture. This is precisely what we derive at the level of the 3d mirror via a structured calculation.

\subsection{Three different, but related, $N=5$ examples}
\label{subsec:su5_examples}

In this subsection, we present some selective examples of theories of class $\cS$ of type $\mf{su}(5)$ associated to a sphere with two maximal and two generic regular punctures. In particular, we study three configurations where the gauge symmetry associated with the tube is Higgsed down to $SU(4)$. The difference between the three configurations is the theories $\widetilde S_1$ obtained after the application of the electric dualisation algorithm on the respective bad theories. These are Wess--Zumino models in the first two cases, and an interacting SCFT in the third.

In the language of \cite{Chacaltana:2010ks}, we consider trinions with three different pairs of regular punctures and the same irregular puncture $\{1,2,3,5\}$ and the same conjugate puncture, a maximal puncture. From our perspective, these three examples feature different $\widetilde S_1$ theories, but the same $\dd{\text{Nil}}\times \dd{\text{Ex}}$ and chiral multiplets upon running the electric dualisation algorithm.

\subsubsection{$\rho_1=[2,1^3],\rho_2=[4,1]$}
\label{subsubsec:su5_1minimal1ntmax}

\begin{figure}[ ]
    \centering
    \includegraphics[width=0.9\linewidth]{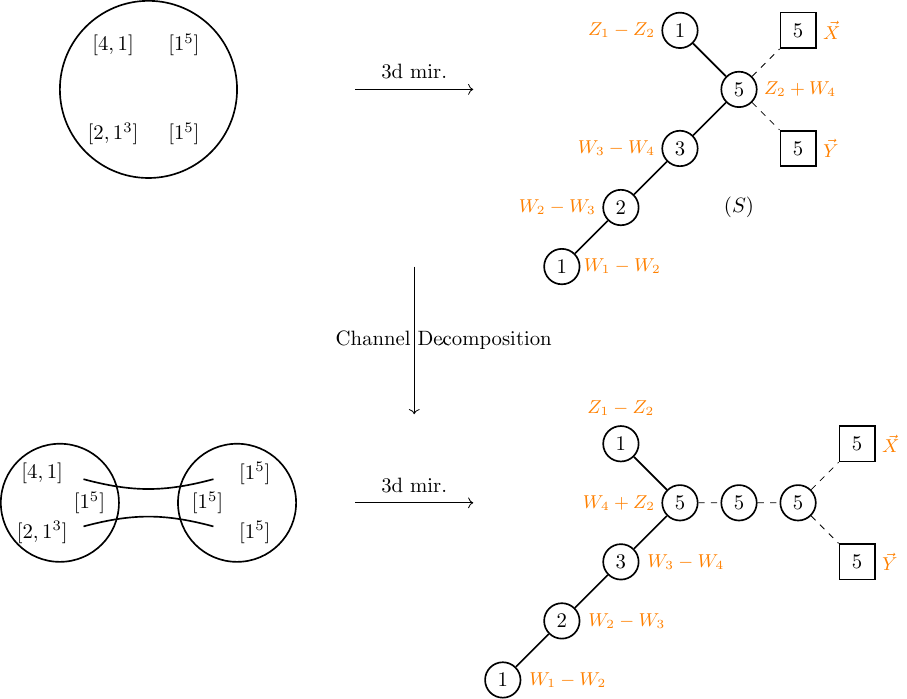}
    \caption{$N=5$ example with one $[2,1^3]$ and one minimal puncture and the associated channel decompositions. These are expressed at the level of the Riemann surfaces associated to the corresponding theory of class $\cS$, and the star-shaped quivers corresponding to their 3d mirror. The original theory $S$ is labelled on the top right corner, while the dashed lines denote the full $T[U(5)]$ tails.}
    \label{fig:su5ex1start}
\end{figure}

We first consider the case $\rho_1=[2,1^3]$ and $\rho_2 = [4,1]$. Once again, we focus on the channel decomposition wherein the two regular punctures are clubbed together in a single trinion. We will not be interested in the other channel decompositions as the two trinions there are both good. We depict the Riemann surface, the chosen channel decomposition, and the associated 3d mirror star-shaped quiver in Figure \ref{fig:su5ex1start}.

\begin{figure}[]
    \centering
    \includegraphics[width=.8\linewidth]{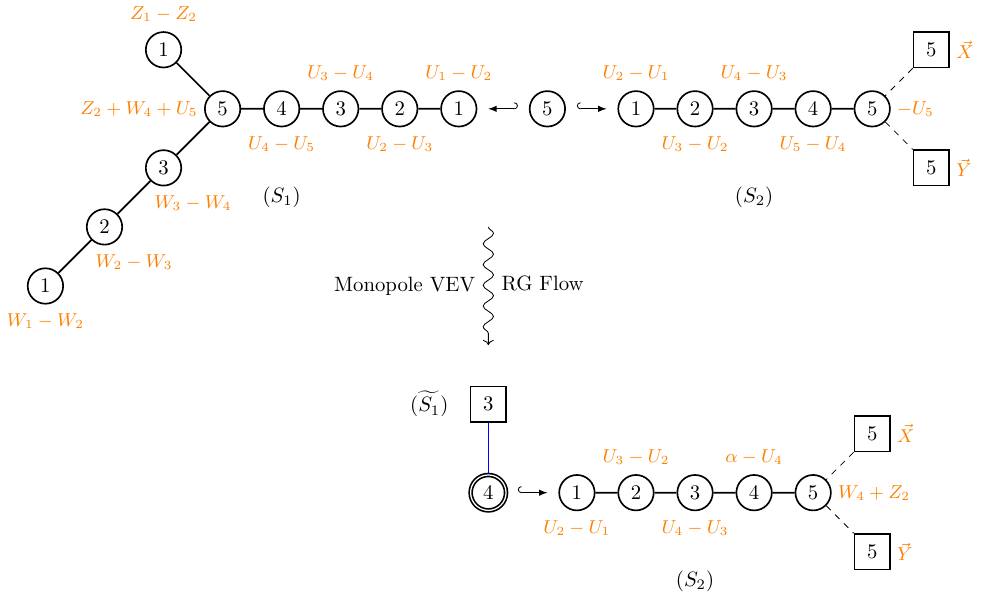}
    \caption{At the top, we depict the quiver corresponding to the 3d mirror of the chosen channel decomposition. Notice that we use a gauge node with hook arrows to indicate the gauging of a $U(5)$ symmetry on the Coulomb branch of the two connected full tails. As pointed out in the main text, $S_1$ is a bad theory, and there is an RG flow driven by monopole VEVs leading to the quiver theory depicted on the bottom. In the interest of brevity, we denote the following combinations of FI parameters as $\alpha=\tfrac14\left(\sum_{i=1}^3W_i-3W_4+Z_1-Z_2\right)$. Note that $U_1,U_2,U_3,U_4$ are not independent but are related by the $\d$ function appearing in \eqref{N=5eq3}. Hypers shown in blue are twisted with respect to those in the original theory.}
    \label{fig:su5ex1finish}
\end{figure}

The partition function of the original theory $S$ can be expressed as
\begin{equation}\label{N=5eq1}
\begin{split}
    \widehat{\PFS{S}}\left(\vec{X}, \vec{Y}, \vec{Z}_{ [4,1] }, \vec{W}_{ [2,1^3] } ; m \right)=\intd C \; \DCB{1}(C;m) \; \PFS{S}(\vec{X}, \vec{Y}, \vec{Z}_{ [4,1] }, \vec{W}_{ [2,1^3] } ; m ) ~,
\end{split}
\end{equation}
where the combination $C$, that is integrated over in order to decouple the redundant $U(1)$ gauge group, is
\begin{equation}
    C=\sum_{i=1}^5(X_i+Y_i)+Z_1+4Z_2+W_1+W_2+W_3+2W_4 ~.
\end{equation}
The partition function associated with this decomposition channel is:
\begin{equation}\label{eq:05064}
   \PFS{S}\left(\vec{X}, \vec{Y}, \vec{Z}_{ [4,1] }, \vec{W}_{ [2,1^3] } ; m \right)=\intd\vec{U}_{5} \; \DCB{5}(\vec{U},m) \; \PFS{S_1}\left(\vec{W}_{ [2,1^3] }, \vec{Z}_{ [4,1] }, \vec{U};m\right) \PFS{S_2}(-\vec{U}, \vec{X}, \vec{Y}; m) \,.
\end{equation}
For convenience, we report the 3d quiver corresponding to the channel decomposition again at the top of Figure \ref{fig:su5ex1finish}.

By design, the theory $S_1$ is a bad theory as the central $U(5)$ node only sees eight fundamental flavours. The application of the electric dualisation algorithm yields
\begin{equation}\label{N=5eq2}
    \begin{split}
        &\PFS{S_1}\left(\vec{W}_{ [2,1^3] }, \vec{Z}_{ [4,1] }, \vec{U};m\right) = \delta(U_1+W_4+Z_2)\;\delta\left(\sum_{i=1}^4W_i+\sum_{j=2}^5U_j+Z_1+3Z_2\right) \\
        & \hspace{1cm} \times \; \prod_{i=2}^5\bigg[s_b\left(\frac{iQ}{2}\pm(U_i+W_4+Z_2)\right)s_b\left(2m-\frac{iQ}{2}\pm(U_i+W_4+Z_2)\right)\bigg] \\
        & \hspace{1cm} \times \; s_b\left(2m-\frac{iQ}{2}\right)^2\prod_{j=2}^5\prod_{l=1}^3 s_b(m\pm(U_j+W_l+Z_2))\;+\text{ 4 permutations} ~.
    \end{split}
\end{equation}

From the argument of the Dirac deltas in \eqref{N=5eq2}, we can deduce that the $U(5)$ flavour symmetry group is Higgsed down to $SU(4)$, so the 5 different frames can be understood as the action of $S_5/S_4$. Making contact with \eqref{eq: stsh_distrib_gen}, we can group the terms of the first frame as follows:
\begin{equation}\label{eq: su5ex1_ch}
\left.
\begin{aligned}
    &\PFS{\widetilde{S}_1} =\; \prod_{j=2}^5\prod_{l=1}^3 s_b(m\pm(U_j+W_l+Z_2)) \\[0.69mm]
    &\dd{\rm{Nil}} =\; 1  \\[1.1mm]
    &\dd{\rm{Ex}} =\; \delta(U_1+W_4+Z_2) \; \delta\left(\sum_{i=1}^4W_i+\sum_{j=2}^5U_j+Z_1+3Z_2\right) \\[2.1mm]    &\left.\frac{\DCB{SU(4)}\left(U_i+\frac{\sum_{j=1}^4W_j+Z_1+3Z_2}{4};m\right)}{\DCB{5}(\vec{U};m)} \right\vert_{\dd{\text{Ex}}} = s_b\left(2m-\frac{iQ}{2}\right)^2\prod_{j=2}^5\bigg[s_b\left(\frac{iQ}{2}\pm(U_j+W_4+Z_2)\right) \\
    & \qquad\qquad\qquad\qquad\qquad\qquad\qquad\qquad\qquad\left.\times s_b\left(2m-\frac{iQ}{2}\pm(U_j+W_4+Z_2)\right)\bigg]\right\vert_{\dd{\text{Ex}}}\\[3mm]
\end{aligned}
\right\vert\;\text{\small Frame 1}
\end{equation}
Evidently, the set $\dd{\text{Nil}}$ is empty; therefore, there is no nilpotent VEV for the moment map of $U(5)$, while the set $\dd{\text{Ex}}$ contains two Dirac deltas in total. While the first one freezes a single Cartan, effectively breaking $U(5)$ to $U(4)$, the second one constrains the sum of the remaining parameters, breaking the $U(4)$ further down to $SU(4)$.

Having analysed $S_1$ in isolation, let us study the partition function when $S_1$ is glued to $S_2$. As in the previous examples, after substituting \eqref{N=5eq2} back in \eqref{eq:05064} and making the shift: $U_j\rightarrow U_j-\frac{\sum_{i=1}^4W_i+Z_1+3Z_2}{4}$ we obtain:
\begin{equation}\label{N=5eq3}
\begin{split}
    \PFS{S}&\left(\vec{X}_{ [1^5] }, \vec{Y}_{ [1^5] }, \vec{W}_{ [4,1] }, \vec{Z}_{ [2,1^3] } ; m\right) = \intd\vec{U}_{4} \; \DCB{SU(4)}(\vec{U},m) \; \delta\left(\sum_{i=1}^4U_i\right) \\
    & \hspace{3cm} \times \; \prod_{i=1}^4\bigg[s_b(m\pm(U_i+W_1+W_2-3W_3+W_4+Z_1-Z_2)) \\
    & \hspace{3cm} \times \; s_b(m\pm(U_i+W_1-3W_2+W_3+W_4+Z_1-Z_2)) \\
    & \hspace{3cm} \times \; s_b(m\pm(U_i-3W_1+W_2+W_3+W_4+Z_1-Z_2))\bigg] \\
    & \hspace{3cm} \times \; \PFS{S_2}(-\vec{U}^{\text{spec}}, \vec{X}, \vec{Y}; m) ~,
\end{split}
\end{equation}
where
\begin{equation}
    \begin{split}
        \vec{U}^{\rm spec}&=(U_1,U_2,U_3,U_4,U_5)|_{\text{$\dd{\rm{Ex}}$+shift}} \\[2mm]
        & =\left( U_1-\frac{\sum_{i=1}^4W_i+Z_1+3Z_2}{4} , \; U_2-\frac{\sum_{i=1}^4W_i+Z_1+3Z_2}{4}, \right. \\
        & \qquad\quad \left. U_3-\frac{\sum_{i=1}^4W_i+Z_1+3Z_2}{4}, \; U_4-\frac{\sum_{i=1}^4W_i+Z_1+3Z_2}{4}, \; -W_4-Z_2 \right) ~.
    \end{split}
\end{equation}
Here $U_1,U_2,U_3,U_4$ can be any of the starting $U_i$, and are gauge parameters of $SU(4)$, constrained by the Dirac deltas appearing in \eqref{N=5eq3}. Once again, the diagonal redundant $U(1)$ gauge group in $S_2$ can be decoupled by the integral over $C$ in $\widehat{\PFS{S}}$.

The final result in this channel corresponds to an $SU(4)$ gauging of three hypermultiplets in the fundamental of $SU(4)$ and of an $SU(4)$ subgroup of the $U(5)$ symmetry associated to the full tail in $S_2$ (which in this case is the 3d mirror of the $T_5$ theory). We depict the resulting theory in Figure \ref{fig:su5ex1finish}. 
Notice that the hypers (in blue) gauged to the $SU(4)$ node are twisted with respect to the hypers in the quivers shown in the first line of Figure \ref{fig:su5ex1finish}, indeed if we mirror back $S_2$ we obtain the direct (non-mirror) 3d reduction of the 4d theory associated to this channel. Once again, we match the prediction under this channel decomposition in \cite{Chacaltana:2010ks}. In their language, this corresponds to a $\{1,2,3,5\}$ irregular puncture, that is glued to a maximal puncture on the other trinion via an $SU(4)$ gauging. Moreover, our theory labelled $\widetilde S_1$ containing three free fundamental hypers of $SU(4)$ matches their prediction regarding the trinion containing a $[2,1^3]$, a minimal puncture, and the $\{1,2,3,5\}$ irregular puncture. This is precisely what we derive at the level of the 3d mirror via a structured calculation.

\subsubsection{$\rho_1=[3,1^2],\rho_2=[3,2]$}
\label{subsubsec:su5_311_32}

\begin{figure}
    \centering
    \includegraphics[width=0.9\linewidth]{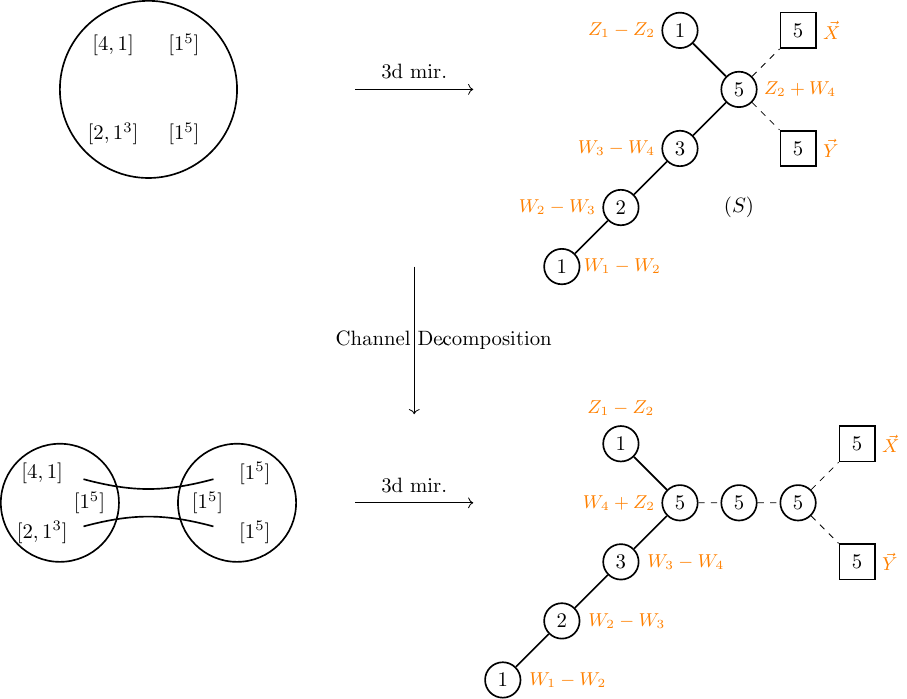}
    \caption{$N=5$ example with a $[3,1^2]$ and a $[3,2]$ puncture, and the associated channel decompositions. These are expressed at the level of the Riemann surfaces associated to the corresponding theory of class $\cS$, and the star-shaped quivers corresponding to their 3d mirror. The original theory $S$ is labelled on the top right corner, while the dashed lines denote the full $T[U(5)]$ tails.}
    \label{fig:su5ex2start}
\end{figure}

The second case that we study for theories of class $\cS$ of type $\mf{su}(5)$ is the one associated to a sphere with two maximal and two regular punctures: $[3,1^2]$ and $[3,2]$. Once again, we focus on the channel decomposition wherein the two regular punctures are clubbed together in a single trinion. We will not be interested in the other channel decompositions as the two trinions there are both good. We depict the Riemann surface, the chosen channel decomposition, and the associated 3d mirror star-shaped quiver in Figure \ref{fig:su5ex2start}.

\begin{figure}[ ]
    \centering
    \includegraphics[width=.9\linewidth]{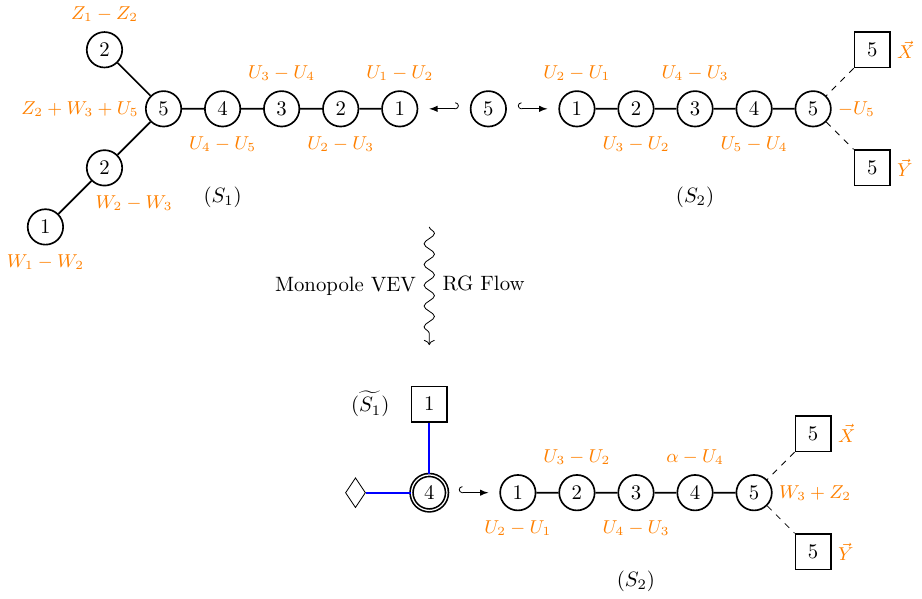}
    \caption{At the top, we depict the quiver corresponding to the 3d mirror of the chosen channel decomposition. In the second line, the diamond stands for a hypermultiplet in the antisymmetric representation of $SU(4)$. Notice that we use a gauge node with hook arrows to indicate the gauging of an $U(5)$ symmetry on the Coulomb branch of the two connected full tails. As pointed out in the main text, $S_1$ is a bad theory, and there is an RG flow driven by monopole VEVs leading to the quiver theory depicted on the bottom. In the interest of brevity, we denote the following combinations of FI parameters as $\alpha=\frac14\left(W_1+W_2-2W_3+2Z_1-2Z_2\right)$. Note that $U_{1,2,3,4}$ are not independent but are related by the $\d$ function appearing in \eqref{N=5t6}. Recall that the hypers shown in blue are twisted with respect to those in the original theory.}
    \label{fig:su5ex2finish}
\end{figure}

The partition function of the original theory $S$ can be expressed as
\begin{equation}\label{N=5eq5}
\begin{split}
    \widehat{\PFS{S}}&\left(\vec{X}, \vec{Y}, \vec{W}_{ [3,1^2] }, \vec{Z}_{ [3,2] } ; m \right)=\intd C \; \DCB{1}(C;m) \; \PFS{S}\left(\vec{X}, \vec{Y}, \vec{W}_{ [3,1^2] }, \vec{Z}_{ [3,2] } ; m \right) ~,
\end{split}
\end{equation}
where the combination $C$, that is integrated over in order to decouple the redundant $U(1)$ gauge group, is
\begin{equation}
    C=\sum_{i=1}^5(Y_i+X_i)+2Z_1+3Z_2+W_1+W_2+3W_3 ~.
\end{equation}
The partition function associated with this decomposition channel is:
\begin{equation}\label{N=5eq5decomp}
\begin{split}
    \PFS{S}&\left(\vec{X}, \vec{Y}, \vec{W}_{ [3,1^2] }, \vec{Z}_{ [3,2] } ; m \right) \\
    & \qquad\qquad=\intd \vec{U}_{5} \; \DCB{5}(\vec{U};m) \; \PFS{S_1}\left(\vec{Z}_{ [3,2] }, \vec{W}_{ [3,1^2] }, \vec{U};m\right) \PFS{S_2}(-\vec{U}, \vec{X}, \vec{Y}; m) ~.
\end{split}
\end{equation}
For convenience, we report the 3d quiver corresponding to the channel decomposition again at the top of Figure \ref{fig:su5ex2finish}.

By design, the theory $S_1$ is a bad theory as the central $U(5)$ node only sees eight fundamental flavours. The application of the electric dualisation algorithm yields
\begin{equation}\label{N=5eq234}
    \begin{split}
        \PFS{S_1}&\left(\vec{Z}_{ [3,2] }, \vec{W}_{ [3,1^2] }, \vec{U}; m\right) = \\[2mm]
        & \quad = \; \delta\left(U_1+W_3+Z_2\right) \delta \left(U_2+U_3+U_4+U_5+W_1+W_2+2 W_3+2 Z_1+2 Z_2\right) \\
        & \qquad \times \; \prod_{j=2}^5\bigg[s_b\left(\frac{iQ}{2} \pm(U_j+W_3+Z_2)\right)s_b\left(2m-\frac{iQ}{2}\pm(U_j+W_3+Z_2)\right)\bigg] \\
        & \qquad \times \; s_b\left(2m-\frac{iQ}{2}\right)^2\prod_{j=2}^5 \; \bigg[s_b(m\pm(U_j+W_3+Z_1))\bigg] \\
        & \qquad \times \; \prod_{i=2}^4\prod_{j=1}^{2}s_b\left(m\pm (U_i+U_5+W_j+W_3+Z_1+Z_2)\right)
        \; +\text{ 4 permutations} ~.
    \end{split}
\end{equation}
From the argument of the Dirac deltas in \eqref{N=5eq234}, we can deduce that the $U(5)$ flavour symmetry group is Higgsed down to $SU(4)$, so the 5 different frames can be understood as the action of $S_5/S_4$. Making contact with \eqref{eq: stsh_distrib_gen}, we can group the terms of the first frame as follows:
\begin{equation}\label{eq: su5ex2_ch}
\left.
\begin{aligned}
    &\PFS{\widetilde{S}_1} =\; \prod_{j=2}^5s_b(m\pm(U_j+W_3+Z_1))\prod_{i=2}^4\prod_{j=1}^{2}s_b(m\pm (U_i+U_5+W_j+W_3+Z_1+Z_2)) \\[0.69mm]
    &\dd{\rm{Nil}} =\; 1  \\[1.1mm]
    &\dd{\rm{Ex}} =\; \delta \left(U_1+W_3+Z_2\right) \delta \left(U_2+U_3+U_4+U_5+W_1+W_2+2 W_3+2 Z_1+2 Z_2\right) \\[2.1mm]
    &\left.\frac{\DCB{SU(4)}\left(U_i+\frac{W_1+W_2+2 W_3+2 Z_1+2 Z_2}{4};m\right)}{\DCB{5}(\vec{U};m)} \right\vert_{\dd{\text{Ex}}} = \prod_{j=2}^5\bigg[s_b\left(\frac{iQ}{2}\pm(U_j+W_3+Z_2)\right)\times \\
    & \qquad\qquad\qquad\qquad\qquad \left.\times \; s_b\left(2m-\frac{iQ}{2}\pm(U_j+W_3+Z_2)\right)\bigg] s_b\left(2m-\frac{iQ}{2}\right)^2\right\vert_{\dd{\text{Ex}}} ~. \\[1.5mm]
\end{aligned}
\right\vert\;\text{\small Frame 1}
\end{equation}
Evidently, the set $\dd{\text{Nil}}$ is empty; therefore, there is no nilpotent VEV for the moment map of $U(5)$, while the set $\dd{\text{Ex}}$ contains two Dirac deltas in total. While the first one freezes a single Cartan, effectively breaking $U(5)$ to $U(4)$, the second one constrains the sum of the remaining parameters, breaking the $U(4)$ further down to $SU(4)$.

Having analysed $S_1$ in isolation, let us study the partition function when $S_1$ is glued to $S_2$. As in the previous examples, after substituting \eqref{N=5eq234} back in \eqref{N=5eq5decomp} and making the shift: $U_i\rightarrow U_i-\frac{W_1+W_2+2W_3+2Z_1+2Z_2}{4}$ we obtain:
\begin{equation}\label{N=5t6}
    \begin{split}
         \PFS{S}&(\vec{X}, \vec{Y}, \vec{W}_{ [3,1^2] }, \vec{Z}_{ [3,2] } ; m ) \\
         & =\intd\vec{U}_{4}\DCB{SU(4)} \; (\vec{U};m) \; \delta\left(\sum_{i=1}^4U_i\right)\prod_{j=1}^4\bigg[s_b\left(m\pm\left(U_j+\frac{W_1+W_2+2Z_2-2W_3-2Z_1}{4}\right)\right)\bigg] \\
         & \quad \times \; \prod_{i=1}^4\bigg[s_b\left(m\pm\left(U_i+U_5\pm\frac{W_1-W_2}{2}\right)\right)\bigg] \; \PFS{S_2}(-\vec{U}^{\text{spec}}, \vec{X}, \vec{Y}; m) ~,
\end{split}
\end{equation}
where
\begin{equation}
    \begin{split}       
        \vec{U}^{\rm spec}&=(U_1,U_2,U_3,U_4,U_5)|_{\text{$\dd{\rm{Ex}}$+shift}} \\[3mm]
        & = \left(U_1-\frac{W_1+W_2+2W_3+2Z_1+2Z_2}{4}, \; U_2-\frac{W_1+W_2+2W_3+2Z_1+2Z_2}{4}, \right. \\
        & \;  \qquad U_3-\frac{W_1+W_2+2W_3+2Z_1+2Z_2}{4}, \; U_4-\frac{W_1+W_2+2W_3+2Z_1+2Z_2}{4}, \; -W_3-Z_2 \bigg) ~.
    \end{split}
\end{equation}
Here $U_1,U_2,U_3,U_4$ can be any of the starting $U_i$, and are gauge parameters of $SU(4)$, constrained by the Dirac deltas appearing in \eqref{N=5t6}. Once again, the diagonal redundant $U(1)$ gauge group in $S_2$ can be decoupled by the integral over $C$ in $\widehat{\PFS{S}}$.

The theory we obtain in this channel is depicted at the bottom of Figure \ref{fig:su5ex2finish}. It is given by four hypermultiplets transforming in the representation $(1,4)$ of $SU(2)\times SU(4)$ and six half-hypermultiplets transforming in the $\frac{1}{2}(2,6)$ of $SU(2)\times SU(4)$ (note that they are half-hypers since $\frac{1}{2}(2,6)$ is a pseudo-real representation) glued via an $SU(4)$ gauging to theory $S_2$, which, upon removing the overall $U(1)$ (via the $C$ integration) coincides with the mirror of the $T_5$ theory. Notice that in Figure \ref{fig:su5ex2finish} the hypers (in blue) gauged to the $SU(4)$ node are twisted with respect to the hypers in the quivers shown in the first line of the Figure, indeed if we mirror back $S_2$ we obtain the direct (non-mirror) 3d reduction of the 4d theory associated to this channel. Once again, we match the prediction under this channel decomposition in \cite{Chacaltana:2010ks}. In their language, this corresponds to a $\{1,2,3,5\}$ irregular puncture, that is glued to a maximal puncture on the other trinion via an $SU(4)$ gauging. Moreover, our theory labelled $\widetilde S_1$ matches their prediction regarding the trinion containing a $[3,1^2]$ and a $[3,2]$ puncture, and the $\{1,2,3,5\}$ irregular puncture. This is precisely what we derive at the level of the 3d mirror via a structured calculation.

\subsubsection{$\rho_1=[3,1^2],\rho_2=[3,1^2]$}
\label{subsubsec:su5_2hooks}

\begin{figure}[ ]
    \centering
    \includegraphics[width=0.9\linewidth]{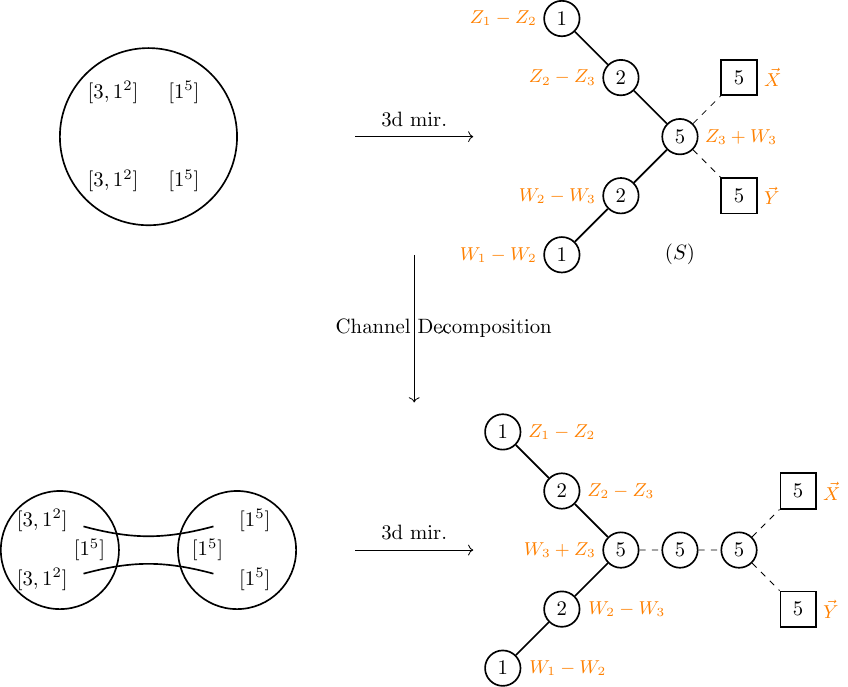}
    \caption{$N=5$ example with two regular punctures of type $[3,1^2]$, and the associated channel decompositions. These are expressed at the level of the Riemann surfaces associated to the corresponding theory of class $\cS$, and the star-shaped quivers corresponding to their 3d mirror. The original theory $S$ is labelled on the top right corner, while the dashed lines denote the full $T[U(5)]$ tails.}
    \label{fig:su5ex3start}
\end{figure}

The third and final case we consider is the theory of class $\cS$ of type $\mf{su}(5)$ associated to a sphere with two maximal and two identical regular punctures of the type $[3,1^2]$. Once again, we focus on the channel decomposition wherein the two regular punctures are clubbed together in a single trinion. We will not be interested in the other channel decompositions as the two trinions there are both good. We depict the Riemann surface, the chosen channel decomposition, and the associated 3d mirror star-shaped quiver in Figure \ref{fig:su5ex3start}.

\begin{figure}[]
    \centering
    \includegraphics[width=.8\linewidth]{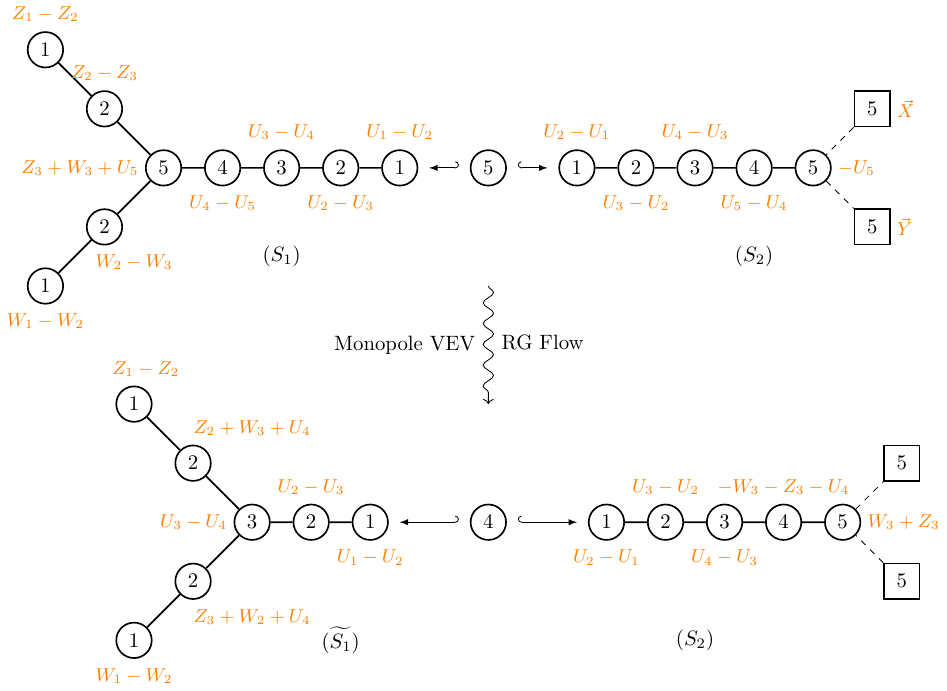}
    \caption{At the top, we depict the quiver corresponding to the 3d mirror of the chosen channel decomposition. Notice that we use a gauge node with hook arrows to indicate the gauging of an $U(5)$ symmetry on the Coulomb branch of the two connected full tails. As pointed out in the main text, $S_1$ is a bad theory, and there is an RG flow driven by monopole VEVs leading to the quiver theory depicted on the bottom. 
    }
    \label{fig:su5ex3finish}
\end{figure}

The partition function of the original theory $S$ can be expressed as
\begin{equation}\label{N=5ex3eq1}
\begin{split}
    \widehat{\PFS{S}}&\left(\vec{X}, \vec{Y}, \vec{W}_{ [3,1^2] }, \vec{Z}_{ [3,1^2] } ; m \right) = \intd C \; \DCB{1}(C;m) \; \PFS{S}\left(\vec{X}, \vec{Y}, \vec{W}_{ [3,1^2] }, \vec{Z}_{ [3,1^2] } ; m \right) ~,
\end{split}
\end{equation}
where the combination $C$, that is integrated over in order to decouple the redundant $U(1)$ gauge group, is
\begin{equation}\label{eq:170710}
    C=\sum_{i=1}^5(Y_i+X_i)+Z_1+Z_2+3Z_3+W_1+W_2+3W_3 ~.
\end{equation}
The partition function associated with this decomposition channel is:
\begin{equation}\label{N=5eqex3channel}
\begin{split}
    \PFS{S}&\left(\vec{X}, \vec{Y}, \vec{W}_{ [3,1^2] }, \vec{Z}_{ [3,1^2] } ; m \right) \\
    & \hspace{1.5cm} =\intd\vec{U}_{5} \; \DCB{5}(\vec{U};m) \; \PFS{S_1}\left(\vec{Z}_{ [3,1^2] }, \vec{W}_{ [3,1^2] }, \vec{U};m\right) \PFS{S_2}(-\vec{U}, \vec{X}, \vec{Y}; m) ~.
\end{split}
\end{equation}
For convenience, we report the 3d quiver corresponding to the channel decomposition again at the top of Figure \ref{fig:su5ex3finish}.

By design, the theory $S_1$ is a bad theory as the central $U(5)$ node only sees eight fundamental flavours. The application of the electric dualisation algorithm yields
\begin{equation}\label{eqN=587473}
    \begin{split}
        \PFS{S_1}&(\vec{Z}_{ [3,1^2] }, \vec{W}_{ [3,1^2] }, \vec{U}; m) = \\
        & = \delta \left(U_5+W_3+Z_3\right)\prod_{j=1}^4 \bigg[s_b\left(\frac{iQ}{2}\pm(U_j+W_3+Z_3)\right)s_b\left(2m-\frac{iQ}{2}\pm(U_j+W_3+Z_3)\right)\bigg] \\
        & \quad \; \times s_b\left(2m-\frac{iQ}{2}\right) \; \PFS{\widetilde{S}_1}
        \; + \text{ 4 permutations} ~,
    \end{split}
\end{equation}
where $\PFS{\widetilde S_1}$ is the partition function of the star-shaped quiver at the bottom left of Figure \ref{fig:su5ex3finish}. From the arguments of the Dirac deltas, we can deduce that the $U(5)$ flavour symmetry group is Higgsed down to $U(4)$, so the 5 different frames can be understood as the action of $S_5/S_4$. The $U(4)$ symmetry is then further reduced to $SU(4)$ when the redundant $U(1)$ gauge symmetry is removed from the $\tilde{S}_1$ theory. We will comment in detail this at the end of the subsection.
Making contact with \eqref{eq: stsh_distrib_gen}, we can group the terms of the first frame as follows:
\begin{equation}\label{eq: su5ex3_ch}
\left.
\begin{aligned}
    &\PFS{\widetilde{S}_1} =(\text{Quiver in the bottom left corner of Figure \ref{fig:su5ex3finish}})\\[0.69mm]
    &\dd{\rm{Nil}} =\; 1  \\[1.1mm]
    &\dd{\rm{Ex}} =\;\delta \left(U_5+W_3+Z_3\right) \\[2.1mm]    &\left.\frac{\DCB{4}\left(\vec{U};m\right)}{\DCB{5}(\vec{U};m)} \right\vert_{\dd{\text{Ex}}} = \prod_{j=1}^4\bigg[s_b\left(\frac{iQ}{2}\pm(U_j+W_3+Z_3)\right) \\ 
    & \hspace{4cm} \; \times \; \left. s_b\left(2m-\frac{iQ}{2}\pm(U_j+W_3+Z_3)\right)\bigg]s_b\left(2m-\frac{iQ}{2}\right)\right\vert_{\dd{\text{Ex}}} ~.\\[3mm]
\end{aligned}
\right\vert\;\text{\small Frame 1}
\end{equation}
Evidently, the set $\dd{\text{Nil}}$ is empty; therefore, there is no nilpotent VEV for the moment map of $U(5)$, while the set $\dd{\text{Ex}}$ contains only one Dirac delta. This effectively Higgses the $U(5)$ global symmetry to $U(4)$.

Having analysed $S_1$ in isolation, let us study the partition function when $S_1$ is glued to $S_2$. As in the previous examples, after substituting \eqref{eqN=587473} back in \eqref{N=5eqex3channel}, 
\begin{equation}\label{N=5penultimorisultato}
    \begin{split}
         \PFS{S}&\left(\vec{X}, \vec{Y}, \vec{W}_{ [3,1^2] }, \vec{Z}_{ [3,1^2] } ; m \right)=\intd \vec{U}_{4} \; \DCB{U(4)}(\vec{U};m) \; \PFS{\widetilde{S}_1}(\ldots;m) \; \PFS{S_2}(-\vec{U}^{\text{spec}}, \vec{X}, \vec{Y}; m) ~,
\end{split}
\end{equation}
where
\begin{equation}
    \begin{split}       
        \vec{U}^{\rm spec}&=(U_1,U_2,U_3,U_4,U_5)|_{\text{$\dd{\rm{Ex}}$}}=(U_1,U_2,U_3,U_4,-W_3-Z_3) ~.
    \end{split}
\end{equation}

Note that this example is different from the rest that we have considered so far, since the theory $\widetilde{S}_1$ here is an interacting theory, the 3d mirror of the $T_3$ theory. We report the quiver for the theory with the appropriate FI parametrisation in the bottom left corner of Figure \ref{fig:su5ex3finish}.

Note that both $\widetilde{S_1}$ and $S_2$ are star-shaped quivers, and thus we must decouple the redundant $U(1)$ gauge transformation from each of them. To do so, we begin by redefining the gauge Cartans $L_i$ of the  $\widetilde{S_1}$ quiver as $ L_i = L_i' + Z$ with  $\sum_i L_i' = 0$ (the sum over $i$ then runs over all the components of all the gauge groups of  $\widetilde S_1$). This redefinition produces a term in the partition function of the form 
\begin{equation}
    e^{2\pi i Z(U_1+U_2+U_3+U_4+Z_1+Z_2+2Z_3+W_1+W_2+2W_3)}
\end{equation}
which indeed represents the contribution of the  FI coupling for the diagonal $U(1)$ of the $\widetilde{S}_1$ theory. The integral over $Z$ has the effect of producing the following $\d$ function:\footnote{Note that the argument of this $\d$ function is clearly consistent with the rule given in \eqref{eqtomodU1}.}
\begin{equation}\label{eq:080710}
    \begin{split}
        &\d(U_1+U_2+U_3+U_4+Z_1+Z_2+2Z_3+W_1+W_2+2W_3) \,.
    \end{split}
\end{equation}
We then  shift $U_i$ variables by:
\begin{equation}
    U_i\rightarrow U_i-\frac{Z_1+Z_2+2Z_3+W_1+W_2+2W_3}{4} \,.
\end{equation}
Thus, the partition function in \eqref{N=5penultimorisultato} becomes:
\begin{equation}\label{eq:07061}
    \begin{split}
        \PFS{S}&\left(\vec{X}, \vec{Y}, \vec{W}_{ [3,1^2] }, \vec{Z}_{ [3,1^2] } ; m \right) \\
        &=\intd\vec{U}_{4} \; \DCB{SU(4)}(\vec{U};m) \; \d(U_1+U_2+U_3+U_4) \; \widehat{\PFS{\widetilde{S}_1}}(\ldots;m) \; \PFS{S_2}(-\vec{U}^{\text{spec}'}, \vec{X}, \vec{Y}; m) ~,
    \end{split}
\end{equation}
Notice the presence of the delta constraining the sum of all the gauge $U_j$ parameters, so that the gauge group is effectively $SU(4)$ instead of $U(4)$.

We still have to decouple the diagonal redundant $U(1)$ gauge group in $S_2$. According to the rule given in \eqref{eqtomodU1}, this would require integrating over the following combination of FIs of the $S_2$ theory as given in \eqref{eq:07061}:
\begin{equation}\label{eq:170711}
    C_2= \sum_{i=1}^5(X_i+Y_i)+\sum_{j=2}^5(U_j)+Z_1+Z_2+3Z_3+W_1+W_2+3W_3=C\,,
\end{equation}
where in the last equality we use the $\d$ function in \eqref{eq:07061}. Hence, the integral over $C$ in $\widehat{\PFS{S}}$ exactly decouples the overall $U(1)$ in ${S_2}$.

The final result in this channel corresponds to a diagonal gauging of an $SU(4)$ subgroup of the CB symmetry of (the 3d mirror of) the $T_5$ theory, and of an $SU(4)$ subgroup of the CB symmetry of (the 3d mirror of) the $T_3$ theory. Once again, we match the prediction under this channel decomposition in \cite{Chacaltana:2010ks}. In their language, this corresponds to a $\{1,2,3,5\}$ irregular puncture, which is glued to the rank-1 $E_6$ SCFT via an $SU(4)$ gauging. Moreover, our theory labelled $\widetilde S_1$, which is the 3d mirror of the rank-1 $E_6$ SCFT, matches their prediction regarding the trinion containing two $[3,1^2]$ regular punctures along with the $\{1,2,3,5\}$ irregular puncture. This is precisely what we derive at the level of the 3d mirror via a structured calculation.

\subsection{$N=6$ example: $\rho_1=[3^2],\rho_2=[5,1]$}
\label{subsec:su6_example_1minimal32}

\begin{figure}[h!]
    \centering
    \includegraphics[width=0.9\linewidth]{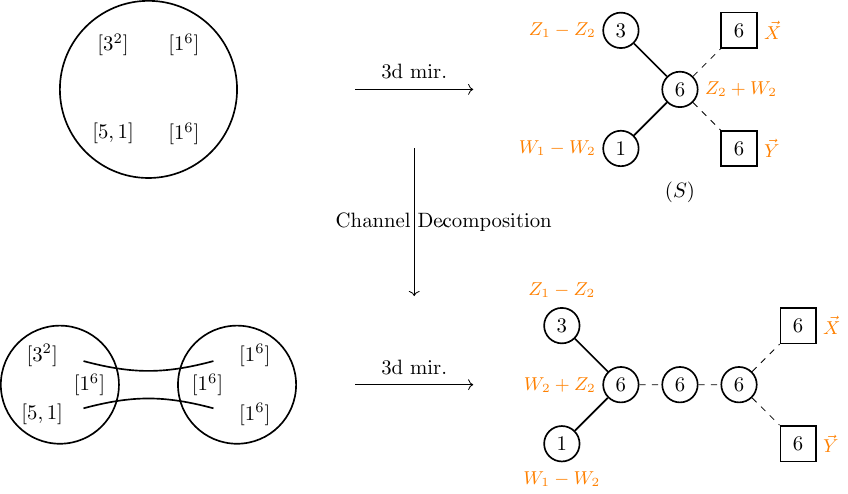}
    \caption{$N=6$ example with one $[3^2]$ and one minimal puncture, and the associated channel decompositions. These are expressed at the level of the Riemann surfaces associated to the corresponding theory of class $\cS$, and the star-shaped quivers corresponding to their 3d mirror. The original theory $S$ is labelled on the top right corner, while the dashed lines denote the full $T[U(6)]$ tails.}
    \label{fig:su6ex1start}
\end{figure}

In this section, we study the 3d mirror of the theory of class $\cS$ of type $\mf{su}(6)$ associated to a sphere with two maximal punctures and two regular punctures: $[3^2]$ and $[5,1]$. We focus on the channel decomposition wherein the two regular punctures are clubbed together in a single trinion. The other channel decomposition yields two good trinions. The trinion with the two regular punctures is a broken bad theory, and as usual, from the set of Dirac deltas obtained by the electric dualisation algorithm, we can 
read the nilpotent VEV for the CB moment map, which in this case has two Jordan blocks. This is distinct from the examples discussed so far, as those had either no nilpotent VEV or only a single non-trivial Jordan block (see Subsection \ref{subsubsec:su4_2minimal}). We depict the Riemann surface, the chosen channel decomposition, and the associated 3d mirror star-shaped quiver in Figure \ref{fig:su6ex1start}.

\begin{figure}[]
    \centering
    \includegraphics[width=.9\linewidth]{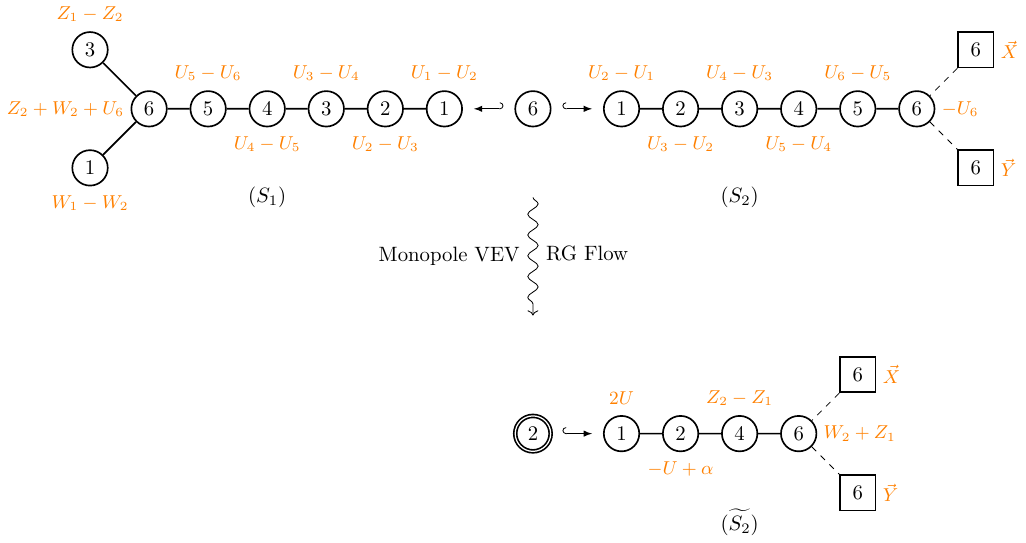}
    \caption{At the top, we depict the quiver corresponding to the 3d mirror of the chosen channel decomposition. Notice that we use a gauge node with hook-arrows to indicate the gauging of a $U(6)$ symmetry on the Coulomb branch of the two connected full tails. As pointed out in the main text, $S_1$ is a bad theory, and there is an RG flow driven by monopole VEVs leading to the quiver theory depicted on the bottom. In the interest of brevity, we denote the following combinations of FI parameters as $\alpha=\frac12(W_1+Z_1-W_2-Z_2)$.}
    \label{fig:su6ex1finish}
\end{figure}

The partition function of the original theory $S$ can be expressed as
\begin{equation}\label{eq:08062}
    \begin{split}
        \widehat{\PFS{S}}&\left(\vec{X},\vec{Y},\vec{W}_{[5,1]},\vec{Z}_{[3^2]};m\right) = \intd C \; \DCB{1}(C;m) z; \PFS{S}\left(\vec{X},\vec{Y},\vec{W}_{[5,1]},\vec{Z}_{[3^2]};m\right) ~,
    \end{split}
\end{equation}
where the combination $C$, that is integrated over in order to decouple the redundant $U(1)$ gauge group, is
\begin{equation}
    C=\sum_{i=1}^6(F_i+C_i)+W_1+5W_2+3Z_1+3Z_2 ~.
\end{equation}
The partition function associated with this decomposition channel is:
\begin{equation}\label{Neq6equation}
    \begin{split}
        \PFS{S}&\left(\vec{X},\vec{Y},\vec{W}_{[5,1]},\vec{Z}_{[3^2]};m\right) \\
        & = \intd \vec{U}_6 \; \DCB{6}(\vec{U};m) \; \PFS{S_1}\left(\vec{U},\vec{W}_{[5,1]},\vec{Z}_{[3^2]};m\right) \PFS{S_2}(-\vec{U},\vec{X},\vec{Y};m)= \\
        & = \intd \vec{U}_6 \; \DCB{6}(\vec{U};m) \; \PFS{S_1}\left(\vec{U},\vec{W}_{[5,1]},\vec{Z}_{[3^2]};m\right)\intd \vec{V}_6\Delta^{HB}_6(\vec{V};m) \\
        & \quad \times \; \PFT{}{6}(-\vec{U},\vec{V};m) \; \PFT{}{6}(\vec{X},\vec{V};m) \; \PFT{}{6}(\vec{Y},\vec{V};m) ~.
    \end{split}
\end{equation}
For convenience, we report the 3d quiver corresponding to the channel decomposition again at the top of Figure \ref{fig:su6ex1finish}.

By design, the theory $S_1$ is a bad theory as the central $U(6)$ node only sees nine fundamental flavours. The application of the electric dualisation algorithm yields
\begin{equation}\label{Neq6equation1}
    \begin{split}
        &\PFS{S_1}(\vec{U},\vec{W}_{[5,1]},\vec{Z}_{[3^2]};m)=\delta \left(U_5+U_6+W_1+W_2+Z_1+Z_2\right) \\
        & \hspace{2cm} \times \; \delta\left(m-\frac{i Q}{2}+U_3+W_2+Z_1\right) \delta \left(-m+\frac{i Q}{2}+U_4+W_2+Z_1\right) \\
        & \hspace{2cm} \times \; \delta \left(m-\frac{i Q}{2}+U_1+W_2+Z_2\right) \delta \left(-m+\frac{i Q}{2}+U_2+W_2+Z_2\right) \\
        & \hspace{2cm} \times \; \prod_{j=1,2}\prod_{k=1,2}\bigg[ s_b(iQ-m\pm(U_6+W_j+Z_k))s_b(m\pm(U_6+W_j+Z_k)) \\ 
        & \hspace{2cm} \times \; s_b(3m-iQ\pm(U_6+W_j+Z_k)) \bigg] s_b\left(4m-\frac{3iQ}{2}\pm(Z_1-Z_2)\right) s_b\left(4m-\frac{3iQ}{2}\right)^2 \\
        & \hspace{2cm} \times \; s_b\left(2m-\frac{iQ}{2}\pm(Z_1-Z_2)\right)^2 s_b\left(-2m+\frac{3iQ}{2}\pm(Z_1-Z_2)\right)  \\
        & \hspace{2cm} \times \; s_b\left(\frac{iQ}{2}\pm(Z_1-Z_2)\right)^2 s_b\left(2m-\frac{iQ}{2}\right)^5  s_b\left(-2m+\frac{3iQ}{2}\right) \\[2mm]
        & \hspace{2cm} + 359\text{ Permutations} ~.
    \end{split}
\end{equation}
From the argument of the Dirac deltas in \eqref{Neq6equation1}, we can deduce that the $U(6)$ global symmetry is broken down to $SU(2)$. The number of frames is in accordance with this prediction, as the 360 distinct frames are rotated by the permutation group $S_6/\mathbb Z_2$. Making contact with \eqref{eq: stsh_distrib_gen}, we can group the terms of the first frame as follows:
\begin{equation}\label{eq: su6ex1_ch}
\left.
\begin{aligned}
    & \PFS{\widetilde{S}_1} =1\\[0.69mm]
    & \dd{\rm{Nil}} =\; \delta \left(m-\frac{i Q}{2}+U_3+W_2+Z_1\right) \delta \left(-m+\frac{i Q}{2}+U_4+W_2+Z_1\right) \\
    & \qquad \quad \times \; \delta \left(m-\frac{i Q}{2}+U_1+W_2+Z_2\right) \delta \left(-m+\frac{i Q}{2}+U_2+W_2+Z_2\right)  \\[1.1mm]
    & \dd{\rm{Ex}} =\;\delta \left(U_5+U_6+W_1+W_2+Z_1+Z_2\right)  \\[2.1mm]
    & \frac{1}{\text{Ch}^{[2^2,1^2]}_{\text{Nil}}}\left.\frac{\DCB{SU(2)}\left(U_6\rightarrow U_6+\frac{W_1+W_2+Z_1+Z_2}{2};m\right)}{\DCB{6}(\vec{U};m)} \right\vert_{\dd{\text{Ex}},\dd{\rm{Nil}}} = \\
    & \hspace{2cm} = \; \prod_{j,k=1}^2\bigg[s_b(iQ-m\pm(U_6+W_j+Z_k)) s_b(m\pm(U_6+W_j+Z_k)) \\
    & \hspace{2cm} \quad \times \; s_b(3m-iQ\pm(U_6+W_j+Z_k)) \bigg] s_b\left(4m-\frac{3iQ}{2}\pm(Z_1-Z_2)\right) \\
    & \hspace{2cm} \quad \times \; s_b\left(2m-\frac{iQ}{2}\pm(Z_1-Z_2)\right)^2 s_b\left(-2m+\frac{3iQ}{2}\pm(Z_1-Z_2)\right) \\
    & \hspace{2cm} \quad \times \; s_b\left(\frac{iQ}{2}\pm(Z_1-Z_2)\right)^2 s_b\left(-2m+\frac{3iQ}{2}\right) \\
    & \hspace{2cm} \quad \times \; \left. s_b\left(2m-\frac{iQ}{2}\right)^5 s_b\left(4m-\frac{3iQ}{2}\right)^2 \right\vert_{\dd{\text{Ex}},\dd{\rm{Nil}}} ~. \\[1.5mm]
\end{aligned}
\right\vert\;\text{\small Frame 1}
\end{equation}

Having analysed $S_1$ in isolation, let us study the partition function when $S_1$ is glued to $S_2$. Since the $S_1$ theory is bad, it generates a VEV that breaks the $U(6)$ gauge symmetry to $SU(2)$. The four Dirac deltas in $\dd{\text{Nil}}$ in \eqref{eq: su6ex1_ch} correspond to a nilpotent VEV that Higgses the $U(6)$ gauge symmetry down to $U(2)$. Additionally, the $T[U(6)]$ theory associated to the full tail of $S_2$ is Higgsed to a $T_{[2^2,1^2]}[U(6)]$ theory due to the gluing superpotential imposing the same nilpotent VEV for the CB moment map of $T[U(6)]$ in $S_2$. Based on the general discussion in Subsection \ref{subsec:channel_decomposition} (see also Appendices \ref{app:TSUN_TRHOSIGMA} and \ref{app:higgsing_TSUN}), we have that:
\begin{equation}\label{eq: su6_TtoTr_ch}
    \PFT{}{6}(-\vec{U},\vec{V};m)\vert_{\dd{\rm{Nil}}} = \PFT{[2^2,1^2]}{6}\left(-\vec{U}_{[2^2,1^2]},\vec{V};m\right) \times \text{Ch}^{[2^2,1^2]}_{\text{Nil}} \,,
\end{equation}
where $\text{Ch}_{\text{Nil}}^{[2^2,1^2]}$ is precisely the contribution of chirals appearing in \eqref{eq: su6ex1_ch}. Finally, the single Dirac delta corresponding to $\dd{\text{Ex}}$ simply has the effect of freezing the diagonal $U(1)$ of the unbroken $U(2)$ gauge group, thus breaking it to $SU(2)$.

In summation, by substituting \eqref{Neq6equation1} back in \eqref{Neq6equation}, along with \eqref{eq: su6_TtoTr_ch}, a shift $U_6\rightarrow U_6-\frac{1}{2}(W_1+W_2+Z_1+Z_2)$ (in order to reproduce the correct $SU(2)$ measure), and relabelling $U_6=U$, we obtain:
\begin{equation}\label{eq:08061}
    \begin{split}
        & \PFS{S}(\vec{X},\vec{Y},\vec{W}_{[5,1]},\vec{Z}_{[3^2]};m) = \intd\vec{U}_{SU(2)} \; \DCB{SU(2)} (\vec{U};m)\intd\vec{V}_6 \; \Delta^{HB}_6(\vec{V};m) \\
        & \hspace{3cm} \times \; \PFT{[2^2,1^2]}{6}\left(-\vec{U}_{[2^2,1^2]}^{\text{spec}},\vec{V};m\right) \; \PFT{}{6}(\vec{X},\vec{V};m) \; \PFT{}{6}(\vec{Y},\vec{V};m) ~,
    \end{split}
\end{equation}
where
\begin{equation}\label{FIPARAMETRIZATIONNequal4channel2}
    \begin{split}
        \vec{U}^{\rm spec}_{[2^2,1^2]} & = \vec{U}_{[2^2,1^2]}|_{\text{$\dd{\rm{Ex}}$+shift}}=(U_5,U_6,-W_2-Z_2,-W_2-Z_1)\vert_{\text{$\dd{\rm{Ex}}$+shift}} \\
        & = \left(-U-\frac{W_1+W_2+Z_1+Z_2}{2} , \; U-\frac{W_1+W_2+Z_1+Z_2}{2} \right. \\
        & \qquad\quad , \; -W_2-Z_2 , \; -W_2-Z_1 \bigg) ~,
    \end{split} 
\end{equation}
where $U$ can be any of the $U_i$ of the initial $U(6)$ symmetry. Note that the diagonal redundant $U(1)$ gauge symmetry in $\widetilde S_2$ can be decoupled by performing the integral over $C$ in $\widehat{\PFS{S}}$.

The final result in this channel corresponds to a gauging of an $SU(2)$ subgroup of the CB symmetry associated with the $T_{[2^2,1^2]}$ tail in $S_2$. We depict the resulting theory in Figure \ref{fig:su6ex1finish}. Once again, we match the prediction under this channel decomposition in \cite{Chacaltana:2010ks}. In their language, this corresponds to a $\{1,3,4,6,7\}$ irregular puncture, that is glued to a maximal puncture on the other trinion via an $SU(2)$ gauging. Moreover, our theory labelled $\widetilde S_1$ being empty matches their prediction regarding the trinion containing a $[3^2]$ and a minimal puncture, and the $\{1,3,4,6,7\}$ irregular puncture. This is precisely what we derive at the level of the 3d mirror via a structured calculation.

\subsection{Generic $N$ example: \texorpdfstring{$\rho_{1,2}=[N-1,1]$}{rho12=[N-1,1]}}
\label{subsec:genresult_2minimal}

\begin{figure}[]
    \centering
    \includegraphics[width=0.8\linewidth]{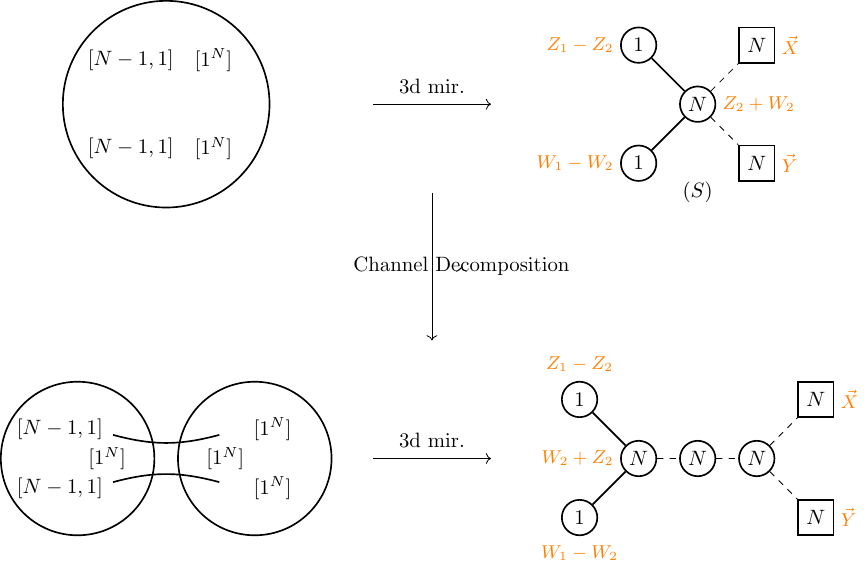}
    \caption{Generic $N$ example with two minimal punctures, and the associated channel decompositions. These are expressed at the level of the Riemann surfaces associated to the corresponding theory of class $\cS$, and the star-shaped quivers corresponding to their 3d mirror. The original theory $S$ is labelled on the top right corner, while the dashed lines denote the full $T[U(N)]$ tails.}
    \label{eq:ngen2minimal}
\end{figure}

In this subsection, we finally study the 3d mirror of the theory of class $\cS$ of type $\mf{su}(N)$, for generic $N$, associated to a sphere with two maximal and two minimal punctures. This directly generalises the detailed calculations already presented for $N=3$ in Subsection \ref{subsec:su3_example} and $N=4$ in Subsection \ref{subsubsec:su4_2minimal}. Based on these and our calculations for some other values of $N$, we now extrapolate the result for generic $N$. Once again, we focus on the channel decomposition wherein the two regular punctures are clubbed together in a single trinion. We will not be interested in the other channel decomposition as it corresponds to the 4d $\cN=2$ $SU(N)$ SQCD with $2N$ fundamental hypermultiplets. We depict the Riemann surface, the chosen channel decomposition, and the associated 3d mirror star-shaped quiver in Figure \ref{eq:ngen2minimal}.

\begin{figure}[]
    \centering
    \includegraphics[width=.8\linewidth]{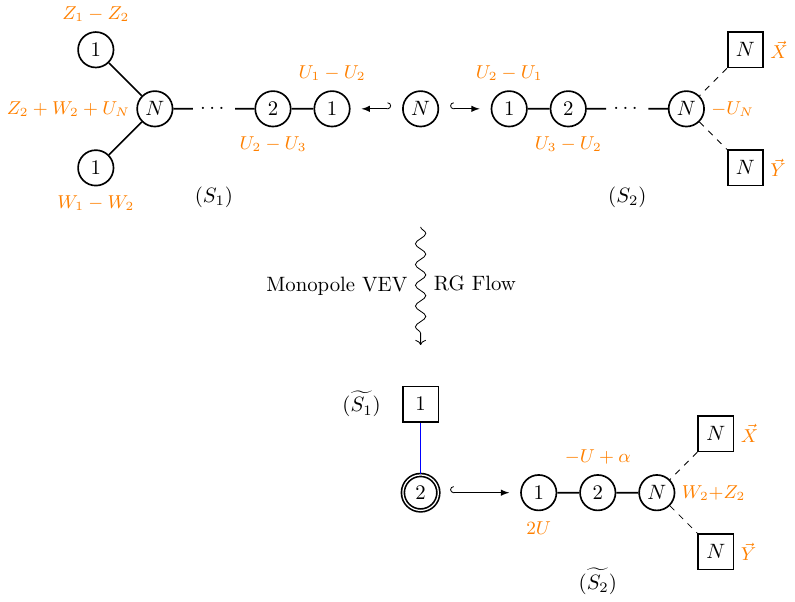}
    \caption{At the top, we depict the quiver corresponding to the 3d mirror of the chosen channel decomposition. Notice that we use a gauge node with hook arrows to indicate the gauging of a $U(N)$ symmetry on the Coulomb branch of the two connected full tails. As pointed out in the main text, $S_1$ is a bad theory, and there is an RG flow driven by monopole VEVs leading to the quiver theory depicted on the bottom. In the interest of brevity, we denote the following combination of FI parameters as $\alpha=\tfrac12(W_1+Z_1-W_2-Z_2)$. Note that the hypers shown in blue are twisted with respect to those in the original theory.}
    \label{fig:sunex1finish}
\end{figure}

The partition function of the original theory $S$ can be expressed as
\begin{equation}\label{eq:13062}
    \begin{split}
        \widehat{\PFS{S}}&\left(\vec{X}, \vec{Y}, \vec{W}_{ [N-1,1] }, \vec{Z}_{ [N-1,1] } ; m\right) \\[1mm]
        & \hspace{2cm} = \intd C \; \DCB{N}(C;m) \; \PFS{S}\left(\vec{X}, \vec{Y}, \vec{W}_{ [N-1,1] }, \vec{Z}_{ [N-1,1] } ; m\right) ~,
    \end{split}
\end{equation}
where the combination $C$, that is integrated over in order to decouple the redundant $U(1)$ gauge group, is
\begin{equation}
    C=\sum_{i=1}^N(X_i+Y_i)+W_1+Z_1+(N-1)(W_2+Z_2)
\end{equation}
The partition function associated with this channel decomposition is:
\begin{equation}
    \begin{split}
        \PFS{S}&\left(\vec{X}, \vec{Y}, \vec{W}_{ [N-1,1] }, \vec{Z}_{ [N-1,1] } ; m\right)= \\
        & = \intd\vec{U}_{N} \; \D^{CB}_{N}(\vec{U};m) \; \PFS{S_1}\left(\vec{Z}_{[N-1,1]}, \vec{W}_{ [N-1,1] }, \vec{U}; m\right) \PFS{S_2}(-\vec{U}, \vec{Y}, \vec{X}; m)
    \end{split}
\end{equation}
For convenience, we report the 3d quiver corresponding to the channel decomposition again at the top of Figure \ref{fig:sunex1finish}.

By design, the trinion containing the two minimal and a maximal puncture corresponds to a bad theory since the central $U(N)$ node only sees $2N-2$ fundamental flavours. Thus, upon running the electric dualisation algorithm on the theory $S_1$, we obtain a Wess--Zumino model with $N-1$ Dirac deltas. In particular, $N-2$ Dirac deltas out of these specify a Nilpotent VEV, whereas the last one converts the measure from $U(2)$ to $SU(2)$:
\begin{equation}\label{DELTA}
    \begin{split}
        & \dd{\rm{Nil}}=\prod_{i=1}^{N-2}\d \left((N-3)(m-\frac{iQ}{2})+(i-1)(iQ-2m)+U_i+W_2+Z_2\right) ~,\\
        & \dd{\rm{Ex}}=\d(U_{N-1}+U_N+W_1+W_2+Z_1+Z_2) ~.
    \end{split}
\end{equation}
These are the expressions for the Dirac deltas in a single frame. In this case, there are a total of $\frac{N!}{2}$ frames that are permuted by the permutation group $S_N/\mathbb Z_2$. The arguments of the Dirac deltas allow us to predict that the $U(N)$ gauge group of the Identity wall is Higgsed down to $SU(2)$ because of a nilpotent VEV corresponding to the freezing conditions:
\begin{equation}
   \begin{cases}
        U_1 = -W_2-Z_2+(N-3)\left(\frac{i Q}{2}-m\right) ~, \\
        U_2 = -W_2-Z_2+(N-1)\left(\frac{i Q}{2}-m\right) ~. \\
        \qquad \vdots \\
        U_{N-2} = -W_2-Z_2-(N-3)\left(\frac{i Q}{2}-m\right) ~,
    \end{cases}
\end{equation}
that is compatible with $U_{i+1}=U_i-iQ+2m$, for $i=1,2,\dots,N-3$.

Such an FI parameter specification corresponds to a VEV for the Coulomb branch moment map given by the partition $[N-2,1^2]$. Following the explanation in Subsection \ref{subsec:channel_decomposition} (see also Appendix \ref{app:TSUN_TRHOSIGMA}), we have that:
\begin{equation}
    \PFT{}{N}(-\vec{U},\vec{V},m)\vert_{\dd{\rm{Nil}}} = \PFT{[N-2,1^2]}{N}\left(-\vec{U}_{[N-2,1^2]},\vec{V};m\right)\times \text{Ch}_{\rm{Nil}}^{[N-2,1^2]} ~,
\end{equation}
where $\text{Ch}_{\text{Nil}}^{[N-2,1^2]}$ are the massless modes produced by the nilpotent vev.
Before presenting the partition function for the final result, we first have to make a shift in the integration variables in order to have the correct measure for $SU(2)$. This can be simply deduced by looking at $\dd{\rm{Ex}}$ in \eqref{DELTA}. Since we need $\d(U_{N-1}+U_N)$ in order to have the correct $SU(2)$ measure, therefore, we make the shift $U_i\rightarrow U_i-\frac{W_1+W_2+Z_1+Z_2}{2}, i=N-1,N$.

Thus, we finally have the following expressions for the partition function:
\begin{equation}\label{eq:13061}
    \begin{split}
    &\PFS{S}(\vec{X}, \vec{Y}, \vec{W}_{ [N-1,1] }, \vec{Z}_{ [N-1,1] } ; m ) = \\
    & \quad = \intd \vec{U}_{SU(2)} \;  \DCB{SU(2)} (\vec U;m) \; s_b\left(m\pm\left(U_N+\frac{W_1-W_2-Z_1+Z_2}{2}\right)\right) \\
    & \hspace{4.5cm} \times \; s_b\left(m\pm\left(U_N+\frac{-W_1+W_2+Z_1-Z_2}{2}\right)\right) \\
    & \qquad \times \; \intd \vec{V}_N \; \Delta^{HB}_N(\vec{V};m) \; \PFT{[N-2,1^2]}{N}\left(-\vec{U}^{\text{spec}}_{[N-2,1^2]},\vec{V};m\right) \\[2mm]
    & \hspace{4cm} \times \; \PFT{}{N}(\vec{X},\vec{V};m) \; \PFT{}{N}(\vec{Y},\vec{V};m) ~,
    \end{split}
\end{equation}
where
\begin{equation}
    \begin{split}
        \vec{U}^{\rm spec}_{[N-2,1^2]} & = \vec{U}_{[N-2,1^2]}|_{\text{$\dd{\rm{Ex}}$+shift}} = (U_{N-1},U_N,-W_2-Z_2)|_{\text{$\dd{\rm{Ex}}$+shift}} \\
        & = \left(-U_N-\frac{W_1+W_2+Z_1+Z_2}{2} , \; U_N-\frac{W_1+W_2+Z_1+Z_2}{2} , \; W_2+Z_2\right) ~.     
    \end{split} 
\end{equation}
Note that the diagonal redundant $U(1)$ gauge symmetry in $\widetilde S_2$ can be decoupled by performing the integral over $C$ in $\widehat{\PFS{S}}$.
The final result in this channel consists in an $SU(2)$ gauging of a single fundamental hypermultiplet to an $SU(2)$ subgroup of the CB symmetry of the $T_{[N-2,1^2]}[U(N)]$ tail in $\widetilde S_2$. We depict the resulting theory in Figure \ref{fig:sunex1finish}. Notice that the hypers (in blue) gauged to the $SU(2)$ node are twisted with respect to the hypers in the quivers shown in the first line of Figure \ref{fig:sunex1finish}, indeed if we mirror back $\widetilde{S}_2$ we obtain the direct (non-mirror) 3d reduction of the 4d theory associated to this channel.
What we obtain in this way is the 3d (mirror) realisation of the Argyres--Seiberg dual of the 4d $\cN=2$ $SU(N)$ SQCD with $2N$ fundamental hypermultiplets, as expected. 

\section{Further interesting features and examples}
\label{sec:further_results}

In the previous section, we focused on spheres with two regular and two maximal punctures. In this section, we turn to a selection of more general configurations that allow us to discuss some further interesting features and advantages of our approach. 

\subsection{$N=6$ example: \texorpdfstring{$\rho_1=[1^6],\;\rho_2=[2^3]\;,\rho_3=[4,2]\;,\rho_4=[5,1]$}{rho1=[1,1,1,1],rho2=[2,2,2],rho3=[4,2],rho4=[5,1]}}
\label{subsec:su6_1max1minimal1box42}

We begin by considering the following good theory: the theory of class $\cS$ of type $\mf{su}(6)$ associated to a sphere with a single maximal puncture and three regular punctures of the type: $[2^3],[4,2],[5,1]$. This theory is already known to correspond to the $SU(6)$ SQCD with nine flavours and matter in the three-index antisymmetric representation of $SU(6)$ \cite{Tachikawa:2011yr,Bhardwaj:2013qia}. This description arises in the channel decomposition that clubs the maximal and minimal puncture together. We will show that our approach easily reproduces this slightly more exotic matter content.

\begin{figure}[]
    \centering
    \includegraphics[width=0.9\linewidth]{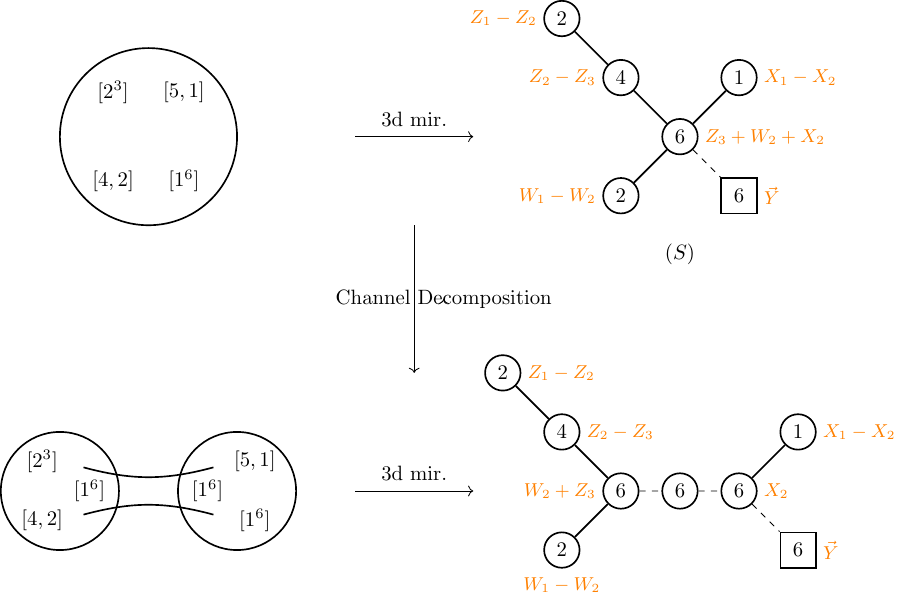}
    \caption{$N=6$ example with one maximal, and three regular punctures of types: $[2^2],[4,2],[5,1]$, and the associated channel decompositions. These are expressed at the level of the Riemann surfaces associated to the corresponding theory of class $\cS$, and the star-shaped quivers corresponding to their 3d mirror. The original theory $S$ is labelled on the top right corner, while the dashed lines denote the full $T[U(6)]$ tails. }
    \label{fig:sec6ex3start}
\end{figure}

\begin{figure}[ ]
    \centering
    \includegraphics[width=1\linewidth]{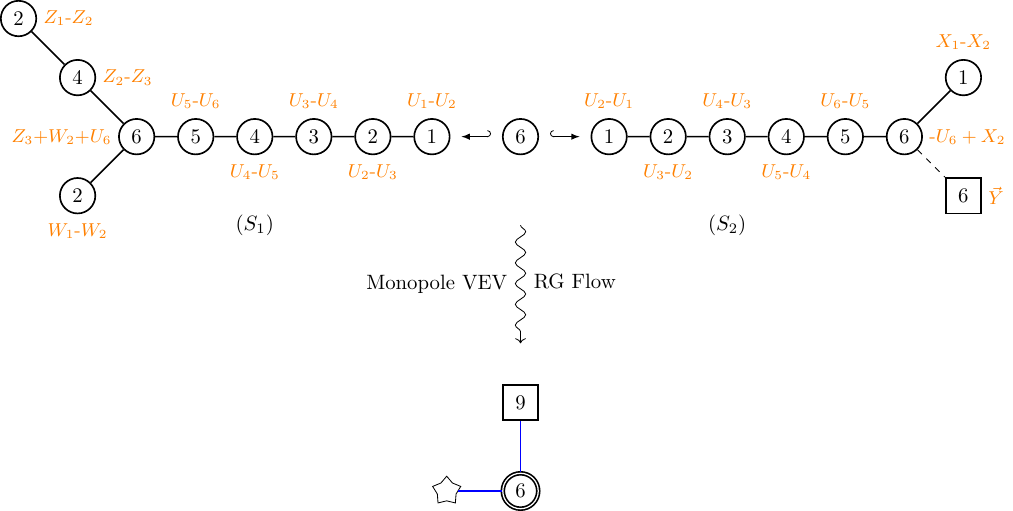}
    \caption{At the top, we report the quiver corresponding to the 3d mirror of the chosen channel decomposition. Notice that we use a gauge node with hook-arrows to indicate the gauging of a $U(6)$ symmetry on the Coulomb branch of the two connected full tails. As pointed out in the main text, $S_1$ and $S_2$ are ugly theories, and there is an RG flow driven by spontaneous monopole VEVs which leads to the quiver theory
depicted on the bottom. The star stands for matter transforming in the 3-index antisymmetric representation of $SU(6)$. Note that the hypers shown in blue are twisted with respect to those in the original theory.}
    \label{fig:sec6ex3finish}
\end{figure}

In the described channel decomposition, the two isolated trinions are both ugly. The Riemann surface, the chosen channel decomposition, and the associated 3d mirror star-shaped quiver are depicted in Figure \ref{fig:sec6ex3start}.
The partition function of the original theory $S$ can be expressed as
\begin{equation}
    \begin{split}
        & \widehat{\PFS{S}}\left(\vec{X}_{[5,1]}, \vec{Y}, \vec{W}_{ [4,2] }, \vec{Z}_{ [2^3] } ; m \right) = \\[1.5mm]
        & \qquad\qquad\qquad\qquad = \; \intd C \; \DCB{1}(C;m) \; \PFS{S}\left(\vec{X}_{[5,1]}, \vec{Y}, \vec{W}_{ [4,2] }, \vec{Z}_{ [2^3] } ; m \right) ~,
    \end{split}
\end{equation}
where the combination $C$, that is integrated over in order to decouple the redundant $U(1)$ gauge group, is
\begin{equation}
    C=\sum_{i=1}^6Y_i+2\sum_{i=1}^3 Z_j+X_1+5X_2+2W_1+4W_2 ~.
\end{equation}
The partition function associated with this decomposition channel is:
\begin{equation}\label{riferimentoNeq6}
\begin{split}
    \widehat{\PFS{S}}&\left(\vec{X}_{[5,1]}, \vec{Y}, \vec{W}_{ [4,2] }, \vec{Z}_{ [2^3] } ; m \right) = \\[1.5mm]
    & \qquad\quad \intd C \; \DCB{1}(C;m) \; \PFS{S_1}\left(\vec{W}_{ [4,2] }, \vec{Z}_{ [2^3] },\vec{U} ; m \right) \; \PFS{S_2}\left(\vec{X}_{[5,1]}, \vec{Y},-\vec{U}; m \right) ~.
\end{split}
\end{equation}
For convenience, we report the 3d quiver corresponding to the channel decomposition again at the top of Figure \ref{fig:sec6ex3finish}.

Note that both the theories, depicted at the top of Figure \ref{fig:sec6ex3finish}, are ugly and thus we can dualise them via the electric algorithm. The partition function of $S_1$ after this dualisation takes the form
\begin{equation}
    \begin{split}
        \PFS{S_1}&\left(\vec{W}_{ [4,2] }, \vec{Z}_{ [2^3] },\vec{U}; m \right) = \delta \left(\sum_{i=1}^6 U_i+2 W_1+4 W_2+2 Z_1+2 Z_2+2Z_3\right) \\
        & \qquad\quad \times \; \prod_{i<j}^5s_b(m\pm (U_i+U_j+U_6+W_1+2W_2+Z_1+Z_2+Z_3)) \\
        & \qquad\quad \times \; \prod_{i=1}^6\prod_{j=1}^3s_b(m\pm(U_i+W_2+Z_j)) ~.
    \end{split}
\end{equation}
Since there is only one Dirac delta, we do not have a nilpotent VEV, and hence we label it as:
\begin{equation}
    \begin{split}
        & \dd{\rm{Nil}}=1 ~, \\
        & \dd{\rm{Ex}}=\delta \left(\sum_{i=1}^6 U_i+2 W_1+4 W_2+2 Z_1+2 Z_2+2 Z_3\right) ~.
    \end{split}
\end{equation}
On the other hand, for the partition function of the theory $S_2$, we have
\begin{equation}
    \begin{split}
        \PFS{S_2}(\vec{X}_{[5,1]}, \vec{Y},-\vec{U} ; m ) = & \delta \left(-\sum_{i=1}^6 U_i+X_1+5 X_2+\sum_{i=1}^6Y_i\right) \\
        & \times \; \prod_{i=1}^6\prod_{j=1}^6 \bigg[s_b\left(m\pm(U_i-X_2-Y_j)\right) \bigg] \; s_b \left(2m-\frac{iQ}{2}\right) \,.
    \end{split}
\end{equation}
Once again, we label the Dirac deltas as:
\begin{equation}
    \begin{split}
        & \dd{\rm{Nil}} = 1 ~, \\
        & \dd{\rm{Ex}}=\delta \left(-\sum_{i=1}^6 U_i+X_1+5 X_2+\sum_{i=1}^6Y_i\right) ~.
    \end{split}
\end{equation}

Thus, in total, we have two Dirac deltas in \eqref{riferimentoNeq6}. As in the previous examples, one of the two Dirac deltas can be written as $\d(C)$, thereby absorbing the integration over $C$, which decouples the diagonal redundant $U(1)$ gauge symmetry. On the other hand, the second Dirac delta converts the gauge group of the Identity wall from $U(6)$ to $SU(6)$. In order to have the correct parametrisation, we shift the integration variables in the following way:
\begin{equation}
    U_j\rightarrow U_j+\frac{\sum_{i=1}^6X_i+X_1+5X_2}{6}, \ j=1,...,6 ~,
\end{equation}
or equivalently,
\begin{equation}
   U_j\rightarrow U_j-\frac{2W_1+4W_2+2Z_1+2Z_2+2Z_3}{6}, \ j=1,...,6 ~.
\end{equation}
Therefore, the partition function in \eqref{riferimentoNeq6} transforms to the following:
\begin{equation}
    \begin{split}
        \widehat{\PFS{S}}&\left(\vec{X}_{[5,1]}, \vec{Y}, \vec{W}_{ [4,2] }, \vec{Z}_{ [2^3] } ; m \right) = \intd\vec{U}_6 \; \DCB{SU(6)}(\vec{U};m) \; \d\left(\sum_{l=1}^6 U_l\right) \\
        & \qquad\quad \times \; \prod_{i=1}^6 \prod_{j=1}^6 s_b\left(m\pm\left(U_i+\frac{W_1+2W_2+\sum_{l=1}^3Z_l}{3}-X_2-Y_j\right)\right) \\
        & \qquad\quad \times \; \prod_{i=1}^6 \prod_{j=1}^3 s_b\left(m\pm\left(U_i-\frac{\sum_{i=1}^6X_i+X_1+5X_2}{6}+W_2+Z_j\right)\right) \\
        & \qquad\quad \times \; \prod_{i<j}^5s_b(m\pm (U_i+U_j+U_6)) ~.
    \end{split}
\end{equation}
As anticipated at the beginning of the subsection, we have obtained a 3d $\cN=4$ $SU(6)$ SQCD with nine flavours and matter in the three-index antisymmetric representation of $SU(6)$. The resulting theory is depicted in the second line of Figure \ref{fig:sec6ex3finish}, and is in fact the direct 3d reduction of the 4d $\cN=2$ $SU(6)$ SQCD with the matter in the same representations.

\subsection{$N=4$ example: \texorpdfstring{$\rho_1=[1^4],\;\rho_{2,3,4}=[3,1]$}{rho1=[1,1,1,1],rho234=[3,1]}}
\label{subsec:su4_1max3minimal}

In this subsection, we study the example associated with the compactification of a 6d $\cN=(2,0)$ theory of type $\mf{su}(4)$ on a sphere with one maximal puncture and three minimal punctures, all labelled by the partition $[3,1]$. This corresponds to a bad theory, both in four and in three dimensions. Note that, one can think of this theory as arising from a channel decomposition of a good theory with more punctures.

The atomic approach of \cite{Chacaltana:2010ks} allows us to understand the theory associated with this pathological configuration, only after a full channel decomposition down to the constituent trinions. The purpose of this section is to show that our approach allows us to bypass this additional step. We can apply the electric dualisation algorithm directly to the sphere with one maximal and three minimal punctures, and read off the resulting theory. In addition to this, we also analyse the channel decomposition explicitly to confirm the consistency of the two approaches.

\begin{figure}[]
    \centering
    \includegraphics[width=0.9\linewidth]{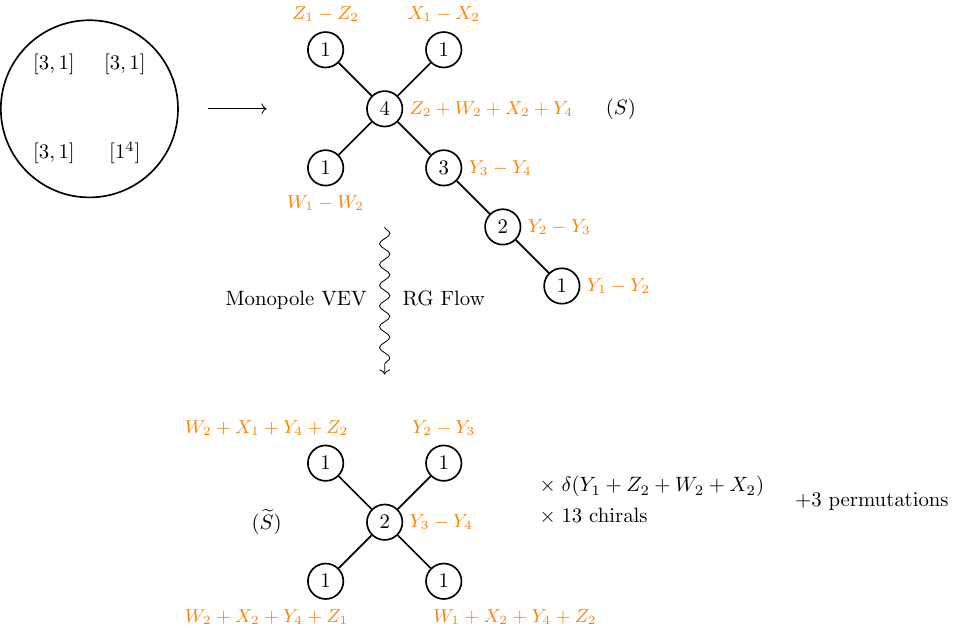}
    \caption{Bad theory associated to a sphere with a full and three minimal punctures and corresponding bad 3d mirror $S$. The RG flow triggered by the monopole VEV leads to theory $\widetilde S$ together with seven twisted hypermultiplets.}
    \label{fig:sec6ex2.pdf}
\end{figure}

The partition function of the original theory $S$, depicted at the top of Figure \ref{fig:sec6ex2.pdf}, can be expressed as
\begin{equation}
    \begin{split}
        & \widehat{\PFS{S}}\left(\vec{X}_{[3,1]}, \vec{Y}, \vec{W}_{ [3,1] }, \vec{Z}_{ [3,1] } ; m \right) = \\[1.5mm]
        & \qquad\qquad\qquad\qquad = \; \intd C \; \DCB{1}(C;m) \; \PFS{S}\left(\vec{X}_{[3,1]}, \vec{Y}, \vec{W}_{ [3,1] }, \vec{Z}_{ [3,1] } ; m \right) ~,
    \end{split}
\end{equation}
where the combination $C$, that is integrated over in order to decouple the redundant $U(1)$ gauge group, is
\begin{equation}\label{eq:17071}
    C=\sum_{i=1}^4Y_i+X_1+3X_2+W_1+3W_2+Z_1+3Z_2 ~.
\end{equation}

The theory $S$ is bad as the central node is underbalanced. Therefore, upon running the electric algorithm, we find:
\begin{equation}\label{eq:sec6secex_fin}
\begin{split}
    \PFS{S}&\left(\vec{X}_{[3,1]}, \vec{Y}, \vec{W}_{ [3,1] }, \vec{Z}_{ [3,1] } ; m \right) = \\
    & = \; \delta \left(Y_1+Z_2+W_2+X_2\right)s_b\left(2m-\frac{iQ}{2}\right) \\
    & \quad \times \; \prod_{j=2}^4\bigg[s_b\left(\frac{iQ}{2}\pm (W_2+X_2+Z_2+Y_j)\right)s_b\left(2m-\frac{iQ}{2}\pm (W_2+X_2+Z_2+Y_j)\right)\bigg] \\[1.5mm]
    & \quad \times \PFS{\widetilde{S}} + \text{ 3 permutations} ~.
\end{split}
\end{equation}
Where $\PFS{\tilde{S}}$ is the partition function of the interacting star-shaped quiver depicted at the bottom of Figure \ref{fig:sec6ex2.pdf}. From the argument of the Dirac delta function, we understand that the $U(4)$ symmetry that is emergent on the Coulomb branch of the full-tail is broken to $U(3)$. Therefore, the 4 frames can be understood as permuted by the action of $S_4/S_3$. Making contact with \eqref{eq: 2reg2fullGenericFeature}, we can group the terms of the first frame as follows:
\begin{equation}\label{refdelta}
\left.
\begin{aligned}
    &\PFS{\widetilde{S}} =(\text{Quiver in the bottom of Figure \ref{fig:sec6ex2.pdf}})\\[0.69mm]
    &\dd{\rm{Nil}} =\; 1  \\[1.1mm]
    &\dd{\rm{Ex}} =\;\delta \left(Y_1+Z_2+W_2+X_2\right) \\[2.1mm]    &\left.\frac{\DCB{3}\left(\vec{Y};m\right)}{\DCB{4}(\vec{Y};m)} \right\vert_{\dd{\text{Ex}}} = \prod_{j=2}^4\bigg[s_b\left(\frac{iQ}{2}\pm (W_2+X_2+Z_2+Y_j)\right) \\ 
    & \hspace{4cm} \; \times \; \left. s_b\left(2m-\frac{iQ}{2}\pm (W_2+X_2+Z_2+Y_j)\right)\bigg]s_b\left(2m-\frac{iQ}{2}\right)\right\vert_{\dd{\text{Ex}}} \\[3mm]
\end{aligned}
\right\vert\;\text{\small Frame 1}
\end{equation}
We recall that the $S$ theory should be thought of as arising from a channel decomposition of a good theory with more punctures. From this point of view, the result in \eqref{refdelta} is signalling that the $U(4)$ gauge symmetry on the tube connecting $S$ to the rest of the theory is Higgsing from $U(4)$ to $U(3)$.
The $U(3)$ symmetry is then also further reduced to $SU(3)$ when one decouples the redundant $U(1)$ gauge symmetry in $\tilde{S}$. This is done by integrating over the combination
\begin{equation}
        \widetilde{C} = Y_2+Y_3-Y_4+2W_2+X_1+2X_2+2X_4+2Z_2+Z_1+W_1 = C ~,
\end{equation}
where we have also used the Dirac delta in \eqref{refdelta} to show the equivalence with the combination in \eqref{eq:17071}. Hence, the overall $U(1)$ in the theory $\widetilde{S}$ is modded out by the original integration over $C$:
\begin{equation}\label{riferiem}
    \begin{split}
        \widehat{\PFS{S}}&\left(\vec{X}_{[3,1]}, \vec{Y}, \vec{W}_{ [3,1] }, \vec{Z}_{ [3,1] } ; m \right) \\
        & = \; \delta \left(Y_1+Z_2+W_2+X_2\right) \, \times \, \prod_{i=2}^4 \bigg[s_b\left(\frac{iQ}{2}\pm (W_2+X_2+Y_i+Z_2)\right) \\
        & \quad\; \times \; s_b\left(2m-\frac{iQ}{2}\pm (W_2+X_2+Y_i+Z_2)\right) \bigg] \; s_b\left(2m-\frac{iQ}{2}\right) \; \widehat{\PFS{\widetilde{S}}} \\
        & \quad + \text{ 3 permutations} ~.   
    \end{split}
\end{equation}
Note that $\widetilde{S}$ is the affine $D_4$ quiver, which is the mirror dual of the 3d $\mathcal{N}=4$ $SU(2)$ SQCD with 4 flavours.

The interpretation of this result should be the following. When the bad $S$ theory appears in the channel decomposition of a good theory, there is a dynamical process that Higgses the $U(4)$ gauge symmetry on the tube attached to $S$ down to $SU(3)$. The $S$ theory is then replaced by an $SU(2)$ SQCD with four flavours, which is connected to the rest of the theory via the diagonal gauging of an $SU(3)$ symmetry. As in all the examples proposed in the paper, the ratio between determinants in \eqref{refdelta} and the multiplicity of frames serves to properly convert the integration measure of $U(4)$ to that of $SU(2)$.

\begin{figure}[]
    \centering
    \includegraphics[width=0.9\linewidth]{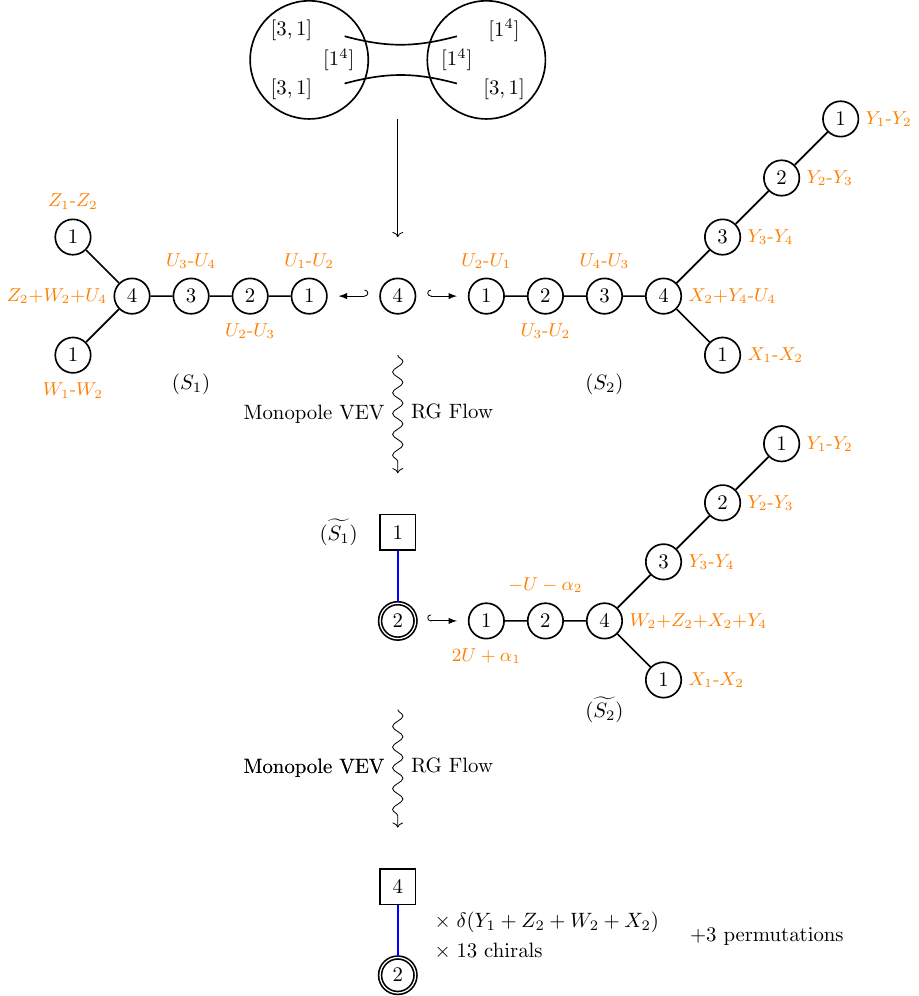}
    \caption{
    At the top and middle, we depict the surface and the quiver corresponding to the 3d mirror of the chosen channel decomposition. Note that we use a gauge node with hook arrows to indicate the gauging of a $U(4)$ symmetry on the Coulomb Branch of the two connected full tails. As pointed out in the main text, $S_1$ is a bad theory, and there is an RG flow driven by monopole VEVs leading to the quiver theory depicted on the third line. These monopole vevs trigger a Nilpotent vev of partition $[2,1^2]$ in the tail of theory $S_2$, leading to the quiver $\widetilde S_2$ on the third line. Now, the theory $\widetilde{S}_2$ becomes a bad theory, and thus there is an RG flow that leads to the quiver at the bottom. In the interest of brevity, we denote the following combinations of FI parameters as $\alpha_1=2W_1+2Z_1+2Z_2$ and $\alpha_2=\frac12(W_1+Z_1+3W_2+3Z_2)$. Note that the hypers shown in blue are twisted with respect to those in the original theory.}
    \label{fig:extraex2finish}
\end{figure}

As a consistency check, we will now explain that the same result can be obtained by performing a channel decomposition into trinions. We consider the channel decomposition at the top of Figure \ref{fig:extraex2finish}, where we club two minimal punctures together.
At the level of the partition function, we have:
\begin{equation}
\begin{split}
    \PFS{S} \left(\vec{X}_{[3,1]}, \vec{Y}, \vec{W}_{ [3,1] }, \vec{Z}_{ [3,1] } ; m \right) = & \intd \vec U_4 \Delta_4^{\rm CB}(\vec U;m) \PFS{S_1}(\vec{W}_{ [3,1] }, \vec{Z}_{ [3,1] },\vec{U} ; m ) \\
    & \quad \times \PFS{S_2}(-\vec{U},\vec{X}_{[3,1]}, \vec{Y}; m ) ~.
\end{split}
\end{equation}
In this decomposition, the $S_1$ theory is bad while $S_2$ is good. Notice in particular that the $S_1$ theory, which has one full and two minimal punctures, is precisely the bad theory studied in Subsection \ref{subsubsec:su4_2minimal}. Following the discussion from \eqref{eqN=42min2fullchannel2} to \eqref{eqriferimentoultimasezione}, we conclude that the $U(4)$ gauge group on the tube is Higgsed to $SU(2)$. This is due to a nilpotent VEV that partially closes one full-puncture in $S_2$ to a regular puncture labelled by $[2,1^2]$.
We thus obtain the partition function
\begin{equation}\label{su2fin}
    \begin{split}
        \PFS{S}&(\vec{X}_{[3,1]}, \vec{Y}, \vec{W}_{ [3,1] }, \vec{Z}_{ [3,1] } ; m ) = \\
        & = \; \intd \vec{U}_{SU(2)} \; \DCB{SU(2)} (\vec U;m) \;  s_b\left(m\pm U \pm \frac{W_1-W_2-Z_1+Z_2}{2} \right) \\
        & \quad \times \; \, \PFS{\widetilde{S}_2}(-\vec{U}^{\text{spec}}_{[2,1^2]},\vec{X}_{[3,1]},\vec{Y};m) ~,
    \end{split}
\end{equation}
which describes the third line in Figure \ref{fig:extraex2finish}. 

After implementing the nilpotent VEV in the $S_2$ theory, which was a good theory, we obtain the $\tilde{S}_2$ theory that is a bad trinion having one maximal, one minimal and a $[2,1^2]$ puncture. The $\tilde{S}_2$ theory is therefore the same bad theory studied in Subsection \ref{subsubsec:su4_1minimal1box}. Following the discussion from \eqref{eqnequal4punct212} to \eqref{eq:05061}, we obtain:
\begin{equation}\label{fin2}
    \begin{split}
        \widehat{\PFS{S}}&(\vec{X}, \vec{Y}, \vec{W}_{ [3,1] }, \vec{Z}_{ [3,1] } ; m ) \; = \; \delta \left(Y_1+Z_2+W_2+X_2\right) \; s_b\left(2m-\frac{iQ}{2}\right) \; \times \\
        & \quad \times \; \prod_{j=2}^4 \bigg [s_b\left(\frac{iQ}{2}\pm(W_2+X_2+Y_j+Z_2)\right) \, s_b\left(2m-\frac{iQ}{2}\pm(W_2+X_2+Y_j+Z_2)\right) \bigg] \\
        & \quad \times \; \intd \vec{U}_{SU(2)} \; \DCB{SU(2)} (\vec U;m) \; \prod_{j=2}^4 \bigg[s_b\left(m\pm\left(U-X_2-Y_j+\frac{W_1+W_2+Z_1+Z_2}{2}\right)\right)  \\
        & \quad \times \; s_b\left(m\pm U \pm \frac{W_1-W_2-Z_1+Z_2}{2}\right)  \\[1.5mm]
        & \; + \text{ 3 permutations} ~.
    \end{split}
\end{equation}
Which describes the last line in Figure \ref{fig:extraex2finish}.
By comparing this result with that in \eqref{eq:sec6secex_fin} we observe that the two expression contain exactly the same delta function and chiral multiplets. The only apparent difference is that in \eqref{eq:sec6secex_fin} the partition function of the remaining $\tilde{S}$ theory is that of the affine $D_4$ quiver, while in \eqref{fin2} the interacting part is that of an $SU(2)$ SQCD with four flavours. The two theories are indeed mirror dual, therefore establishing the perfect agreement of \eqref{fin2} with \eqref{eq:sec6secex_fin}.

In conclusion, we have seen that applying the electric algorithm directly to the bad theory with three minimal and one maximal puncture yields the same result as the longer procedure involving channel decomposition into trinions. Clearly, our approach—that bypasses the trinion decomposition—substantially streamlines the procedure for obtaining the final theory, particularly in the case of spheres with many generic regular punctures.

As a final remark, let us point out a technical detail. Note that in the channel decomposition shown at the top of Figure~\ref{fig:extraex2finish}, the $S_2$ theory is ugly. This means we could have also applied the algorithm directly to $S_2$, resulting in a set of 16 free twisted hypermultiplets.
This alternative route ultimately leads to the same final result. In this case, the four inequivalent frames appear as the four distinct ways to implement a VEV, which can be identified by examining the charges of the chiral multiplets after imposing the delta-function constraints. For the sake of generality, we did not use this observation in the presented computation.

\subsection{$N=4$ example: \texorpdfstring{$\rho_1=[2,1^2],\;\rho_{2,3}=[2^2],\;\rho_4=[3,1]$}{rho1=[2,1,1],rho23=[2,2],rho4=[3,1]}}
\label{subsec:su4_1minimal2box1ntmax}

\begin{figure}[]
    \centering
    \includegraphics[width=0.7\linewidth]{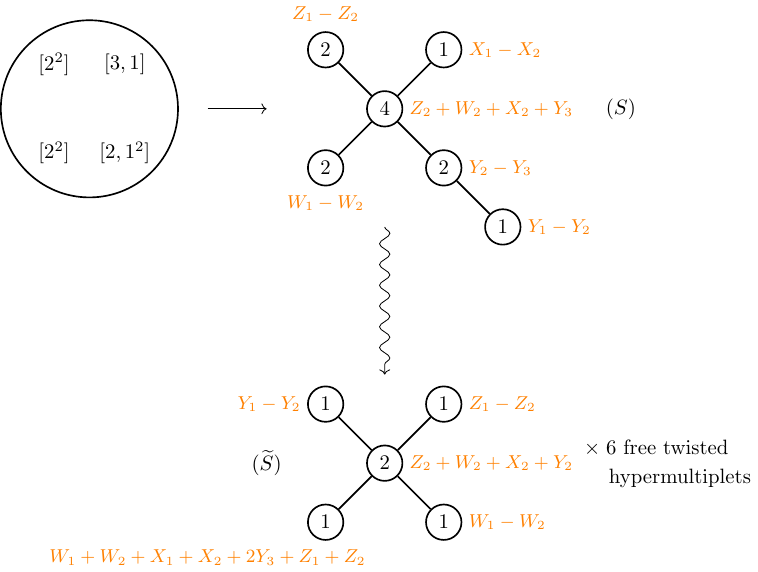}
    \caption{$N=4$ example with four punctures of types: $[2,1^2],[2^2],[2^2],[3,1]$ expressed as the punctured Riemann surface along with the corresponding 3d mirror. The original theory $S$ is labelled on the top right corner. Note that $S$ is ugly and hence dual to the theory $\widetilde S$ in which six free twisted hypermultiplets are released. }
    \label{fig:extraex1}
\end{figure}

We will now consider the 3d mirror of the theory of class $\cS$ of type $\mf{su}(4)$ associated to a sphere with four regular punctures, labelled by the partitions: $[2,1^2],[2^2],[2^2],[3,1]$. We depict the Riemann surface, the chosen channel decomposition, and the associated 3d mirror star-shaped quiver in Figure \ref{fig:extraex1}. 

The star-shaped quiver $S$ is ugly; therefore, as explained in the previous section, we can bypass the channel decomposition and directly apply the electric algorithm, avoiding the need to analyse the intermediate steps. However, we will study the channel decomposition as an exercise, not only to confirm the consistency of the procedure, but also to explore the interesting situation in which both trinions appearing in the decomposition are bad. 
Moreover, in the language of \cite{Chacaltana:2010ks}, this situation corresponds to having two trinions carrying irregular punctures that are connected by a tube. We therefore aim to also show consistency with this result.

The partition function of the original theory $S$ can be expressed as
\begin{equation}
    \begin{split}
         \widehat{\PFS{S}}\left(\vec{X}_{[3,1]}, \vec{Y}_{[2,1^2]}, \vec{W}_{ [2^2] }, \vec{Z}_{ [2^2] } ; m \right)=\intd C \; \DCB{1}(C;m) \; \PFS{S}\left(\vec{X}_{[3,1]}, \vec{Y}_{[2,1^2]}, \vec{W}_{ [2^2] }, \vec{Z}_{ [2^2] } ; m \right) ~,
    \end{split}
\end{equation}
where the combination $C$, that is integrated over in order to decouple the redundant $U(1)$ gauge group, is
\begin{equation}\label{eq:02071}
    C=Y_1+Y_2+2Y_3+X_1+3X_2+2Z_1+2Z_2+2W_1+2W_2 ~.
\end{equation}

By running the algorithm on the ugly theory $S$, we find:
\begin{equation}\label{eq:08071}
    \begin{split}
        \PFS{S}&\left(\vec{X}_{ [3,1] }, \vec{Y}_{ [2,1^2] }, \vec{W}_{ [2^2] }, \vec{Z}_{ [2^2] } ; m \right)= \\
        & \qquad = \; \prod_{i=1}^2[s_b(m\pm(W_1+W_2+2X_2+Y_i+Y_3+Z_1+Z_2))] \\
        & \qquad\quad \times \; \prod_{i=1}^2\prod_{j=1}^2[s_b(m\pm(W_i+X_2+Y_3+Z_j))] \; \PFS{\widetilde{S}} ~,
    \end{split}
\end{equation}
where $\widetilde{S}$ is the partition function of the quiver depicted at the bottom of Figure \ref{fig:extraex1}. Note that $\widetilde{S}$, after the decoupling of the overall $U(1)$, is the affine $D_4$ theory, which is the mirror of the $SU(2)$ SQCD with four flavours. In addition, we also have the contribution of six free twisted hypermultiplets.

\begin{figure}[]
    \centering
    \includegraphics[width=.9\linewidth]{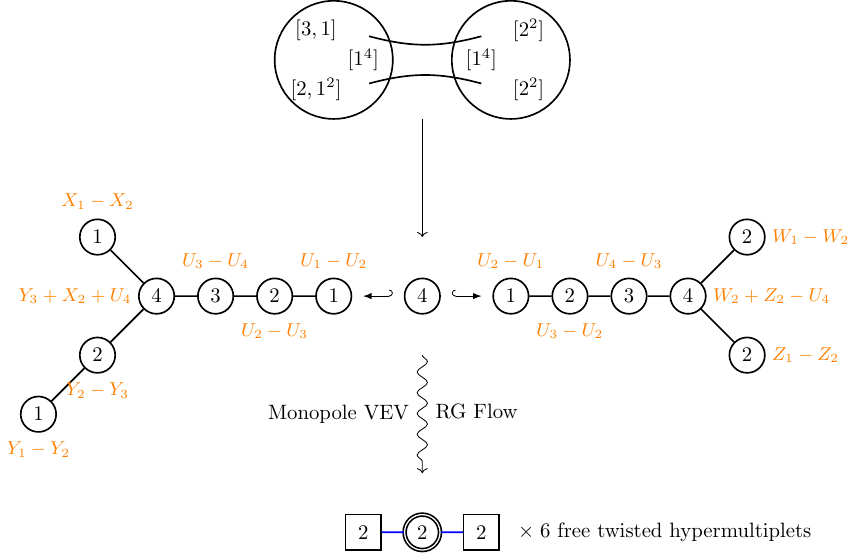}
    \caption{At the top and middle, we depict the surface and the quiver corresponding to the 3d mirror of the chosen channel decomposition. Note that we use a gauge node with hook arrows to indicate the gauging of a $U(4)$ symmetry on the Coulomb Branch of the two connected full tails. As pointed out in the main text, $S_1$ and $S_2$ are both bad theories, and thus trigger an RG flow driven by monopole VEVs leading to the quiver theory depicted on the bottom. Note that the hypers shown in blue are twisted with respect to those in the original theory.}
    \label{fig:extraex1finish}
\end{figure}

We now consider the channel decomposition in which we club the two $[2^2]$ punctures in a single trinion and the $[2,1^2]$ and minimal puncture in the other. For convenience, we report the 3d quiver corresponding to the channel decomposition again at the top of Figure \ref{fig:extraex1finish}. 
The partition function associated with this channel decomposition is:
\begin{equation}\label{2irrpf}
\begin{split}
    \widehat{\PFS{S}}&\left(\vec{X}_{[3,1]}, \vec{Y}_{[2,1^2]}, \vec{W}_{ [2^2] }, \vec{Z}_{ [2^2] } ; m \right) = \intd C \; \DCB{1}(C;m) \\
    & \qquad \times \; \intd\vec{U}_4 \; \DCB{4}(\vec{U};m) \; \PFS{S_1}(\vec{X}_{ [3,1] }, \vec{Y}_{ [2,1^2] }, \vec{U}; m) \; \PFS{S_2}(-\vec{U}, \vec{W}_{ [2^2] }, \vec{Z}_{ [2^2] }; m)
\end{split}
\end{equation}
Notice that both $S_1$ and $S_2$ are bad theories; thus, we run the electric algorithm on both of them. Note that these theories already appeared in Sections \ref{subsubsec:su4_1minimal1ntmax} and \ref{subsubsec:su4_2boxes}, respectively. The difference in this example is that these are now glued to each other.

Recalling and adapting the result in \eqref{ZS1lastcaseN=4} to this case, we get the following partition function identity for $S_1$:
\begin{equation}\label{eq:08072}
    \begin{split}
    \PFS{S_1}&\left(\vec{X}_{ [3,1] }, \vec{Y}_{ [2,1^2] }, \vec{U}; m\right) \\[1.5mm]
   & \quad = \delta(U_4+X_2+Y_3)\delta(U_1+U_2+U_3+X_1+2X_2+Y_1+Y_2+Y_3) \\
    & \qquad \times \; \prod_{j=1}^3 \bigg[s_b\left(\frac{iQ}{2}\pm(U_j+X_2+Y_3)\right)s_b\left(2m-\frac{iQ}{2}\pm(U_j+X_2+Y_3)\right) \bigg] \\
    & \qquad \times \; s_b\left(2m-\frac{iQ}{2}\right)^2\prod_{k=1}^3 \bigg[ s_b\left(m\pm\left(U_k-\frac{Y_1-2Y_2+Y_3+X_1-X_2}{3}\right)\right) \\
    & \qquad \times \; s_b\left(m\pm\left(U_k+\frac{2Y_1-Y_2-Y_3-X_1+X_2}{3}\right)\right) \bigg] \\
    & \qquad + \text{ 3 permutations} ~.
    \end{split}
\end{equation}
These terms can be grouped as in \eqref{eq: su4ex4_ch}, which we do not repeat here for brevity. On the other hand, the partition function for the theory $S_2$ can be adapted to this example from \eqref{eqN=42squared2squaredbadS1}, and takes the following form:
\begin{equation}\label{eq:08073}
    \begin{split}
        \PFS{S_2}&(-\vec{U}, \vec{W}_{ [2^2] }, \vec{Z}_{ [2^2] }; m) \\[1.5mm]
        & \quad = \delta(-U_2-U_3+W_1+W_2+Z_1+Z_2)\delta(-U_1-U_4+W_1+W_2+Z_1+Z_2) \\
        & \qquad \times \; \prod_{j=3}^4 \bigg[s_b(m\pm (-U_j+W_2+Z_2))s_b(m\pm (-U_j+W_1+Z_2))s_b(m\pm (-U_j+W_2+Z_1)) \bigg] \\
        & \qquad \times \; s_b\left(2m-\frac{iQ}{2}\right)^2 s_b\left(\frac{iQ}{2}\pm(U_3-U_4)\right)s_b\left(2m-\frac{iQ}{2}\pm(U_3-U_4)\right) \\
        & \qquad \times \; s_b\left(\frac{iQ}{2}\pm(-U_3-U_4+W_1+W_2+Z_1+Z_2)\right) \\
        & \qquad \times \; s_b\left(2m-\frac{iQ}{2}\pm(-U_3-U_4+W_1+W_2+Z_1+Z_2)\right) \\
        & \qquad + \text{ 2 permutations}
    \end{split}
\end{equation}
These terms can be grouped as in \eqref{eq: su4ex2_ch}, which we once again do not repeat here for brevity. To see the effect of gauging them together, we substitute \eqref{eq:08072} and \eqref{eq:08073} in \eqref{2irrpf}. This way, we have four Dirac deltas and an integral over the four variables $\vec{U}$ associated with the $U(4)$ gauge group on the tube. However, we have an additional integral over $C$ (in the ``hatted" partition function) that decouples the overall diagonal $U(1)$ from the starting theory $S$. Therefore, we have to be careful while solving the Dirac deltas.

Using the various Dirac deltas, we can rewrite one of the four Dirac deltas as $\delta(C)$, where $C$ is given by \eqref{eq:02071}, while using the other three Dirac deltas to Higgs the $U(4)$ gauge group down to $SU(2)$. Notice that the product of the multiplicity of frames is $4 \times 3 = 12$, which is in agreement with the size of the $S_4/Z_2$ permutation symmetry, {\it i.e.} the quotient of the Weyl groups of $U(4)$ and $SU(2)$, rotating these frames. At the level of the partition function, we have:
\begin{equation}
    \begin{split}
        &\widehat{\PFS{S}}\left(\vec{X}_{ [3,1] }, \vec{Y}_{ [2,1^2] }, \vec{W}_{ [2^2] }, \vec{Z}_{ [2^2] } ; m \right) = \intd C \; \DCB{1}(C;m) \PFS{S}\left(\vec{X}_{ [3,1] }, \vec{Y}_{ [2,1^2] }, \vec{W}_{ [2^2] }, \vec{Z}_{ [2^2] } ; m \right) = \\[1.5mm]
        & \qquad = \; \intd C \; \DCB{1}(C;m) \intd\vec{U}_{4} \; \DCB{4}( \vec{U}; m) \; \delta(C) \; \delta(U_1+X_2+Y_3) \\[1.5mm]
        & \qquad\quad \times \; \delta(-U_2-U_3+W_1+W_2+Z_1+Z_2) \; \delta(-U_1-U_4+W_1+W_2+Z_1+Z_2) \\[1.5mm]
        & \qquad\quad \times \; \prod_{i=2}^4\bigg[s_b\left(\frac{iQ}{2}\pm(U_i+X_2+Y_3)\right) s_b\left(2m-\frac{iQ}{2}\pm(U_i+X_2+Y_3)\right) \bigg] \\
        & \qquad\quad \times \; \prod_{i=2}^3 \bigg[s_b\left(\frac{iQ}{2}\pm(U_i-U_4)\right) s_b\left(2m-\frac{iQ}{2}\pm(U_i-U_4)\right) \bigg] \\
        & \qquad\quad \times \; \prod_{i=3}^4\bigg[s_b\left(m\pm (U_i-W_1-Z_2)\right) \,s_b\left(m\pm (U_i-W_2-Z_1)\right) \bigg] \\
        & \qquad\quad \times \; \prod_{i=1}^4\bigg[s_b(m\pm (U_i-W_2-Z_2))\bigg] s_b\left(2m-\frac{iQ}{2}\right)^4 ~.
    \end{split}
\end{equation}
We implement Dirac deltas and make a shift of the integration variable $U_3\rightarrow U_3+\frac{W_1+W_2+Z_1+Z_2}{2}$ (in order to reproduce the correct $SU(2)$ measure) to obtain the following final partition function:
\begin{equation}\label{tube between two irr, final result}
    \begin{split}
        & \widehat{\PFS{S}}(\vec{X}_{ [3,1] }, \vec{Y}_{ [2,1^2] }, \vec{W}_{ [2^2] }, \vec{Z}_{ [2^2] } ; m ) = \prod_{i=1}^2 \bigg[s_b(m\pm(W_1+W_2+2X_2+Y_i+Y_3+Z_1+Z_2))\bigg] \\
        & \qquad \times \; \prod_{i=1}^2\prod_{j=1}^2 \bigg[s_b(m\pm(W_i+X_2+Y_3+Z_j)) \bigg] \times \intd\vec{U}_{SU(2)} \; \DCB{SU(2)}(\vec{U};m) \\
        & \qquad \times \; s_b\left(m\pm\left(U\pm\left(X_1+2 X_2+Y_2+2 Y_3 + \frac32(W_1+W_2+Z_1+Z_2)\right)\right)\right) \\
        & \qquad \times \; s_b\left(m\pm\left(U+\frac12(W_1-W_2+Z_1-Z_2)\right)\right) \; s_b(m\pm (U-W_1-Z_1)) \\
        & \qquad \times \; s_b\left(m\pm\left(U+X_2+Y_2+\frac{1}{2} \left(W_1+W_2+Z_1+Z_2\right)\right)\right) \\
        & \qquad \times \; s_b\left(m\pm\left(U-X_2-Y_2-\frac12(W_1+W_2+Z_1+Z_2)\right)\right) \\
        & \qquad \times \; s_b\left(m\pm\left(U\pm\frac12(W_1-W_2)+\frac12(Z_2-Z_1)\right)\right) ~.
    \end{split}
\end{equation}

Thus, the final result is an $SU(2)$ gauge theory with four fundamental hypermultiplets, along with six free hypermultiplets. We depict this final theory at the bottom of Figure 
\ref{fig:extraex1finish}.
 Notice that the interacting part of this result is exactly the 3d mirror dual of the theory $\widetilde{S}$ (with the overall $U(1)$ decoupled). Additionally, the free twisted hypers appearing in \eqref{tube between two irr, final result} and \eqref{eq:08071}  exactly match. Thus, we demonstrate the consistency of the procedure at the level of the four-punctured sphere and the corresponding channel decomposition, such that the 3d mirror association to the former is ugly (see Figure \ref{fig:extraex1}), while those associated with the theories appearing in the channel decomposition are both bad (see Figure \ref{fig:extraex1finish}).

Finally, we match the prediction in \cite{Chacaltana:2010ks}. In their language, this situation corresponds to having two spheres carrying one irregular puncture. One carries the $[3,1]$ and $[2,1^2]$ regular punctures and the irregular puncture labelled by the pole structure $\{ 1,3,4 \}$, while the other carries two $[2^2]$ regular punctures and an irregular puncture labelled by the pole structure $\{ 1,3,3 \}$.
The two spheres are also glued together via a tube supporting an $SU(2)$ gauge group.


\acknowledgments

We thank Chiung Hwang, Shlomo Razamat and  Matteo Sacchi  for discussions and comments on the draft.
We also thank William Harding, Fabio Marino, Noppadol Mekareeya, Simone Rota and Alberto Zaffaroni for interesting discussions.
RC, SP and PS are partially supported by the MUR-PRIN grant No. 2022NY2MXY (Finanziato dall’Unione europea- Next Generation EU, Missione 4 Componente 1 CUP H53D23001080006).

\appendix

\section{$\mathbf{S}^3_b$ partition function notation}
\label{app:S3b_partfn}

In this appendix, we introduce the notation for the 3d $\mathcal{N}=2$ partition function computed on the squashed three-sphere $S^3_b$ \cite{Hama:2011ea}. 

Consider a theory with gauge group $G$ and chiral multiplets in the representation $R_G$ and $R_F$ of the gauge and flavour symmetry groups, respectively. Its $S^3_b$ partition function is given by the following integral on $\mathbb{R}^{{\rm rank}(G)}$:
\begin{equation}\label{conventionpf}
\mathcal{Z} (Y,k,\vec{X}) =  \intd \vec{Z} \PFS{\text{Cl}}(Y,k) \frac{\prod_{\vec{\sigma}_G \in R_G} \prod_{\vec{\sigma}_F \in R_F} s_b \left( \frac{iQ}{2}(1-r) - \vec{\sigma}_G(\vec{Z}) - \vec{\sigma}_F(\vec{X}) \right)}{\prod_{\vec{\rho} \in G} s_b \big( \frac{iQ}{2} - \vec{\rho}(\vec{Z}) \big) }    
\end{equation}
where the integration measure is given by:
\begin{equation}
    \text{d}\vec{Z}=\frac{1}{|W(G)|}\prod_{i=1}^{Rank(G)}\text{d}Z_i ~.
\end{equation}
Here $|W(G)|$ is the dimension of the Weyl group of $G$. Meanwhile, the classical term contains the contribution of the FI term with parameter $Y$:
\begin{align}
    \PFS{\text{Cl}}(Y,k) = \exp \left[ 2\pi i Y \sum_{a=1}^{Rank (G)} Z_a \right]  \,.
\end{align}

The remaining terms appearing in (\ref{conventionpf}) have the following meanings: $\vec{\r}_G(\vec{Z})$ are the roots of G, while $\vec{\s}_G(\vec{Z})$ and $\vec{\s}_F(\vec{X})$ are respectively the weights of the representation $R_G$ and $R_F$. The parameters $\vec{Z}$ and $\vec{X}$ are indeed in the Cartan of the gauge and flavour group. The parameter $Q$ is equal to $b+b^{-1}$ with $b$ the squashing parameter of the sphere. Finally, the function $s_b$ that appears is the double-sine function and is given by:
\begin{equation}
    s_b(x)=\prod_{l,n\geq 0}\frac{l\,b+n\,b^{-1}+\frac{Q}{2}-i\,x}{l\,b+n\,b^{-1}+\frac{Q}{2}+i\,x} ~.
\end{equation}

Throughout this paper, we work with $\mathcal{N}=4$ theories in the $\mathcal{N}=2^*$ setup \cite{Tong:2000ky}. We do this by turning on a real mass $Re(m)$ for the axial symmetry $U(1)_A$, which is the commutant of the chosen $U(1)_R$ $\mathcal{N}=2$ $R$-symmetry subgroup in the $\mathcal{N}=4$ $R$-symmetry $SU(2) \times SU(2)$. The $U(1)_R$ symmetry in the IR can only mix with other Abelian symmetries, and the only Abelian symmetry in this case is $U(1)_A$. Therefore, we denote the mixing between these two symmetries with $r$, and the charge under $U(1)_A$ by $q_A$. Therefore, the IR $R$-charge is given by:
\begin{equation}
    R=R_0+q_A\,r ~,
\end{equation}
where $R_0$ is the UV $U(1)_R$ charge. Note that the IR mixing in $\cN=4$ theories is such that the $R$-charge of a single hypermultiplet becomes $\frac{1}{2}$ and for an adjoint chiral becomes $1$, as expected.

It is useful to define the following holomorphic combination of parameters to keep track of the mixing of $U(1)_R$ and $U(1)_A$:
\begin{equation}\label{eq:massm}
    m=Re(m)+\frac{iQ}{2}r ~.
\end{equation}
Then we should assign charges under $U(1)_A$ and $U(1)_R$ consistently with the $\mathcal{N}=4$ superpotential, which for a single gauge node is given by:
\begin{equation}
    W=\text{Tr}\left(\mathbf\Phi \; Q \, \widetilde{Q}\right) ~,
\end{equation}
where $\mathbf\Phi$ is the adjoint chiral inside the $\mathcal{N}=4$ vector multiple, and $Q/\widetilde{Q}$ is the chiral/antichiral multiplet inside the hypermultiplet charged under the gauge node under consideration. In particular, we assign the charges for fields in a 3d $\cN=4$ theory and in its mirror dual as listed in Table \ref{chargeN2star}. From the point of view of parameters in the partition function, mirror symmetry amounts to the following shift:
\begin{equation}
    m\rightarrow \frac{iQ}{2}-m ~.
\end{equation}

\begin{table}[]
\centering
\begin{tabular}{c|c|c}
{} & $U(1)_{A}$ & $U(1)_{R_0}$  \\ \hline
$Q$ & $1$ & $0$ \\
$\widetilde{Q}$ & $1$ & $0$ \\
$\mathbf\Phi$& $-2$ & $2$ 
\end{tabular}
\qquad
\begin{tabular}{c|c|c}
{} & $U(1)_{A}$ & $U(1)_{R_0}$  \\ \hline
$Q$ & $-1$ & $1$ \\
$\widetilde{Q}$ & $-1$ & $1$ \\
$\mathbf\Phi$& $2$ & $0$ 
\end{tabular}
\caption{Charges and representations of the chiral fields in the $\mathcal{N}=2^*$ setup. On the LHS, we list the charges conventionally picked for a generic theory, meanwhile on the RHS, we list those appearing in its mirror theory. The charges in the mirror theory follow from the $SU(2)_H \longleftrightarrow SU(2)_C$ swapping. }
\label{chargeN2star}
\end{table}

It is convenient to report the measure for the $U(N)$ gauge group in the $\mathcal{N}=2^*$ setup. In particular we group together the contribution of the $\mathcal{N}=2$ vector and adjoint in the following term.
\begin{equation}
    \D^{HB}_N(\vec{z};m)=\frac{\prod_{i,j=1}^Ns_b(2m-\frac{iQ}{2}+z_i-z_j)}{\prod_{i\neq j}^Ns_b(\frac{iQ}{2}+z_i-z_j)}
\label{HBmeasure}
\end{equation}
The label ``HB" stands for the fact that this is a gauging of a Higgs branch, i.e.~flavour, global symmetry in the electric theory. This respects the assignment of charges in Table \ref{chargeN2star}, then the hypermultiplets charged under this node have a contribution of the form:
\begin{equation}
    \prod_{i=1}^{N-1}\prod_{\r\in G}s_b({\frac{iQ}{2}-m\pm(z^{(N-1)}_i-X_\r)}) \,.
\end{equation}
We report also the measure for a $U(N)$ gauging of a Coulomb branch, i.e.~topological, symmetry in the electric theory:
\begin{equation}\label{eq:CBmeasure}
    \D^{CB}_N(\vec{z};m)=\frac{\prod_{i,j=1}^Ns_b(-2m+\frac{iQ}{2}+z_i^{(m)}-z_j^{(m)})}{\prod_{i\neq j}^Ns_b(\frac{iQ}{2}+z_i^{(m)}-z_j^{(m)})} \,.
\end{equation}
It is convenient also to report the measure for a $USp(2N)$ gauging on the CB of the electric theory. It is given by the following formula:
\begin{equation}
    \D^{CB}_{USp(2N)}(\vec{z};m)=\frac{s_b(-2m+\frac{iQ}{2})^N\prod_{i<j}^Ns_b(-2m+\frac{iQ}{2}\pm z_i \pm z_j))\prod_{i=1}^N s_b(-2m+\frac{iQ}{2}\pm (2z_i))}{\prod_{i< j}^Ns_b(\frac{iQ}{2}\pm z_i \pm z_j)\prod_{i=1}^N s_b(\frac{iQ}{2}\pm (2z_i))}
\end{equation}
\section{Linear unitary quivers}
\label{app:TSUN_TRHOSIGMA}
In this section we review the $T^\sigma_\rho[U(N)]$ theory, that are linear unitary quivers that appear in theories in this paper.
\subsection*{$T[SU(N)]$ and $T[U(N)$] theory}
\begin{figure}
    \centering
    \includegraphics[]{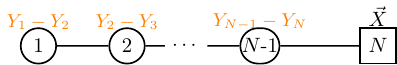}
    \caption{T[SU(N)] quiver. In orange are reported FI parameters for each gauge node and $\vec{X}$ are the flavour parameters}
    \label{fig:tsuNappendix1}
\end{figure}

The $T[SU(N)]$ theory admits a Lagrangian description in terms of the quiver reported in Figure \ref{fig:tsuNappendix1}.
The gauge group is $\prod_{i=1}^{N-1}U(i)$ and the last node has $N$ flavours transforming under an $SU(N)_H$ global symmetry.
\\It is an $\mathcal{N}=4$ theory, so the superpotential is given by:
\begin{equation}
W_{T[SU(N)]}= \sum_{i=1}^{N-1}  \text{Tr}_i \left[ \mathbf{\Phi}^{(i)} \left( \text{Tr}_{i+1} \mathbf{Q}^{(i,i+1)}-\text{Tr}_{i-1} \mathbf{Q}^{(i-1,i)} \right) \right]\, ,
\end{equation}
with $\mathbf{\Phi}^{(i)},i=1,...,N$ chiral multiplet in the adjoint representation of $U(N)$, that is part of the $\mathcal{N}=4$ vector multiplet, and with $\mathbf{Q}^{(i,i+1)}$ the matrix of bifundamentals connecting $U(i)$ to the $U(i+1)$ gauge node.
\\ The CB symmetry in the UV is $U(1)^{N-1}$, but there's a symmetry enhancement in the IR to $SU(N)_C$.

For each Cartan in the two $SU(N)$ global symmetries we can turn on real masses. The parametrization we choose for them in this section consists in two set of parameters $X_i, i=1,...,N$ and $Y_i, i=1,...,N$, such that they respect the tracelessness condition $\sum_{i=1}^NX_i=\sum_{i=1}^NY_i=0$. We work in the $\mathcal{N}=2^*$ setup (introduced in Appendix \ref{app:S3b_partfn}).
\\The global symmetry is then given by $SU(N)_H\times SU(N)_C\times U(1)_A$. The assignment of charges to the fields in the theory is reported in Table \ref{chargestsun}.
\begin{table}[t]
\centering
\scalebox{1}{
\begin{tabular}{c|ccc|c}
{} & $SU(N)_H$ & $SU(N)_{C}$ & $U(1)_{A}$ & $U(1)_{R_0}$  \\ \hline
$Q^{(i-1,i)}$ & $\bullet$ & $\bullet$ & $1$ & $0$ \\
$\widetilde{Q}^{(i-1,i)}$ & $\bullet$ & $\bullet$ & $1$ & $0$ \\
$Q^{(N-1,N)}$ & $ N$ & $\bullet$ & $1$ & $0$ \\
$\widetilde{Q}^{(N-1,N)}$ & $\bar{N}$ & $\bullet$ & $1$ & $0$ \\ 
$\mathbf{\Phi} ^{(i)}$ & $\bullet$ & $\bullet$ & $-2$ & $2$ 
\end{tabular}}
    \caption{Charges and representations of the chiral fields under the global symmetries. The R-charge assignment is in the UV, in the IR it flows to the usual $\mathcal{N}=4$ values, mixing with $U(1)_A$ charge. In the table $i=1,\cdots,N-1$ and $Q^{(0,1)}=\widetilde{Q}^{(0,1)}=0$.}
 \label{chargestsun}
\end{table}
\\The generators of the chiral ring of the $T[SU(N)]$ theory are the Higgs branch and Coulomb branch moment maps, labelled respectively by $\mathcal{H}$ and $\mathcal{C}$.
\\The HB moment map is given by:
\begin{equation}
    \mathcal{H}=\mathcal{Q}-\frac{1}{N}\text{Tr}_X\mathcal{Q}
\end{equation}
with $\mathcal{Q}$ the meson given by:
\begin{equation}
    \mathcal{Q}_{ij}=Tr_{N-1}\mathbf{Q}^{(N-1,N)} \,.
\end{equation}
The CB moment map is given by $Tr_i\mathbf{\Phi}^{(i)}$ and bare monopole operators $\mathfrak{M}^{\vec{m}}$ with fluxes of the form $\vec{m}=(0^i,(\pm 1)^j,0^k)$ with $i+j+k=N-1$. 
These monopoles have the same $U(1)_R$ and $U(1)_A$ charges. We can collect these $N(N-1)$ monopoles and the traces of the N-1 adjoint chirals into an $N\times N$ traceless matrix.
We report as an example the case $N=3$.
\begin{equation}
\mathcal C \equiv 
\left(\begin{array}{ccccccc}
0									&\ & \mathfrak{M}^{(1,0)} 							&\ & \mathfrak{M}^{(1,1)} 	\\[.15cm]	
\mathfrak{M}^{(\textrm{-}1,0)}      				&\ & 0									&\ & \mathfrak{M}^{(0,1)} \\[.15cm]	
\mathfrak{M}^{(\textrm{-}1,\textrm{-}1)}			&\ &  	\mathfrak{M}^{(0,\textrm{-}1)}				&\ & 	0				 \\[.15cm]		

\end{array}\right)+ \sum_{i=1}^{2} \text{Tr}_i\mathbf\Phi^{(i)} \mathcal{D}_i \, ,
\end{equation}
The charges of the moment map under symmetries in the theory are reported in Table \ref{chargemomentmapstsun}.
\begin{table}[t]
\centering
\scalebox{1}{
\begin{tabular}{c|ccc|c}
{} & $SU(N)_H$ & $SU(N)_{C}$ & $U(1)_{A}$ & $U(1)_{R_0}$  \\ \hline
$\mathcal{H}$ & $\bf N^2-1$ & $\bullet$ & $2$ & $0$ \\
$\mathcal{C}$  & $\bullet$ & $\bf N^2-1$ & $-2$ & $2$
\end{tabular}}
    \caption{Charges and representations of the chiral ring generators of T[SU(N)] under the global symmetries. In the table $i=1,\cdots,N-1$ and $Q^{(0,1)}=\widetilde{Q}^{(0,1)}=0$.}
 \label{chargemomentmapstsun}
\end{table}
We can then write the partition function for the $T[SU(N)]$ theory recursively as:
\begin{equation}
\begin{split}
    \mathcal{Z}_{T[SU(N)]}(\vec X,\vec Y;m)&=\intd \vec{z}^{(N-1)}_{N-1} e^{2\pi i(Y_{N-1}-Y_N)\sum_{i=1}^{N-1}z_i^{(N-1)}}\D^{HB}_N(\vec{z}^{(N-1)}_{N-1};m)\\
    &\quad\times\prod_{i=1}^{N-1}\prod_{n=1}^Ns_b({i\frac{Q}{2}\pm(z^{(N-1)}_i-X_n)-m}) \\
    &\quad\times\mathcal{Z}_{T[SU(N-1)]}\left(\vec z^{(N-1)},Y_1, \cdots , Y_{N-1};m\right) \,.
\label{eq:tsunPF}
\end{split}
\end{equation}
The $T[SU(N)]$ theory is self-mirror under mirror duality. In particular it involves the following operator map:
\begin{equation}
    \begin{split}
        &\mathcal{H}\longleftrightarrow \mathcal{C}^V\\&\mathcal{C}\longleftrightarrow \mathcal{H}^V
    \end{split}
\end{equation}
At the level of $S_b^3$ partition function (in $\mathcal{N}=2^*$ setup) mirror duality for $T[SU(N)]$ theory amounts to the following non-trivial integral identity:
\begin{equation}
    \mathcal{Z}_{T[SU(N)]}(\vec X,\vec Y;m)=\mathcal{Z}_{T[SU(N)]}\left(\vec Y,\vec X;i\frac{Q}{2}-m\right)
\end{equation}
So far we have talked about $T[SU(N)]$ theory. It is possible to introduce also the $T[U(N)]$ theory, that is obtained from the $T[SU(N)]$ theory with the addition of a contact term for the flavour symmetry. The partition function for the $T[U(N)]$ theory is given by:
\begin{equation}
    \mathcal{Z}_{T[U(N)]}(\vec X,\vec Y;m)=\mathcal{Z}_{T[SU(N)]}(\vec X,\vec Y;m)e^{2\pi i(\sum_{i=1}^N X_i)Y_N} \,.
\end{equation}
In this paper we label $\mathcal{Z}_{T[U(N)]}$ as $\mathcal{Z}_{T^N}$ in order to simplify the notation. Then mirror duality for $T[U(N)]$ theories at the level of the partition function becomes:
\begin{equation}
\begin{split}
    \mathcal{Z}_{T^N}(\vec X,\vec Y;m) = & \mathcal{Z}_{T[SU(N)]}(\vec X,\vec Y;m)e^{2\pi i(\sum_{i=1}^N X_i)Y_N} = \\
     = & \mathcal{Z}_{T^N}\left(\vec Y,\vec X;i\frac{Q}{2}-m\right) = \\
     = & \mathcal{Z}_{T[SU(N)]}\left(\vec Y,\vec X;i\frac{Q}{2}-m\right)e^{2\pi i(\sum_{i=1}^N Y_i)X_N} \,.
\end{split}
\end{equation}

\begin{figure}[ ]
    \centering
    \includegraphics[width=0.25\linewidth]{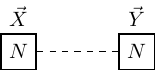}
    \caption{Compact notation for  the $T[U(N)]$ quiver theory.}
    \label{fig:tsundashedline}
\end{figure}

We will ofter represent this theory in compact form
as a dashed line connecting two squared nodes as in
 Figure \ref{fig:tsundashedline}. This notation highlights the presence of the two non-abelian $U(N)$ global  symmetries.

\subsection*{$T_\r^\s[U(N)]$ theories}
\begin{figure}[ ]
    \centering
    \includegraphics[width=0.5\linewidth]{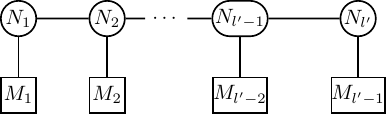}
    \caption{$T_\r^\s$ quiver}
    \label{fig:trsquiver}
\end{figure}

In this section we review $T_\r^\s[SU(N)]$ theories. 
\\$T_\r^\s[SU(N)]$ has a Lagrangian description in term of a linear quiver as reported in Figure \ref{fig:trsquiver}.
To define a $T_\r^\s[SU(N)]$ theory is enough to give $\r$ and $\s$ that are non-increasing partitions of $N$, so they are given by:
\begin{equation}
    \rho=(\rho_1,...,\rho_{\widehat{k}}), \ \rho_1\geq ...\geq \rho_{\widehat{k}}>0, \ \sum_{i=1}^{\widehat{k}}\rho_i=N
\end{equation}
and
\begin{equation}
    \sigma=(\sigma_1,...,\sigma_{k}), \ \sigma_1\geq ...\geq \sigma_{k}>0, \ \sum_{i=1}^k\sigma_i=N \,.
\end{equation}
The gauge group is $G=U(N_1)\times U(N_2)\times ... \times U(N_{\widehat{k}-1})$, and $M_j$ is the number of flavours of the $j$-th gauge node. Given $\rho,\sigma$ all $M_i,N_j$ are completely fixed. The ranks of flavour and gauge groups are obtained in the following way. First, let us rewrite the partitions as:
\begin{equation}
\begin{split}
&\sigma= (1^{M_1},2^{M_2},\ldots,N^{M_N}), \text{ with} \sum_{l} l M_l = N,\\&
\r= (1^{\widehat{M_1}},2^{\widehat{M_2}},\ldots,N^{\widehat{M_N}}), \text{ with} \sum_{l} l \widehat{M_l} = N
\end{split}
\end{equation}
The values $M_j$ are precisely the number of hypermultiplets in the fundamental representation of the $j$-th gauge group.\\
The ranks of the gauge group are obtained from:
\begin{equation}\label{equation2.4TRHOSIGMA}
\begin{split}
    & N_1=k-\rho_1 \\
    & N_j=N_{j-1}+m_j-\rho_j \quad , \quad j=2,..,\hat{k}-1
\end{split}
\end{equation}
with $m_j$ that counts the number of terms bigger or equal than $j$ in $\sigma = (\sigma_1,\ldots,\sigma_l)$. Thus $m_1=k$ and $m_{l+1} = m_l -M_l$.
$T_{\rho}^{\sigma}[SU(N)]$ flows to a non-trivial fixed point in the IR if:
\begin{equation}
    \sigma^T > \rho \Leftrightarrow \rho^T > \sigma
\end{equation}
$\sigma^T > \rho$ contains the following set of inequalities:
\begin{equation}
    \sigma^T > \rho \Leftrightarrow \sum_{s=1}^i m_s > \sum_{s=1}^i \rho_s, \  \forall i=1,...,l_1
    \label{TRHOSIGMA CONdition on partitions}
\end{equation}
According to \eqref{equation2.4TRHOSIGMA}, this condition—originally stated in \eqref{TRHOSIGMA CONdition on partitions}-is equivalent to requiring that the gauge group ranks \( N_i \) are all positive integers. This requirement also leads to \( M_l = 0 \) for all \( l \geq \hat{k} \), meaning there are no hypermultiplets linked to gauge group factors that do not actually exist (i.e., those that are empty). Conditions of \eqref{TRHOSIGMA CONdition on partitions} are essential in order for the quiver theory defined by the triple \( (\rho, \sigma, N) \) to be well-defined. Mirror symmetry acts on $T_\r^\s[SU(N)]$, by exchanging partitions $\r$ and $\s$.
As in the case of $T[SU(N)]$, where we can add contact terms and obtain a $T[U(N)]$ theory, for $T_\r^\s[SU(N)]$ theories we can also consider the contact terms of flavour symmetry and thus obtain a $T_\r^\s[U(N)]$ theory.

\begin{figure}[ ]
    \centering
    \includegraphics[width=0.25\linewidth]{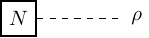}
    \caption{Compact notation for a $T_\r$ theory. The $N$ squared box denotes the $SU(N)$ HB global symmetry of the theory.
    }
    \label{fig:trsdashed}
\end{figure}

In Figure \ref{fig:trsdashed} we introduce the compact notation for 
$T_\r[U(N)]$ theories.

From $T[U(N)]$ theory it's possible to obtain $T_\r^\s[U(N)]$ theories by giving Nilpotent vevs labelled by $\r$ and $\s$ respectively to the CB and HB moment maps:
\begin{equation}
<\mathcal{C}>=J_\s \,, \qquad <\mathcal{H}>=J_\rho 
\end{equation}

with $J_\r$ and $J_\s$ matrices of the type:
\begin{equation}
J_\r=\bigoplus_{i=1}^L\mathcal{J}_{\r_i}=\left( \begin{array}{c|c|c|c}\  \mathcal{J}_{\r_1} \ & \ 0_{\r_1\times \r_2} \ &  \cdots \ & \ 0_{\r_{1}\times \r_L}\\ \hline
\ 0_{\r_2\times \r_1} \ & \  \mathcal{J}_{\r_2} \ &   \cdots \ & \ 0_{\r_{2}\times \r_L}\\ \hline
\  \ &\   \ &   \ddots \ &  \\ \hline
\ 0_{\r_{L}\times \r_1} & \  0_{\r_{L}\times \r_2} \ &   \cdots \ & \ \mathcal{J}_{\r_L}\\
\end{array}\right),\qquad \mathcal{J}_{\r_i}= \underbrace{ \left(\begin{array}{ccccc} 0 &1& \ldots & \ldots & 0 \\ 0 & 0 & 1 & \ldots  & 0  \\ \vdots & \vdots & &  & \vdots  \\ 0 & 0 & \ldots & 0 & 0 \end{array}\right) }_{\r_i} \, .
\label{jordanmatrices}
\end{equation}

%
We can also regard the VEVs as being dynamically generated
as discussed in \cite{Hwang:2020wpd}.
For example the flow to $T_\rho[U(N)]$ can be studied by starting with the $T[U(N)]$ theory and flipping the CB moment map by coupling a chiral field $\Phi_U$, transforming in the adjoint of $SU(N)$, to the Coulomb branch moment map $\mathcal{C}$ of the $T[U(N)]$ tail. We then turn on linearly in the superpotentials some of the components of $\Phi_U$:
\begin{equation}
    \mathcal{W} = Tr[\mathcal{J}_{[\rho]}\Phi_U] + Tr[\Phi_U \, \mathcal{C}] ',,
\end{equation}
where $\mathcal{J}_{[\rho]}$ is a Jordan block
associated to the partition $\rho$ of $N$ (of the form reported in \eqref{jordanmatrices}). We then see from the EOM that  the effect if this deformation is to turn a nilpotent VEV for $\mathcal{C}$:
\begin{align}
    \langle \, \mathcal{C} \, \rangle \sim \mathcal{J}_{[\rho]} \,.
\end{align}

In all our examples the constraints from the set of delta functions
$\dd{\rm{Nil}}$ produced by the algorithm 
as those appearing in 
\eqref{eq:tsuntrho_delta} and 
\eqref{eq: stsh_distrib_gen} can be regarded as indicating that some of the component of the $\Phi_U$ have $R$-charge 2 and then enter the superpotential.

The VEV then triggers a flow resulting in the Higgsing of the $T[U(N)]$ theory as:
\begin{align}
    T[U(N)] \, + \, \Phi_U \quad \text{w/} \, \mathcal{W} = Tr[\mathcal{J}_{[\rho]}\Phi_U] + Tr[\Phi_U \, \mathcal{C}] \quad \rightarrow \quad T_{\rho}[U(N)] \, + \, \text{Free sector} \,.
\end{align}
At the level of partition function we write:
\begin{equation}\label{eq:25061}
    \text{Adj}^{CB}(\vec U,m) \PFT{}{N}(-\vec{U},\vec{V},m)\vert_{\dd{\rm{Nil}}}=
    \PFT{\r}{N}(-\vec{U}_\rho,\vec{V},m)\times \text{Ch}_{\text{Comm}} \,.
\end{equation}
The free sector is encoded in the contribution of a collection of chiral multiplets,
we label their contribution to the partition function as $\text{Ch}_{\text{Comm}}$.
To isolate the interacting $T_\r^\s[U(N)]$ theory we 
can remove them by introducing an $N\times N$ traceless matrix of flipping singlets
$\mathcal{F}_\rho$  whose transpose commutes with $\mathcal{J}_\rho$ and are coupled to $\Phi_U$ so that the complete superpotential reads:
\begin{equation}
    \delta\mathcal{W}=Tr[(\mathcal{J}_{[\rho]} + \mathcal{F}_{[\rho]}) \Phi_U] \,.
\end{equation}

In our discussion we do not introduce these flippers
since the singlets produced in the Higgsing will automatically acquire mass at the end of the RG flow.
So rewrite \eqref{eq:25061} as:
\begin{equation}\label{eq: TtoTr_ch}
    \PFT{}{N}(-\vec{U},\vec{V},m)\vert_{\dd{\rm{Nil}}}=
    \PFT{\r}{N}(-\vec{U}_\rho,\vec{V},m) \times \frac{\text{Ch}_{\text{Comm}}}{\text{Adj}^{CB}(\vec U,m) } \equiv 
    \PFT{\r}{N}(-\vec{U}_\rho,\vec{V},m) \times \text{Ch}^\r_{\text{Nil}}
\end{equation}
The term ``$\text{Ch}^\r_{\text{Nil}}$" is precisely that entering in \eqref{eq: stsh_distrib_gen}.


In Section \ref{app:higgsing_TSUN} we provide the explicit derivation of this result in the case of $T[U(4)]\rightarrow T_{[2,1^2]}[U(4)]$.

\section{Higgsing \texorpdfstring{$T[U(4)]$}{T[SU(N)]} to \texorpdfstring{$T_{[2,1^2]}[U(4)]$}{Trho[SU(N)]}}
\label{app:higgsing_TSUN}

In this section we show the explicit analysis of  the the Nilpotent VEV that appears in Section \ref{subsubsec:su4_2minimal} set by $\dd{\rm{Nil}}$ of \eqref{eq: su4ex_ch}.
We start from the partition function of $T[U(4)]$, with parameters $-\vec{U}$ and $\vec{X}$ for the HB and CB moment maps as in the example in the main text, but we send for simplicity $\vec{U}\rightarrow -\vec{U}$ during computations, so we also invert the condition set by \eqref{eq: su4ex_ch}. 
Thus, we set:
\begin{equation}\label{CondNilVEVB}
   \begin{cases}
  U_1 = W_2+Z_2+m-\frac{i Q}{2} \\
  U_2 = W_2+Z_2-m+\frac{i Q}{2}
\end{cases} 
\end{equation}
We report the partition function of $T[U(4)](\vec{U},\vec{X};m)$:
\begin{equation}\label{eqAfirst}
    \begin{split}
        \PFS{S}=&\PFS{T^4}(\vec{U},\vec{X};m)= \\&=\intd y_1 \D_{1}^{HB}(y_1;m)e^{2\pi i(U_1-U_2)y_1}\intd \vec{y_2}\D_{2}^{HB}(\vec{y_2};m)e^{2\pi i(U_2-U_3)\sum_{i=1}^2y_2^{(i)}}\\&\intd \vec{y_3}\D_{3}^{HB}(\vec{y_3};m)e^{2\pi i(U_3-U_4)\sum_{j=1}^3y_3^{(j)}}
        \prod _{i=1}^2 [s_b(\frac{iQ}{2}-m\pm (y_1-y_2^{(i)}))]\\&\prod _{i=1}^2\prod _{j=1}^3 [s_b(\frac{iQ}{2}-m\pm (y_2^{(i)}-y_3^{(j)}))]
        \prod _{j=1}^3 \prod _{k=1}^4[s_b(\frac{iQ}{2}-m\pm (y_3^{(j)}-X_j))]\\&e^{2\pi iU_4\sum_{k=1}^4X_k}
    \end{split}
\end{equation}
Since it's technically difficult to give VEVs to monopoles and it's easier to give VEVs to mesons, we mirror $T[U(4)](\vec{U},\vec{X};m)$ obtaining $T[U(4)](\vec{X},\vec{U};m)$. At the level of the partition function it means using the equality:
\begin{equation}
    \PFS{T^4}(\vec{U},\vec{X};m)=\PFS{T^4}(\vec{X},\vec{U};\frac{iQ}{2}-m)
\end{equation}
Partition function, reported in  \eqref{eqAfirst}, becomes:
\begin{equation}
    \begin{split}
\PFS{S}=&\intd z_1 \D_{1}^{CB}(z_1;m)e^{2\pi i(X_1-X_2)z_1}\intd \vec{z_2}\D_{2}^{CB}(\vec{z_2};m)e^{2\pi i(X_2-X_3)\sum_{i=1}^2z_2^{(i)}}\intd \vec{z_3}\D_{3}^{CB}(\vec{z_3};m)\\&e^{2\pi i(X_3-X_4)\sum_{j=1}^3z_3^{(j)}}
        \prod _{i=1}^2 [s_b(m\pm (z_1-z_2^{(i)}))]\prod _{i=1}^2\prod _{j=1}^3 [s_b(m\pm (z_2^{(i)}-z_3^{(j)}))]\\&
        \prod _{j=1}^3 \prod _{k=1}^4[s_b(m\pm (z_3^{(j)}-U_j))]e^{2\pi iX_4\sum_{k=1}^4U_k}  
    \end{split}
\end{equation}
Now we impose the condition reported in \eqref{CondNilVEVB} and obtain (labelling $W_2+Z_2\equiv V$):
\begin{equation}
    \begin{split}
\PFS{S}=&\intd z_1 \D_{1}^{CB}(z_1;m)e^{2\pi i(X_1-X_2)z_1}\intd \vec{z_2}\D_{2}^{CB}(\vec{z_2};m)e^{2\pi i(X_2-X_3)\sum_{i=1}^2z_2^{(i)}}\intd \vec{z_3}\D_{3}^{CB}(\vec{z_3};m)\\&e^{2\pi i(X_3-X_4)\sum_{j=1}^3z_3^{(j)}}
        \prod _{i=1}^2 [s_b(m\pm (z_1-z_2^{(i)}))]\prod _{i=1}^2\prod _{j=1}^3 [s_b(m\pm (z_2^{(i)}-z_3^{(j)}))]\\&
        \prod _{j=1}^3 \prod _{k=3}^4[s_b(m\pm (z_3^{(j)}-U_j))] \prod _{j=1}^3[s_b(\frac{iQ}{2}\pm(z_3^{(j)}-V))s_b(2m-iQ\pm (z_3^{(j)}-V))]
        \\&e^{2\pi iX_4(\sum_{k=3}^4U_k+2V)} 
    \end{split}
\end{equation}

A VEV for a meson 
at the level of the partition function appears when two or more poles 
from chirals of R-charge zero,
collide and pinch the integration contour.
We notice indeed that the charges of the meson taking 
In our case, we have:
\begin{equation}
    \intd z_3^{(3)} s_b(\frac{iQ}{2}\pm(z_3^{(3)}-V))
\end{equation}
We see that for $ z_3^{(3)}=V$ both double sine functions are divergent since $s_b(\frac{iQ}{2})=\infty$, and so they pinch the integration countour at that point.
So we set in the integral:
\begin{equation}
    z_3^{(3)}=V
\end{equation}
and obtain (relabelling $z_3^{(1)},z_3^{(2)}$ as $z_{2,1}^{(1)},z_{2,1}^{(2)}$):
\begin{equation}\label{riferimentoeqA}
    \begin{split}
\PFS{S}=&\intd z_1 \D_{1}^{CB}(z_1;m)e^{2\pi i(X_1-X_2)z_1}\intd \vec{z_2}\D_{2}^{CB}(\vec{z_2};m)e^{2\pi i(X_2-X_3)\sum_{i=1}^2z_2^{(i)}}\intd \vec{z_{2,1}}\D_{2}^{CB}(\vec{z_{2,1}};m)\\&e^{2\pi i(X_3-X_4)\sum_{j=1}^2z_{2,1}^{(j)}}
        \prod _{i=1}^2 [s_b(m\pm (z_1-z_2^{(i)}))] \prod _{i=1}^2 [s_b(m\pm (V-z_2^{(i)}))]\\&\prod _{i=1}^2\prod _{j=1}^2 [s_b(m\pm (z_2^{(i)}-z_{2,1}^{(j)}))]
        \prod _{j=1}^2 \prod _{k=3}^4[s_b(m\pm (z_{2,1}^{(j)}-U_j))] \\&
        e^{2\pi iX_4(\sum_{k=3}^4U_k)}e^{2\pi i(X_3+X_4)V} s_b(2m-\frac{iQ}{2})s_b(\frac{iQ}{2})  \prod_{i=3}^4[s_b(-m\pm (V-U_i))]
    \end{split}
\end{equation}
Some contributions have simplified due to the property of the double sine $s_b(x)s_b(-x)=1$.
We recognize the partition function of $T^{[2,1^2]}(\vec{U}^{[2,1^2]},\vec{X};m)$, we mirror it back and obtain $T_{[2,1^2]}(\vec{U}_{[2,1^2]},\vec{X};m)$. 
At the level of the partition function we use the following equality:
\begin{equation}
    \PFS{T^{[2,1^2]}_4}(\vec{X},\vec{U}^{[2,1^2]};\frac{iQ}{2}-m)=\PFT{[2,1^2]}{4}(\vec{U}_{[2,1^2]},\vec{X};m)
\end{equation}
that corresponds to the following equality between integrals:
\begin{equation}\label{eqbetweenintAppendixA}
    \begin{split}
        \intd& z_1 \D_{1}^{CB}(z_1;m)e^{2\pi i(X_1-X_2)z_1}\intd \vec{z_2}\D_{2}^{CB}(\vec{z_2};m)e^{2\pi i(X_2-X_3)\sum_{i=1}^2z_2^{(i)}}\intd \vec{z_{2,1}}\D_{2}^{CB}(\vec{z_{2,1}};m)\\&e^{2\pi i(X_3-X_4)\sum_{j=1}^2z_{2,1}^{(j)}}
        \prod _{i=1}^2 [s_b(m\pm (z_1-z_2^{(i)}))] \prod _{i=1}^2 [s_b(m\pm (V-z_2^{(i)}))]\\&\prod _{i=1}^2\prod _{j=1}^2 [s_b(m\pm (z_2^{(i)}-z_{2,1}^{(j)}))]
        \prod _{j=1}^2 \prod _{k=3}^4[s_b(m\pm (z_{2,1}^{(j)}-U_j))] \\&
        e^{2\pi iX_4(\sum_{k=3}^4U_k)}e^{2\pi i(X_3+X_4)V}=\\&
        \intd y_1 \D_{1}^{HB}(y_1;m)e^{2\pi i(U_3-U_4)y_1}     \intd \vec{y_2}\D_{2}^{HB}(\vec{y_2};m)e^{2\pi i(U_4-V)\sum_{i=1}^2y_2^{(i)}}\\&e^{2\pi iV\sum_{j=1}^4X_j}
        \prod _{i=1}^2 [s_b(\frac{iQ}{2}-m\pm (y_1-y_2^{(i)}))] \prod _{i=1}^2\prod _{j=1}^4 [s_b(\frac{iQ}{2}-m\pm (y_2^{(i)}-X_j)]
    \end{split}
\end{equation}
We use the equality for off-shell theories, meaning that we have contact terms for flavour symmetries. This is due to the fact that we isolated a $T[U(N)]$ theory and worked on it, but in our star-shaped quiver the flavour node of $T[U(N)]$ is gauged so we should keep track of contact terms because for the star-shaped quiver they are FI parameters.
Substituting \eqref{eqbetweenintAppendixA} into \eqref{riferimentoeqA} we obtain:
\begin{equation}
    \begin{split}
\PFS{S}=&\intd y_1 \D_{1}^{HB}(y_1;m)e^{2\pi i(U_3-U_4)y_1}     \intd \vec{y_2}\D_{2}^{HB}(\vec{y_2};m)e^{2\pi i(U_4-V)\sum_{i=1}^2y_2^{(i)}}\\&e^{2\pi iV\sum_{j=1}^4X_j}
        \prod _{i=1}^2 [s_b(\frac{iQ}{2}-m\pm (y_1-y_2^{(i)}))] \prod _{i=1}^2\prod _{j=1}^4 [s_b(\frac{iQ}{2}-m\pm (y_2^{(i)}-X_j)]\\&         
         s_b(2m-\frac{iQ}{2})s_b(\frac{iQ}{2})  \prod_{i=3}^4[s_b(-m\pm (V-U_i))]      
    \end{split}
\end{equation}
So the chirals generated by the Nilpotent vev in this example have the following contribution to the partition function:
\begin{equation}
    Ch_{\text{Nil}}^{[2,1^2]} =s_b(2m-\frac{iQ}{2})s_b(\frac{iQ}{2})  \prod_{i=3}^4[s_b(-m\pm (V-U_i))] 
\end{equation}
These singlets are the ones appearing in \eqref{eq: TtoTr_ch}. Indeed this contribution can be obtained also following the prescription in Appendix \ref{app:TSUN_TRHOSIGMA}.

\section{Relation between  4d and 3d bad-ness}\label{subsec: 4d3d_badness}
\label{app:roadmap_4dto3d}

As discussed in Sections \ref{sec:classS_review} and \ref{sec:brokenbad}, there are two seemingly different notions of badness for a class $\mathcal{S}$ theory. The first one is four-dimensional and refers to the violation of the non-negativity of the dimensions of subspaces $V_k$ of the Coulomb branch, as defined \eqref{eq:graded_CBdim}. Whereas the second notion refers to the badness, in the three-dimensional sense, of the star-shaped quiver associated with the theory of class $\cS$. As pointed out in \cite{Gaiotto:2011xs}, these two criteria are essentially equivalent.\footnote{It is intriguing to note that the 4d CB condition maps directly to the balancing condition in the 3d star-shaped quiver, which determines when CB monopole operators in three dimensions fall below the unitarity bound. Given that the star-shaped quiver is the 3d mirror dual of the direct circle compactification of the 4d theory, one might have instead expected a correspondence between the 4d CB condition and a condition on the HB of the mirror.} The aim of this Appendix is to discuss this link in detail, expanding on the analysis of \cite{Gaiotto:2011xs}.

Let us consider a star-shaped quiver with three tails, each associated with an integer partition of the rank, $N$, of the central node. Notice that starting from a quiver with good tails, the only node which is potentially bad is the central one. The key fact observed in \cite{Gaiotto:2011xs} is that there is a simple relation between the excess number\footnote{The excess number of a $U(N)$ node that has $F$ hypermultiplets charged under it is defined to be $F-2N$.} of the central node of the quiver and the Coulomb branch spectrum in 4d. It can be deduced from \eqref{eq:graded_CBdim} that for the relevant cases of spheres ($g=0$), the contribution from each puncture to the level $N$ dimension, $d_N$, is equal to $N$ minus the first element of the corresponding partition of $N$. However, since this is also just the rank of the first node in the corresponding tail, we have that the total contribution from the punctures to $d_N$ is equal to the number of flavours for the central node in the star-shaped quiver. Thus, given that the excess number is equal to the number of flavours minus twice the rank, we find that the excess number equals $d_N-1$, as defined in \eqref{eq:graded_CBdim}.

Therefore, the central node of the quiver is good if and only if $d_N$ is strictly positive. On the other hand, if the central node is bad in the 3d sense, {\it i.e.}, its excess number is smaller than $-1$, then the CB subspace dimension $d_N$ is negative, and therefore the theory is bad in the 4d sense as well. Instead, if the central node is ugly, then $d_N=0$, which implies that the spectrum of the 4d theory does not contain any Coulomb branch operators with dimension $N$.

In the ugly case, we can dualise the central ugly node by exploiting the duality from \cite{Gaiotto:2008ak}, reducing its rank to $N-1$. Upon such a replacement of the central node, the tails in the star-shaped quiver may now become ugly, in which case we should re-iteratively dualise them as well. This required chain of dualisation required to make each tail good can easily be described at the level of the corresponding partition of $N$. The dualisation of the central node corresponds to decreasing the first element of the corresponding partition by 1, which may result in the resulting sequence of integers no longer being an ordered partition. Whenever this happens, we should simply reshuffle the sequence until we get an ordered partition of $N-1$. As shown in \cite{Giacomelli:2024laq}, every time we transpose two elements of the sequence of integers, we are effectively dualising an ugly node in the quiver (still only using the duality from \cite{Gaiotto:2008ak}).

It is easy to see that the Young tableaux describing the resulting partition of $N-1$ can be obtained from the initial partition of $N$ by removing the last box from the last row. Overall, once we have dualised the central node and all the ugly nodes in the tails, we end up with a star-shaped quiver with $N-1$ in the center and tails described by Young tableaux corresponding to the initial ones with the last box removed. By iterating the earlier argument, we find that the excess number of the central node $U(N-1)$ of this new star-shaped quiver is equal to $d_{N-1}-1$ for the new theory of class $\cS$ of type $\mf{su}(N-1)$. However, the latter is in turn equal to the quantity $d_{N-1}-1$ of the original star-shaped quiver due to the relation between the original and the final Young tableaux!

By iterating the argument so far, we conclude that the star-shaped quiver is bad in the 3d sense only if the corresponding theory of class $\mathcal{S}$ has a Coulomb branch with a negative subspace contribution $d_k$, for some $k$, followed by vanishing contributions for all higher levels. Such a theory is therefore bad from the 4d standpoint as well. 

The implication goes the other way as well, although it does not directly follow from what has been said so far. It can happen that the 4d Coulomb branch spectrum contains a positive contribution for some $k$ (such that the contributions at higher levels all vanish), but there is still a negative contribution for some $k'<k$. This would correspond to a locally good/ugly star-shaped quiver from the 3d standpoint, but to a bad theory according to 4d criterion. It turns out that this happens only for those theories of class $\cS$ whose 3d mirror is described by $n>1$ times an affine Dynkin diagram (which describes the moduli space of $n$ instantons of corresponding Dynkin type). These quivers are locally good (all nodes are balanced), but are known to be globally bad, since they contain monopole operators with vanishing $R$ charge \cite{Gaiotto:2012uq}. Therefore, we conclude that the badness criterion for the star-shaped quivers based on the monopole operators violating the unitary bound in three dimensions, and the four-dimensional criterion using apparent negativity of Coulomb branch subspace dimensions for the corresponding theories of class $\cS$ are equivalent.

\section{Criterion for a single interacting frame}
\label{app:criterion_interacting}

In this work, we have introduced a new family of bad theories, broken theories, and have studied them as they appear in the context of class $\mathcal{S}$. These are bad star-shaped quivers with at least one full tail, which we have conjectured to have a single frame up to permutations (see Subsection \ref{subsec:brokenbad_SSB}).

In this section, we introduce a criterion to determine whether such broken star-shaped quivers are interacting, meaning they do not reduce in the IR to a collection of free hypermultiplets. More generally, we will provide a necessary and sufficient condition for a bad star-shaped theory to have at least one interacting frame. The criterion is the following:\newline

{\it A bad star-shaped quiver has at least one interacting frame if and only if it contains an affine Dynkin diagram.}\newline

We will now prove this in the following two subsections.

\subsection{Bad star-shaped quivers with interacting frames}

First of all, notice that the interacting frame, if it exists, is described by a good quiver obtained from the initial quiver by locally dualising every bad/ugly node, where each local dualization has the effect of reducing the rank of the node. We will now show that if a locally bad quiver contains an affine Dynkin diagram, then it has at least one good frame obtained via the local dualization, such that the good describing the frame contains the same affine Dynkin diagram.

It is easiest to explain what we mean by ``contain" with an example. Let us consider the following quiver:
\begin{equation}\label{exx1}
\begin{tikzpicture}
 \filldraw[fill= white] (0,0) circle [radius=0.1] node[below] {\scriptsize $N_1$};
\filldraw[fill= white] (1,0) circle [radius=0.1] node[below] {\scriptsize $N_2$};
\filldraw[fill= white] (2,0) circle [radius=0.1] node[below] {\scriptsize $N_3$}; 
 \filldraw[fill= white] (3,0) circle [radius=0.1] node[below] {\scriptsize $N_4$};
\filldraw[fill= white] (4,0) circle [radius=0.1] node[below] {\scriptsize $N_5$};
\filldraw[fill= white] (5,0) circle [radius=0.1] node[below] {\scriptsize $N_6$}; 
 \filldraw[fill= white] (6,0) circle [radius=0.1] node[below] {\scriptsize $N_7$};
\filldraw[fill= white] (4,1) circle [radius=0.1] node[left] {\scriptsize $N_8$}; 
\filldraw[fill= white] (4,2) circle [radius=0.1] node[left] {\scriptsize $N_9$};
\draw [thick] (0.1,0)--(0.9,0);
\draw [thick] (1.1,0)--(1.9,0);
\draw [thick] (2.1,0)--(2.9,0);
\draw [thick] (3.1,0)--(3.9,0);
\draw [thick] (4.1,0)--(4.9,0);
\draw [thick] (5.1,0)--(5.9,0);
\draw [thick] (4,0.1)--(4,0.9);
\draw [thick] (4,1.1)--(4,1.9);
\end{tikzpicture}
\end{equation}
This can be good, ugly or bad. We see that this quiver contains an affine $E_6$ diagram, if the following quiver can be subtracted from it:
\begin{equation}\label{exx2}
\begin{tikzpicture}
 \filldraw[fill= red] (0,0) circle [radius=0.1] node[below] {\scriptsize $0$};
\filldraw[fill= red] (1,0) circle [radius=0.1] node[below] {\scriptsize $0$};
\filldraw[fill= white] (2,0) circle [radius=0.1] node[below] {\scriptsize $1$}; 
 \filldraw[fill= white] (3,0) circle [radius=0.1] node[below] {\scriptsize $2$};
\filldraw[fill= white] (4,0) circle [radius=0.1] node[below] {\scriptsize $3$};
\filldraw[fill= white] (5,0) circle [radius=0.1] node[below] {\scriptsize $2$}; 
 \filldraw[fill= white] (6,0) circle [radius=0.1] node[below] {\scriptsize $1$};
\filldraw[fill= white] (4,1) circle [radius=0.1] node[left] {\scriptsize $2$}; 
\filldraw[fill= white] (4,2) circle [radius=0.1] node[left] {\scriptsize $1$};
\draw [thick] (0.1,0)--(0.9,0);
\draw [thick] (1.1,0)--(1.9,0);
\draw [thick] (2.1,0)--(2.9,0);
\draw [thick] (3.1,0)--(3.9,0);
\draw [thick] (4.1,0)--(4.9,0);
\draw [thick] (5.1,0)--(5.9,0);
\draw [thick] (4,0.1)--(4,0.9);
\draw [thick] (4,1.1)--(4,1.9);
\end{tikzpicture}
\end{equation}
This is the affine Dynkin diagram of $E_6$, decorated with the red nodes to depict a quiver with the same shape as \eqref{exx1}, where the rank of decorating nodes is set to zero. Such a ``subtraction" is possible only if the nodes of \eqref{exx1} have larger rank on the corresponding nodes in \eqref{exx2}, leading to the inequalities:
\begin{equation}\label{ineqex}
N_1,N_2\geq0;\quad N_3,N_7,N_9\geq1;\quad N_4,N_6,N_8\geq2;\quad N_5\geq3.
\end{equation}

The logic now proceeds as follows: assume that starting from a bad star-shaped quiver, we find an interacting frame via the electric dualisation algorithm. We want to show that such an interacting frame is described by a good quiver which contains an affine Dynkin diagram. However, the existence of such a frame directly implies that the original bad quiver contains the same affine Dynkin diagram, simply because the sequence of local dualisations that are part of the electric algorithm (that produces the good quiver describing the interacting frame, from the initial bad quiver) can only decrease the rank of the nodes. Thus, this shows that a necessary condition for a bad quiver to have interacting frames is to contain an affine Dynkin diagram. 

We now prove that this condition is also sufficient, which is the same as proving the existence of an interacting frame if the original bad theory contains an affine Dynkin diagram. To prove the existence, we argue as follows: consider a sequence of local dualisations such that every bad node is replaced by its maximal frame (corresponding to the largest rank after being replaced), {\it i.e.}, we consider the move $N_c\rightarrow \left\lfloor\frac{N_f}{2}
\right\rfloor$. After finitely many dualisations, we will necessarily get to one of the frames, which may either be interacting or free.

Let us focus on a single local dualization. Before the dualisation, the node has rank $\bar{N}$ and $N_f=\sum_j N_j$ flavours, where the sum runs over neighbouring nodes whose rank we denote as $N_j$. Since the quiver contains an affine Dynkin diagram, we know from \eqref{ineqex} that $\bar{N}$ is larger than the rank $\bar{N}'$ of the corresponding node in the decorated affine Dynkin diagram. We now denote with $N_j'$ the rank of its neighbouring nodes in the decorated affine Dynkin diagram. The crucial observation is that nodes in the affine Dynkin diagram are balanced (see \eqref{exx2}) and therefore 
\begin{equation}\label{ineq2}
    \bar{N}'\leq\left\lfloor\frac{N_f'}{2}\right\rfloor\;,\quad \text{where}\quad N_f'=\sum_jN_j' \;.
\end{equation}
As a result, after the dualisation, we find 
\begin{equation}
    \bar{N}\rightarrow \left\lfloor\frac{N_f}{2}\right\rfloor\geq \left\lfloor\frac{N_f'}{2}\right\rfloor\geq\bar{N}'\;,
\end{equation}
where we have used \eqref{ineq2} and the fact that $N_j\geq N_j'$ due to \eqref{ineqex}. Thus, the rank of the node after the dualisation is still larger than the rank of the corresponding node in the decorated affine Dynkin diagram, and therefore the quiver obtained after dualisation still contains the same affine Dynkin diagram. By reiterating this argument for each local dualization move, we conclude that the resulting frame is a good quiver containing an affine Dynkin diagram, and will be shown to be interacting as shown in the next section.

To conclude, let us point out that this does not rule out, in general, the existence of other frames which are free. However, applying this result or line of reasoning to broken theories (which by definition have a single frame), we conclude that the single frame must be interacting. This leads to a necessary and sufficient criterion for a single interacting frame in the case of broken theories: the theory must fit an affine Dynkin diagram. 

\subsection{All good quivers fit an affine Dynkin diagram}
\label{app:interacting_3d}

We will now provide the second part of the proof, showing that all good interacting star-shaped quivers contain an affine Dynkin diagram (for a related analysis, see \cite{Zhong:2024jxw}). Let us begin by focusing on good star-shaped quivers with three tails.

Recall that good star-shaped quivers have all nodes (including the central one) balanced or overbalanced, and as we move along a tail starting from the center, the rank of the nodes strictly decreases. We recall that the integer partition associated with a tail can be determined as follows: the first element of the partition is the difference between the rank of the central node and that of the first node in the tail, the second element of the partition is given by the difference between the ranks of the first two nodes, and so on. All nodes in the tail are good if and only if the sequence constructed in this way is an ordered integer partition of $N$, the rank of the central node, with the first element in the sequence being the largest.

Let us start by characterising quivers which do not satisfy the desired property, {\it i.e.}, do not include any exceptional affine Dynkin diagrams. First of all, we can notice that if all the tails have length two or more, then the quiver necessarily contains the $E_6$ affine Dynkin diagram and therefore immediately satisfies this property. We can therefore concentrate on quivers which have a tail of length one. If this tail is abelian, then the central node of the quiver cannot be good: denoting again by $N$ the rank of the central node, the other two tails can contribute at most $2N-2$ flavours since the first node in each tail has rank at most $N-1$. As a result, the central node will be balanced or overbalanced only if the third tail contributes two flavours or more, meaning that it is not abelian. In summary, all quivers with an abelian tail cannot be good, and indeed, none of the affine exceptional Dynkin diagrams can fit in them. 

From this discussion, we see that we should focus on quivers whose tails are all non-abelian, and one tail has length one. Furthermore, if the other two tails both have length three or more, then we can observe that the $E_7$ affine Dynkin diagram fits in the quiver: the central node must have rank at least 4, the first node in the long tails must have rank at least 3, and so on. As a result, the quiver satisfies the desired property. The only case left to analyse is therefore the case in which there is a tail of length one and another of length two. If the third tail has length five or more, it can be shown that the quiver is not good unless the $E_8$ affine Dynkin diagram fits in the quiver. We will see this momentarily, but before doing so, let us focus on the other option, namely, the third tail has length four or less, and show that the quiver cannot be good.

The unitary theory we are interested in is of the form 
\begin{equation}\label{trinion1}
\begin{tikzpicture}
 \filldraw[fill= white] (0,0) circle [radius=0.1] node[below] {\scriptsize A};
\filldraw[fill= white] (1,0) circle [radius=0.1] node[below] {\scriptsize B};
\filldraw[fill= white] (2,0) circle [radius=0.1] node[below] {\scriptsize C}; 
 \filldraw[fill= white] (3,0) circle [radius=0.1] node[below] {\scriptsize D};
\filldraw[fill= white] (4,0) circle [radius=0.1] node[below] {\scriptsize E};
\filldraw[fill= white] (5,0) circle [radius=0.1] node[below] {\scriptsize G}; 
 \filldraw[fill= white] (6,0) circle [radius=0.1] node[below] {\scriptsize H};
\filldraw[fill= white] (4,1) circle [radius=0.1] node[left] {\scriptsize F}; 
\draw [thick] (0.1,0)--(0.9,0);
\draw [thick] (1.1,0)--(1.9,0);
\draw [thick] (2.1,0)--(2.9,0);
\draw [thick] (3.1,0)--(3.9,0);
\draw [thick] (4.1,0)--(4.9,0);
\draw [thick] (5.1,0)--(5.9,0);
\draw [thick] (4,0.1)--(4,0.9);
\end{tikzpicture}
\end{equation}
where the capital letters denote the rank of the corresponding node. The ``goodness" of node $A$ implies that
$$B=2A+a \;,$$
where $a$ is a non-negative integer. With the same logic we find that goodness of nodes $B$, $C$, and $D$ implies that
$$C=3A+2a+b\;,\quad D=4A+3a+2b+c\;,\quad E=5A+4a+3b+2c+d \;,$$
where $b$, $c$, and $d$ are non negative integers. This in turn implies that $D\leq 4E/5$. Finally, by requiring nodes $F$, $G$, and $H$ to be balanced or overbalanced we find $F\leq E/2$ and $G\leq 2E/3$. Since the total number of flavours seen by the central node is $D+F+G$, from the above inequalities, we find that
$$D+F+G\leq\left(\frac{4}{5}+\frac{1}{2}+\frac{2}{3}\right)E=\frac{59}{30}E \;.$$ 
Therefore, the central node cannot be good since the number of flavours is less than $2E$, and thus we conclude that all quivers of the form \eqref{trinion1} are not locally good. 

Let us now go back to the case in which one of the tails has length 5 or more, namely, the quiver is of the form
\begin{equation}\label{trinion2}
\begin{tikzpicture}
\node at (-2,0) {\dots};
 \filldraw[fill= white] (-1,0) circle [radius=0.1] node[below] {\scriptsize A};
 \filldraw[fill= white] (0,0) circle [radius=0.1] node[below] {\scriptsize B};
\filldraw[fill= white] (1,0) circle [radius=0.1] node[below] {\scriptsize C};
\filldraw[fill= white] (2,0) circle [radius=0.1] node[below] {\scriptsize D}; 
 \filldraw[fill= white] (3,0) circle [radius=0.1] node[below] {\scriptsize E};
\filldraw[fill= white] (4,0) circle [radius=0.1] node[below] {\scriptsize F};
\filldraw[fill= white] (5,0) circle [radius=0.1] node[below] {\scriptsize G}; 
 \filldraw[fill= white] (6,0) circle [radius=0.1] node[below] {\scriptsize H};
\filldraw[fill= white] (4,1) circle [radius=0.1] node[left] {\scriptsize L}; 
\draw [thick] (-1.1,0)--(-1.7,0);
\draw [thick] (-0.1,0)--(-0.9,0);
\draw [thick] (0.1,0)--(0.9,0);
\draw [thick] (1.1,0)--(1.9,0);
\draw [thick] (2.1,0)--(2.9,0);
\draw [thick] (3.1,0)--(3.9,0);
\draw [thick] (4.1,0)--(4.9,0);
\draw [thick] (5.1,0)--(5.9,0);
\draw [thick] (4,0.1)--(4,0.9);
\end{tikzpicture}
\end{equation} 
with $A>0$. Goodness of all the nodes on the left of $F$ then requires $F\geq6$. Furthermore, if the quiver is good, then we conclude $L\geq3$. In order to see this, it is enough to notice that if $L=2$, we must have $G=E=F-1$ for the central node to be good, and this directly implies that either $G$ or $H$ are not good. Furthermore, since $E\leq F-1$ we need $L+G\geq F+1$ for the central node to be good. Noticing that $L\leq F/2$ (otherwise the node $L$ is not good), we conclude that $G\geq F/2+1$. This in turn implies that $H\geq 2$ (otherwise the node $G$ is not good) and $G\geq 4$ (since $F\geq 6$). Overall, the bounds we have derived on $F$, $G$, $H$ and $L$ assuming all nodes to be good imply that the affine $E_8$ Dynkin diagram fits in the quiver (\ref{trinion2}), which is the desired result. 

The above argument shows that all good star-shaped quivers with 3 tails satisfy our property: at least one among the affine exceptional Dynkin diagrams fits in the quiver. For example, we conclude that all quivers of the form \eqref{trinion1} reduce to Wess--Zumino models upon local dualisations, regardless of the value of $A,B,C,D,E,F$, and $G$. We would like to stress that the exceptional Dynkin diagram fitting in the quiver is not necessarily unique; for example, the 3d mirror of the trinion $T_6$ includes all of them. For quivers with four tails or more, the analogous statement is that the affine $D_4$ Dynkin diagram fits in the quiver, which is automatically true. As a result, all star-shaped quivers with 4 tails or more do have interacting frames.

\bibliographystyle{JHEP}
\bibliography{ref}

\end{document}